\pdfoutput=1 %arXiv admin added this line

\documentclass[final, 12pt,oneside]{class_diss}
\usepackage{      amsmath} % American Mathematics Society standards
\usepackage{     graphicx} % Extended graphics package.

\usepackage{epsfig}

\usepackage{amsxtra}
\usepackage{amssymb}
\usepackage{amsthm}
\usepackage{latexsym}
\usepackage{bm}
\usepackage{longtable}
\usepackage{color}
\usepackage{listings}
\definecolor{  Pink}{rgb}{1.0, 0.5, 0.5}
\definecolor{Maroon}{rgb}{0.8, 0.0, 0.0}

\usepackage[sort&compress]{natbib}
\bibpunct{[}{]}{,}{n}{}{}
\usepackage{hypernat}

\usepackage[]{hyperref}
\begin{document}

% +--------------------------------------------------------------------+
% | Masters Students -- You Need to Make Some Changes Here

% | The Abstract Title page and Abstract following the Abstract Title
% | page are required only for doctoral dissertations.  For masters
% | theses or reports, comment out or delete the following 7 lines:
% | \input{abstracttitle.tex} through \end{abstract}.  You will also
% | need to uncomment two lines under "Abstract" below.
% |
% +--------------------------------------------------------------------+

% +--------------------------------------------------------------------+
% | Abstract Title Page
% |
% |This page is required only for doctoral dissertations.
% +--------------------------------------------------------------------+

% +--------------------------------------------------------------------+
% | This page should not contain a page number.  We use the
% | \thispagestyle[empty] command below to suppress page numbers
% | and other style elements.
% +--------------------------------------------------------------------+

\thispagestyle{empty}

% +--------------------------------------------------------------------+
% | The Abstract Title page begins here                                |
% +--------------------------------------------------------------------+

\pdfbookmark[0]{Title Page}{PDFTitlePage}

\begin{center}

   \vspace{1cm}

% +--------------------------------------------------------------------+
% | Enter the title of your ETDR below.  Use ALL CAPITAL LETTERS.
% +--------------------------------------------------------------------+

   \Large CONSTRAINING SCALAR FIELD DARK ENERGY WITH COSMOLOGICAL OBSERVATIONS\\

   \vspace{0.5cm}

   by\\

   \vspace{0.5cm}

% +--------------------------------------------------------------------+
% | Enter your name below in ALL CAPITAL LETTERS.
% +--------------------------------------------------------------------+

   LADO SAMUSHIA\\

   \vspace{0.5cm}

% +--------------------------------------------------------------------+
% | List previous degrees in mixed case. Include the abbreviation for  |
% | the degree, the name of the university, and the year separated by  |
% | commas. For example:                                               |
% |                                                                    |
% |    B.A., University of Illinois, 2000                              |
% |                                                                    |
% | If desired, it is acceptable to include a city or country with     |
% | the university name. For example:                                  |
% |                                                                    |
% |    B.S., Jillian University, China, 2002                           |
% +--------------------------------------------------------------------+

   B.S., Tbilisi State University, Georgia, 2002 \\
   M.Sc., Tbilisi State University and National Abastumani Observatory, Georgia, 2004 \\

   \vspace{0.65cm}
   \rule{2in}{0.5pt}\\
   \vspace{0.85cm}

   {\Large AN ABSTRACT OF A DISSERTATION}\\

   \vspace{0.5cm}
   submitted in partial fulfillment of the\\
   requirements for the degree\\

   \vspace{0.5cm}

% +--------------------------------------------------------------------+
% | On the line below, enter the name of your earned degree in ALL
% | CAPITAL LETTERS.  For example: DOCTOR OF PHILOSOPHY
% +--------------------------------------------------------------------+

   {\Large DOCTOR OF PHILOSOPHY}\\
   \vspace{0.5cm}

% +--------------------------------------------------------------------+
% | On the two lines below, enter the name of your department and the
% | name of the college in mixed case.  For example:
% |
% |     Biochemistry Department
% |     College of Arts and Sciences
% +--------------------------------------------------------------------+

   Department of Physics\\
   College of Arts and Sciences\\

   \vspace{0.5cm}
   {\Large KANSAS STATE UNIVERSITY}\\
   Manhattan, Kansas\\

% +--------------------------------------------------------------------+
% | On the line below, enter the year of your graduation.  The year
% | should be the only text on the line.  For example:
% |
% |     2006
% +--------------------------------------------------------------------+

   2009\\
   \vspace{1cm}

\end{center}

\begin{abstract}
   \setcounter{page}{-1}
   \pdfbookmark[0]{Abstract}{PDFAbstractPage}
   % +--------------------------------------------------------------------+
% | Abstract Page
% +--------------------------------------------------------------------+

\pagestyle{empty}
\vspace{1cm}
\setlength{\baselineskip}{0.8cm}

%\indent

% +--------------------------------------------------------------------+
% | Enter the text of your abstract below, maximum of 350 words.
% +--------------------------------------------------------------------+

High precision cosmological observations in last decade suggest that about 70\% of our universe's energy density is in so called ``Dark Energy'' (DE). Observations show that DE has negative effective pressure and therefore unlike conventional energy sources accelerates the cosmic expansion instead of decelerating it. DE is highly uniform and has become a dominant component only recently. 

The simplest candidate for DE is the time-independent cosmological constant $\Lambda$. Although successful in fitting available data, the cosmological constant model has a number of theoretical shortcomings and because of that alternative models of DE are considered. In one such scenario a cosmological scalar field that slowly rolls down its potential acts like a time-dependent cosmological constant. 

I have used different independent cosmological data sets to constrain the time dependence of DE's energy density in the framework of the slowly-rolling cosmological scalar field model. Present data favors a time-independent cosmological constant, but the time-dependent DE can not be ruled out at high confidence level. Ongoing and planned cosmological probes and surveys will provide more and better quality data over the next decade. When the new data sets are available we will be able to either detect the time dependence of DE or constrain it to a very small physically uninteresting value.  

   \vfill
\end{abstract}

% +--------------------------------------------------------------------+
% | Title Page -- Required for both Doctoral and Masters Students
% +--------------------------------------------------------------------+

% +--------------------------------------------------------------------+
% | Title Page
% +--------------------------------------------------------------------+

\newpage

% +--------------------------------------------------------------------+
% | This page should not contain a page number.  We use the
% | \thispagestyle[empty] command below to suppress page numbers
% | and other style elements.
% +--------------------------------------------------------------------+

\thispagestyle{empty}

% +--------------------------------------------------------------------+
% | The Title page begins here.
% +--------------------------------------------------------------------+

\begin{center}

   \vspace{1cm}

% +--------------------------------------------------------------------+
% | On the line below, replace "ENTER YOUR TITLE" with the title of
% | your ETDR.  Use all CAPITAL LETTERS.
% +--------------------------------------------------------------------+

   \Large CONSTRAINING SCALAR FIELD DARK ENERGY WITH COSMOLOGICAL OBSERVATIONS\\

   \vspace{0.5cm}

   by\\

   \vspace{0.5cm}

% +--------------------------------------------------------------------+
% | On the line below, replace "ENTER YOUR NAME" with your name.  Use
% | mixed case, for example, Laura Bush.
% +--------------------------------------------------------------------+

   LADO SAMUSHIA\\

   \vspace{0.5cm}

% +--------------------------------------------------------------------+
% | On the line below, replace List"Enter Your Previous Degrees"
% | with your previous degrees in mixed case. Include the abbreviation
% | for the degree, the name of the university, and the year
% | separated by commas. For example:                                  |
% |                                                                    |
% |    B.A., University of Illinois, 2000                              |
% |                                                                    |
% | If desired, it is acceptable to include a city or country with     |
% | the university name. For example:                                  |
% |                                                                    |
% |    B.S., Jillian University, China, 2002                           |
% +--------------------------------------------------------------------+

   B.S., Tbilisi State University, Georgia, 2002 \\
   M.Sc., Tbilisi State Unviersity and National Abastumani Observatory, Georgia, 2004 \\

   \vspace{0.65cm}
   \rule{2in}{0.5pt}\\
   \vspace{0.85cm}

   {\Large A DISSERTATION}\\

   \vspace{0.5cm}
   submitted in partial fulfillment of the\\
   requirements for the degree\\

   \vspace{0.5cm}

% +--------------------------------------------------------------------+
% | On the line below, replace "ENTER YOUR DEGREE NAME" with the name
% | of your earned degree in ALL CAPITAL LETTERS.
% +--------------------------------------------------------------------+

   {\Large DOCTOR OF PHILOSOPHY}\\
   \vspace{0.5cm}

% +--------------------------------------------------------------------+
% | On the two lines below, replace "Enter Your Department Name" and
% | "Enter Your College Name" with the name of your department and the
% | name of the college in mixed case.  For example:
% |
% |     Biochemistry Department
% |     College of Arts and Sciences
% +--------------------------------------------------------------------+

   Department of Physics\\
   College of Arts and Sciences\\

   \vspace{0.5cm}
   {\Large KANSAS STATE UNIVERSITY}\\
   Manhattan, Kansas\\

% +--------------------------------------------------------------------+
% | On the line below, replace "Graduation Year" with the year of
% | your graduation.  The year should be the only text on the line.
% | For example:
% |
% |     2006
% +--------------------------------------------------------------------+

   2009\\
   \vspace{1cm}

\end{center}

\begin{flushleft}
   \hspace{10cm}Approved by:\\
   \vspace{ 1cm}
   \hspace{10cm}Major Professor\\

% +--------------------------------------------------------------------+
% | On the line below, replace "Enter Your Major Professor's Name"
% | with  the name of your major professor in mixed case.  Use the
% | format Firstname Lastname.  For example:
% |
% |     Lori Goetsch
% |
% +--------------------------------------------------------------------+

   \hspace{10cm}Bharat Ratra\\
\end{flushleft}

% +--------------------------------------------------------------------+
% | Use the section below if you have co-major professors.
% +--------------------------------------------------------------------+

%\begin{flushleft}
%   \hspace{10cm}Approved by:\\
%   \vspace{ 1cm}
%   \hspace{10cm}Co-Major Professor\\
%   \hspace{10cm}Enter Your Co-Major Professor's Name\\
%   \vspace{.5cm}Co-Major Professor\\
%   \hspace{10cm}Enter Your Co-Major Professor's Name\\
%\end{flushleft}

% +--------------------------------------------------------------------+
% | Copyright Page -- Required for both Doctoral and Masters Students
% +--------------------------------------------------------------------+

%\input{copyright.tex}

% +--------------------------------------------------------------------+
% |  Abstract -- Required for both Doctoral and Masters Students
% +--------------------------------------------------------------------+

\begin{abstract}

% +--------------------------------------------------------------------+
% | For masters theses or reports, uncomment the following commands:
% +--------------------------------------------------------------------+

   %\setcounter{page}{-1}
   %\pdfbookmark[0]{Abstract}{PDFAbstractPage}

    % +--------------------------------------------------------------------+
% | Abstract Page
% +--------------------------------------------------------------------+

\pagestyle{empty}
\vspace{1cm}
\setlength{\baselineskip}{0.8cm}

%\indent

% +--------------------------------------------------------------------+
% | Enter the text of your abstract below, maximum of 350 words.
% +--------------------------------------------------------------------+

High precision cosmological observations in last decade suggest that about 70\% of our universe's energy density is in so called ``Dark Energy'' (DE). Observations show that DE has negative effective pressure and therefore unlike conventional energy sources accelerates the cosmic expansion instead of decelerating it. DE is highly uniform and has become a dominant component only recently. 

The simplest candidate for DE is the time-independent cosmological constant $\Lambda$. Although successful in fitting available data, the cosmological constant model has a number of theoretical shortcomings and because of that alternative models of DE are considered. In one such scenario a cosmological scalar field that slowly rolls down its potential acts like a time-dependent cosmological constant. 

I have used different independent cosmological data sets to constrain the time dependence of DE's energy density in the framework of the slowly-rolling cosmological scalar field model. Present data favors a time-independent cosmological constant, but the time-dependent DE can not be ruled out at high confidence level. Ongoing and planned cosmological probes and surveys will provide more and better quality data over the next decade. When the new data sets are available we will be able to either detect the time dependence of DE or constrain it to a very small physically uninteresting value.  

    \vfill

\end{abstract}

% +--------------------------------------------------------------------+
% | We use the following code to suppress page numbers and other
% | style issues we do not want present on a given page.               |
% +--------------------------------------------------------------------+

%\thispagestyle{empty} Looks like it's ok to remove this line
\newpage
\pagenumbering{roman}

% +--------------------------------------------------------------------+
% | On the line below, set the number to represent the page number of
% | the Table of Contents page.  For example, if the Table of Contents
% | page is the 8th page of your document, enter 8 in the brackets.  This
% | number may vary, depending on the length of your abstract.
% |
% | Numbers do not appear on the title and abstract pages, but they are
% | included in the page count.  The Table of Contents page is the
% | first page on which page numbers are displayed.
% +--------------------------------------------------------------------+

\setcounter{page}{5}

% +--------------------------------------------------------------------+
% | Here, we will generate our Table of Contents (TOC) entries.        |
% | This adds the section to the TOC and then generates the indicated  |
% | section.                                                           |
% +--------------------------------------------------------------------+

\phantomsection
\addcontentsline{toc}{chapter}{Table of Contents}

\tableofcontents
\listoffigures
\listoftables

%\hfill  Are these lines necessary?
%\hfill

% +--------------------------------------------------------------------+
% | Acknowledgements Page
% |
% | If you choose not to have an Acknowledgements page, comment out
% | or delete the following 3 lines.
% +--------------------------------------------------------------------+

% +--------------------------------------------------------------------+
% | Acknowledgements Page (Optional)                                   |
% +--------------------------------------------------------------------+

\newpage
\begin{center}
{\bf \Huge Acknowledgments}
\end{center}
\vspace{1cm}
\setlength{\baselineskip}{0.8cm}

%\pdfbookmark[0]{Acknowledgements}{PDF_Acknowledgements}

% +--------------------------------------------------------------------+
% | Enter text for your acknowledgements in the space below this box.  |
% |                                                                    |
% +--------------------------------------------------------------------+

I am grateful to so many people for their help and support while writing this thesis and during my entire graduate studies. I am indebted to my research adviser Professor Bharat Vishnu Ratra who was supervising my research activities. Thank you for your help, guidance and patience, It has been my great honor to have worked with you. Special thanks to Professor Tina Kanhiashvili, its mainly due to her suggestion and help that I joined Kansas State University as a graduate student. Whenever I had a question in physics or mathematics I knew that I could always talk to Professor Larry Weaver, he taught me more than I could read in any reference book on physics. Many thanks to Professors Tim Bolton and Glenn Horton-Smith whom I bugged whenever I had a question about statistics. Professors Victor Turchin and David Aukley were always willing to help and answer my rookie mathematical questions.
 I am very grateful to Dale and Maureen Herspring for hospitality during my first weeks in Manhattan. Special thanks Brad and Susie Shaw, your beautiful friendship made us feel more like home. My dear friends and fellow graduate students from physics department, I will always remember wonderful times spent with you. And of course, many many thanks to my wife Nino and daughter Nutsa, you made my life beautiful.

%\phantomsection
%\addcontentsline{toc}{chapter}{Acknowledgements}

% +--------------------------------------------------------------------+
% | Dedication Page
% |
% | If you choose not to have a Dedication page, comment out
% | or delete the following 3 lines.
% +--------------------------------------------------------------------+

% +--------------------------------------------------------------------+
% | Dedication Page (Optional)
% +--------------------------------------------------------------------+

\newpage
\begin{center}
{\bf \Huge Dedication}
\end{center}
\vspace{1cm}
\setlength{\baselineskip}{0.8cm}

%\pdfbookmark[0]{Dedication}{PDF_Dedication}

% +--------------------------------------------------------------------+
% | Enter the text for your dedication in the space below this box.
% +--------------------------------------------------------------------+

To the memory of my loving mother.

%\phantomsection
%\addcontentsline{toc}{chapter}{Dedication}

% +--------------------------------------------------------------------+
% | Preface Page
% +--------------------------------------------------------------------+

%\input{preface.tex}
%\phantomsection
%\addcontentsline{toc}{chapter}{Preface}

% +--------------------------------------------------------------------+
% | We use arabic (1, 2, 3...) page numbering starting from page 1.    |
% | Note, however, that there are many pages where this is not the     |
% | desired behavior - such as the Title page, or abstract.  In these  |
% | cases, we can use \thispagestyle{empty} to suppress page numbers,  |
% | and other general style issues that we've defined globally.        |
% +--------------------------------------------------------------------+

\newpage
\pagenumbering{arabic}
\setcounter{page}{1}

% +--------------------------------------------------------------------+
% | Here is where we include individual sections of the thesis or
% | dissertation.                                                      |
% +--------------------------------------------------------------------+

% +--------------------------------------------------------------------+
% | Chapters
% +--------------------------------------------------------------------+

% +--------------------------------------------------------------------+
% | Sample Chapter
% |
% | This file provides examples of how to
% | - insert a figure with a caption
% | - construct a table with a caption
% | - create subsections within the chapter
% | - insert a reference to a Figure or Table
% | - make a citation
% +--------------------------------------------------------------------+

\cleardoublepage

% +--------------------------------------------------------------------+
% | Replace "Chapter Title" below with the title of your chapter.  LaTeX
% | will automatically number the chapters.
% +--------------------------------------------------------------------+

\chapter{Introduction}
\label{sec:introduction}
\section{Short overview of relevant topics from theoretical cosmology}

\begin{figure}[htb]%t=top, b=bottom, h=here
\begin{center}
    \includegraphics[height=2.5in]{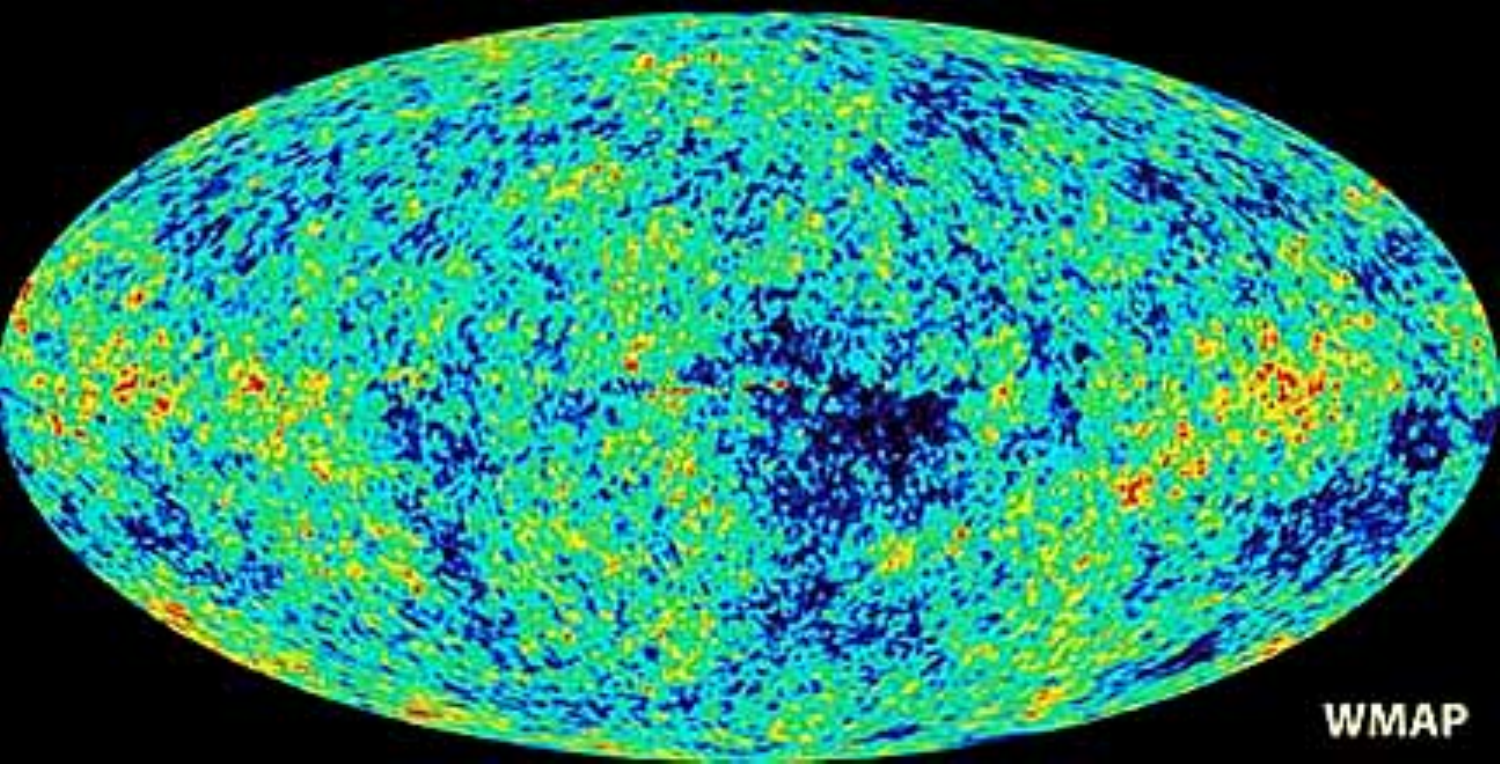}

    \caption[Map of cosmic microwave background radiation]{Map of cosmic microwave background anisotropies as measured by the Wilkinson microwave anisotropy probe (WMAP). Red areas correspond to hotter spots on the sky, while blue areas correspond to colder than average spots. The amplitude of fluctuations relative to the background is less than $10^{-4}$. Source: http://map.gsfc.nasa.gov/.}

    \label{fig:wmapiso}
    \end{center}
\end{figure}

Observations suggest that our Universe started evolving from an extremely hot
and dense state  $13$ to $14\ \rm Gyr$ ago. Since light travels at finite
speed, the distance to which we can see is fundamentally limited. The
observable Universe today has a Hubble radius of about $4000\ \rm Mpc$,
beyond which we can not see. Astronomical observations show that, on average, our Universe is homogeneous and isotropic on large scales.
Observational evidence for isotropy includes the measurement of cosmic microwave background radiation anisotropies \cite{Hinshaw:2008kr}, shown in Fig.~\ref{fig:wmapiso}, that is very isotropic.

Large-scale structure observations confirm that the galaxy distribution
becomes homogeneous beyond about $100\ \rm Mpc$, but even on the largest observed scales inhomogeneities in galaxy density are on the order of a few percent \cite{Hogg:2004vw}. 

Homogeneous and isotropic four dimensional space-times are described by the metric
\begin{equation}
\label{eq:metric}
ds^2=g_{\mu\nu}dx^{\mu}dx^{\nu}=-dt^2+a(t)\gamma_{ij}dx^idx^j,
\end{equation}
\noindent
where $t$ is cosmic time and $\gamma_{ij}$ is the metric of the three-space which can be flat, spherical, or hyperbolic. $a(t)$ is called a scale factor and 
describes how the spatial separation between two noninteracting test particles changes with time. In an expanding Universe the scale factor is an 
increasing function of cosmic time.

In general relativity (GR), the metric obeys Einstein's equations
\begin{equation}
\label{eq:einstein}
R_{\mu\nu}+\frac{1}{2}g_{\mu\nu}R^\lambda_\lambda = \frac{8\pi G}{3}T_{\mu\nu},
\end{equation}
\noindent
where $G$ is Newton's gravitational constant, $R_{\mu\nu}$ is the Ricci tensor,
\begin{equation}
\label{eq:ricci}
R_{\mu\nu}=\frac{\partial\Gamma^\lambda_{\mu\nu}}{\partial x^\lambda} - \frac{\partial\Gamma^\lambda_{\lambda\mu}}{\partial x^\nu} + \Gamma^\lambda_{\lambda\kappa}\Gamma^\kappa_{\mu\nu} - \Gamma^\lambda_{\mu\kappa}\Gamma^\kappa_{\lambda\nu},
\end{equation}
\noindent
and the Christoffel symbols $\Gamma^\lambda_{\mu\nu}$ are given by
\begin{equation}
\label{eq;christoffel}
\Gamma^{\lambda}_{\mu\nu}=\frac{1}{2} g^{\lambda\kappa}\left( \frac{\partial g_{\kappa\mu}}{\partial x^\nu} + \frac{\partial g_{\kappa\nu}}{\partial x^\mu} - \frac{\partial g_{\mu\nu}}{\partial x^\kappa}\right).
\end{equation}
$T_{\mu\nu}$ in Eq.~\eqref{eq:einstein} is the energy-momentum tensor of the
matter components that fill space. The Ricci tensor satisfies the Bianchi identity
$R^{\mu}_{\nu;\mu}-1/2R^\mu_{\mu;\nu}=0$, where the semicolon denotes a covariant derivative, and consequently the energy-momentum tensor satisfies the energy conservation equation
\begin{equation}
\label{eq:econserv}
T^{\mu}_{\nu;\mu}=0.
\end{equation}

\begin{figure}[htb]%t=top, b=bottom, h=here
\begin{center}
    \includegraphics[height=2.5in]{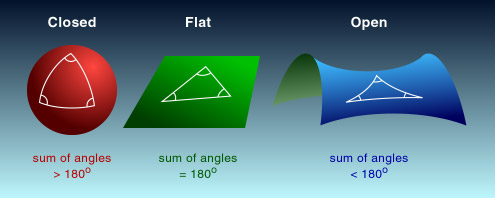}

    \caption[Two-dimensional flat, open, and closed spaces]{Examples of 2 dimensional flat, closed, and open spaces. Source: http://www.lcsd.gov.hk/.}

    \label{fig:curvature}
    \end{center}
\end{figure}

Three-dimensional homogeneous and
isotropic space can be flat, positively curved (surface of a 3-dimensional
``sphere''), or negatively curved (surface of a 3-dimensional ``saddle'');
two dimensional analogs are shown in Fig.~\ref{fig:curvature}.

The metric $\gamma_{ij}$ of homogeneous and isotropic space can be written in Cartesian coordinates as
\begin{equation}
\label{eq:spmetric}
\gamma_{ij}dx^idx^j=\frac{\delta_{ij}dx^idx^j}{\left(1+\frac{1}{4}K\rho^2\right)^2},
\end{equation}
\noindent
and in spherical coordinates as
\begin{equation}
\label{eq:spmetricsp}
\gamma_{ij}=dr^2+\chi^2(r)(d\theta^2+\sin^2(\theta)d\phi^2).
\end{equation}
\noindent
Here $\delta_{ij}$ is the Kroneker delta function and
$\rho^2=\displaystyle\sum{\delta_{ij}x^ix^j}$. $K$ describes the geometry of three-space. For $K<0$ 
the space is negatively curved (open), for $K>0$ the space is positively curved (closed), and for $K=0$ the space is flat. The variable $\chi$ is given by
\begin{equation}
\label{eq:chir}
\chi(r)=\begin{cases}
        \frac{1}{\sqrt{K}}\sin{\sqrt{K}r} &\text{\ if\ } K>0,\\
        r &\text{\ if \ } K=0,\\
        \frac{1}{\sqrt{-K}}\sinh{\sqrt{-K}r} &\text{\ if \ } K<0.
        \end{cases}
\end{equation}
\noindent
The numerical value of $K$ depends on the normalization of the scale factor. In open and closed Universes $a$ is usually normalized in such a way that $K=\pm1$.
In the spatially-flat Universe the scale factor is often normalized so that it's current value $a_0$ is equal to $1$.

If the metric is given by Eqs.~(\ref{eq:metric}) and (\ref{eq:spmetricsp}) than Eq.~\eqref{eq:einstein} forces the energy-momentum tensor to
have the form
\begin{equation}
\label{eq:emdiagonal}
T_{\mu\nu}=\left(\begin{array}{cccc}
           \rho(t) g_{00} & 0 & 0 & 0 \\
           0 &  &  &  \\
           0 &  & P(t) g_{ij} &  \\
           0 &  &  &  \\
           \end{array}
           \right)
\end{equation}
\noindent
where $\rho$ is the mass energy density of matter and $P$ is pressure. 

For barotropic fluids pressure depends only on the local
value of energy density, $P=P(\rho)$. Very often the equation of state $P=\omega\rho$ is used, where $\omega$ is called the equation of state
parameter. For an ultrarelativistic gas (radiation) $\omega_{\rm r}=1/3$, for nonrelativistic matter $\omega_{\rm m}=0$, and for a cosmological constant 
$\omega_\Lambda=-1$.

Energy conservation, Eq.~\eqref{eq:econserv}, in the expanding Universe gives
\begin{equation}
\label{eq:econsgen}
\dot{\rho}=-3(1+\omega)\rho\frac{\dot{a}}{a}.
\end{equation}
\noindent
For a general barotropic fluid Eq.~\eqref{eq:econsgen} can be solved and yields
\begin{equation}
\label{eq:econbar}
\rho=\rho_0\left(\frac{a_0}{a}\right)^{3(1+\omega)} 
\end{equation}
\noindent
which for $\omega=-1$ becomes
\begin{equation}
\label{eq:econl}
\rho=\rho_0 \ \ (\omega=-1),
\end{equation}
\noindent
where $\rho_0$ is the current energy density of the fluid. 

The combination $H=\dot{a}/a$ is known as the Hubble parameter, where an
overdot denotes a derivative with respect to cosmic time $t$. It shows how fast the
Universe is expanding. Its current value $H_0$ is called the Hubble constant and is often quoted in terms of a dimensionless parameter $h=H/(100\rm\ km\ s^{-1} \ Mpc^{-1})$.

In the Universe described by Eqs.~(\ref{eq:metric}) and (\ref{eq:spmetricsp}) and dominated by a single barotropic fluid, Eq.~\eqref{eq:einstein} reduce to two independent equations for the scale factor
\begin{align}
\label{eq:friedman1}
\left(\frac{\dot{a}}{a}\right)^2& = \frac{8\pi G}{3}\rho - \frac{K}{a^2},\\
\label{eq:friedman2}
\frac{\ddot{a}}{a} & = -\frac{4\pi G}{3}(1 + 3\omega)\rho.
\end{align}
\noindent
Equation~\eqref{eq:econsgen} follows from Eqs.~(\ref{eq:friedman1}) and (\ref{eq:friedman2}). In the Universe dominated by a single barotropic fluid, Eqs.~(\ref{eq:friedman1}) and (\ref{eq:friedman2}) give
\begin{equation}
\label{eq:agen}
a=a_0\left(\frac{t}{t_0}\right)^{2/3(1+\omega)} \ \ (\omega\neq -1).
\end{equation}
Equation~\eqref{eq:friedman1} shows that if $\rho=\rho_{\rm cr}\equiv3H^2/8\pi G$ the curvature is zero. This time-dependent energy density
that makes the Universe spatially-flat is called the critical density. Current energy densities of matter components are often given in terms of dimensionless
density parameter, $\Omega_{i}=\rho_{i}/\rho_{\rm cr}$.

In a curved space with radiation, nonrelativistic matter, and cosmological constant Eq.~\eqref{eq:friedman1} can be rewritten as
\begin{equation}
\label{eq:frie}
H=H_0\sqrt{\frac{\Omega_{\rm m}}{a^3}+\frac{\Omega_{\rm r}}{a^4}+\Omega_\Lambda+\frac{\Omega_{\rm K}}{a^2}},
\end{equation}
where $\Omega_{\rm m}$, $\Omega_{\rm r}$, $\Omega_\Lambda$, and $\Omega_{\rm
K}$ are respectively the density parameters of nonrelativistic matter,
radiation, cosmological constant, and curvature, and the scale factor is normalized so that $a_0=1$.

\section{Distance and time measurements in cosmology}

Distances can be measured directly, using the parallax method, only to astronomical objects that are close to us (less than 100\ ${\rm pc}$).
For objects that are farther away indirect methods of distance determination must be used.

When a distant object emits a photon with a wavelength $\lambda_{\rm em}$ at
time $t_{\rm em}$, since light travels at finite speed, we will detect
the photon at a later time $t_{\rm now}$. As the Universe expands, the photon will ``redshift'' and we will detect it at a longer wavelength
\begin{equation}
\label{eq:redshift}
\lambda_{\rm o}=\frac{a_{\rm o}}{a_{\rm em}}\lambda_{\rm em},
\end{equation}
\noindent
where $a_{\rm em}$ and $a_{\rm o}$ are respectively scale factors at the moment of emission and detection (now) of the photon. Redshift of an astronomical object is defined as
\begin{equation}
\label{eq:z}
z_{\rm obj}=\frac{a_0}{a_{\rm obj}}-1
\end{equation}
\noindent
where $a_{\rm obj}$ is the value of the scale factor at the moment when the light that reached us from that body was emitted.  

If the cosmological
expansion history $a(t)$ is known, redshift can be uniquely related to the distance to the object and the time of light's emission. Distances and times in 
cosmology are often quoted in terms of redshifts. Two most frequently used distance measures in cosmology are the luminosity distance and the angular distance.

\subsection{Measuring time intervals}

Light travels on null geodesics $ds^2=0$. When we detect a photon that was
emitted at redshift $z$ it has traveled for time $\Delta T =D/c$, where $D$
is the distance and $c$ is speed of light. Let's set up the coordinate system
so that the photon travels in the radial direction. 
Then from Eqs.~\eqref{eq:metric} and \eqref{eq:spmetricsp} the total elapsed time can be expressed as
\begin{equation}
\label{eq:time}
\Delta T=\displaystyle\int_{t(0)}^{t(z)} dt = \displaystyle\int_{a_0}^{a(z)} a\chi(a) da.
\end{equation}
Using Eq.~\eqref{eq:frie} and the identities $a=a_0/(1+z)$ and
$da=-dz a_0/(1+z)^2$, we can rewrite Eq.~\eqref{eq:time} as
\begin{equation}
\label{eq:time2}
\Delta T = t(z) - t_0 = \frac{1}{a_0H_0}\displaystyle\int_0^z{\frac{dz}{\sqrt{\Omega_{\rm m}(1+z)^3+\Omega_{\rm r}(1+z)^4+\Omega_\Lambda+\Omega_{\rm K}(1+z)^2}}}.
\end{equation}
\noindent
In a given cosmological model Eq.~\eqref{eq:time2} uniquely relates redshift to time.

\subsection{Luminosity distance}
\label{sec:lumdist}
Let's say we have a source of known luminosity (power) $L$ at a redshift $z$. The source emits energy $Ldt$ in a time interval $dt$ and this energy redshifts
by $a/a_0=1/(1+z)$ before it reaches us. The energy is distributed over a
spherical surface with radius $a_0\chi(t_0 - t)$, where $\chi(t_0 - t)$ is the
 distance that a photon travels before detection. The flux that we measure in
unit time interval $dt_0=(a_0/a) dt$ is given by
\begin{equation}
\label{eq:fluxdet}
F=\frac{La^2}{4\pi a_0^4\chi^2(t_0 - t)}.
\end{equation}

Luminosity distance is defined as
\begin{equation}
\label{eq:lumdist}
d_L=\left(\frac{L}{4\pi F}\right)^2=\frac{a_0^2}{a(z)}\chi(t_0 - t_{\rm z}).
\end{equation}
By substituting $t_0 - t_{\rm z}$ from Eq.~\eqref{eq:time2}, the luminosity distance can be rewritten as
\begin{equation}
\label{eq:lumdist2}
d_L=(1+z)\chi\left(\frac{1}{H_0}\displaystyle\int_0^z{\frac{dz}{\sqrt{\Omega_{\rm m}(1+z)^3+\Omega_{\rm r}(1+z)^4+\Omega_\Lambda+\Omega_{\rm K}(1+z)^2}}}\right).
\end{equation}

Luminosities of some astronomical objects can be determined by indirect measurements. Supernovae Type Ia (SNeIa), for example, have very small dispersion in total luminosity. Gamma-ray bursts (GRBs) could be well calibrated in the future with their luminosity 
determined to high accuracy. We can measure the flux of these and other ``standardizable candles'' at different redshifts to determine cosmological parameters
from Eq.~\eqref{eq:lumdist2}.

\subsection{Angular diameter distance}

Let's say an astronomical body at a distance $d$ from us has a physical size $D$, then the observed angle subtended by that body is $\theta=D/d$. The angular diameter distance is defined as
\begin{equation}
\label{eq:angdist}
d_A=\frac{D}{\theta}.
\end{equation}
\noindent
If the object is at redshift $z$ it will span the angle $\theta=a(z)\chi(z)$. Using arguments of Section~\ref{sec:lumdist} we can express the angular diameter distance through cosmological parameters as
\begin{equation}
\label{eq:angdist2}
d_A=\frac{1}{1+z}\chi\left(\frac{1}{a_0H_0}\displaystyle\int_0^z{\frac{dz}{\sqrt{\Omega_{\rm m}(1+z)^3+\Omega_{\rm r}(1+z)^4+\Omega_\Lambda+\Omega_{\rm K}(1+z)^2}}}\right).
\end{equation}
Similarly to standard candles, ``standard rulers'', such as FRII radio galaxies, have small dispersion in physical size. Equation~\eqref{eq:angdist2} then can be used to determine cosmological parameters from the redshifts and apparent angular sizes of those standard rulers.

\section{Short overview of relevant topics from observational cosmology}

\begin{figure}[htb]%t=top, b=bottom, h=here
\begin{center}
    \includegraphics[height=2.5in]{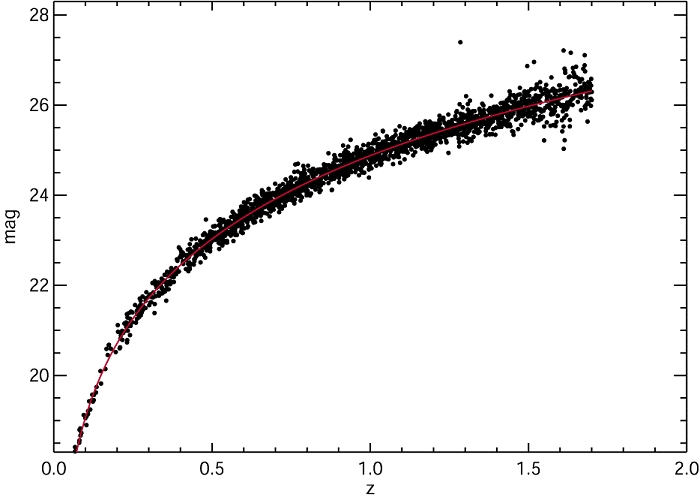}

    \caption[Supernovae Hubble diagram.]{Hubble diagram of SNeIa detected by ESA-WFI. Black dots represent SNeIa with measured absolute magnitude at different redshifts. Source: http://www.sci.esa.int/.}

    \label{fig:supernovae}
    \end{center}
\end{figure}

\begin{figure}[htb]%t=top, b=bottom, h=here
\begin{center}
    \includegraphics[height=2.5in]{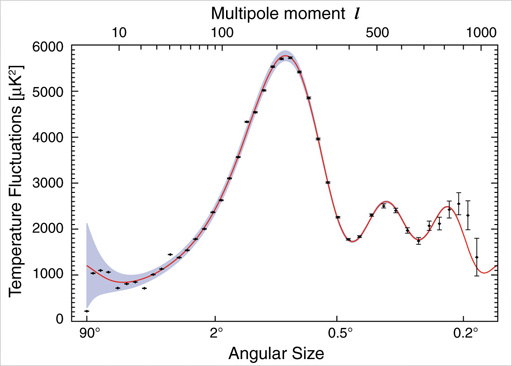}

    \caption[TT power spectrum of CMB anisotropies.]{CMB temperature anisotropy power spectrum. Black dots with error bars denote different measurements, red line shows predictions of the best-fit theoretical model. Source: http://cosmos.astro.uiuc.edu/.}

    \label{fig:cmbpow}
    \end{center}
\end{figure}

\begin{figure}[htb]%t=top, b=bottom, h=here
\begin{center}
    \includegraphics[height=2.5in]{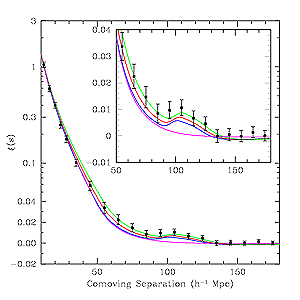}

    \caption[Galaxy power spectrum in redshift space.]{Power spectrum of
giant red galaxies as measured by Sloan Digital Sky Survey. Black dots with
error bars denote measurements, different colored lines show theoretical
predictions. The baryon acoustic peak is zoomed in on in the upper right
corner insert. Source: http://www.sdss3.org/.}

    \label{fig:galpow}
    \end{center}
\end{figure}

In last decade the quality and quantity of astronomical observations relevant to cosmology have greatly improved.
SNeIa can map the expansion of the Universe up to the redshift of 1.7 (see Fig.~\ref{fig:supernovae}).
While the power spectra of CMB anisotropies (see Fig.~\ref{fig:cmbpow}) and
the galaxy distribution (see Fig.~\ref{fig:galpow}) have been measured to a high precision.

The combination of SNeIa absolute magnitude versus redshift data
\cite{Kowalski:2008ez,Rubin:2008wq}, measurements of CMB anisotropy \cite{Komatsu:2008hk},
cluster gas mass fraction versus redshift data \cite{Allen:2007ue},
measurements of the baryon acoustic peak \cite{Eisenstein:2005su, Percival:2007yw}, and
other large-scale structure measurements \cite{Percival:2006gt} can be used to determine cosmological parameters with high precision.

Currently available cosmological data strongly indicate that:
\begin{itemize}
\item{The Universe is highly isotropic and homogeneous.}
\item{The Universe is expanding and distance between gravitationally unbound objects are increasing.}
\item{The cosmic expansion started accelerating at a redshift near $z=0.5$, if the dark energy is a cosmological constant.}
\item{The spatial curvature of the Universe is very close to zero, if the
dark energy is cosmological constant.}
\end{itemize}

The accelerated expansion of the Universe can be understood if we assume that
around $75\%$ of the current cosmological energy budget is made of dark energy (DE). Parameterized as a fluid, DE has
a negative equation of state parameter $\omega_{\rm DE}<-1/3$ and is spatially uniform.

About the other matter components we know that:
\begin{itemize}
\item{The Universe is filled with CMB photons. The CMB has an accurately Plankian spectrum with a temperature of $T=2.73\ K$.
      Radiation today makes a negligible contribution to the total energy density.}
\item{Baryons make up less than 5\% of critical density. About 75\% of baryonic matter is Hydrogen, 25\% Helium. Contribution of heavier elements to 
      the total baryonic energy density is small. The entropy of the Universe is high, around $10^9$ photons per baryon.}
\item{Nonrelativistic, cold dark matter (CDM) contributes around 25\%  of the
critical density. CDM is not baryonic and does not significantly emit, absorb, or scatter light.}
\end{itemize}

A joint analysis of current CMB anisotropy measurements, SNeIa data, and BAO peak measurements results in the following 1$\sigma$ ranges for key cosmological parameters \cite{Web:lambda}:
\begin{itemize}
\item{$h=0.71\pm 0.013$.}
\item{$\Omega_{\rm b}=0.0462\pm 0.0015$.}
\item{$\Omega_{\rm DM}=0.233 \pm 0.013$.}
\item{$\Omega_\Lambda=0.721 \pm 0.15$.}
\item{Age of the Universe = $13.73 \pm 0.12$\ Gyr.}
\end{itemize}

This chapter provides only a concise overview of topics relevant to the rest of this thesis. 
Detailed, in depth treatments of theoretical and observational cosmology can be found in various textbooks \cite{peebles80:lss, Peebles:94, mukhanov:cos,weinberg:cos,durrer:cos}.

% +--------------------------------------------------------------------+
% | Uncomment the lines below to add additional chapters.  Name the
% | files chapter2.tex for Chapter 2, chapter3.tex for Chapter 3, etc.
% +--------------------------------------------------------------------+

% +--------------------------------------------------------------------+
% | Sample Chapter 2
% +--------------------------------------------------------------------+

\cleardoublepage

% +--------------------------------------------------------------------+
% | Replace "This is Chapter 2" below with the title of your chapter.
% | LaTeX will automatically number the chapters.
% +--------------------------------------------------------------------+

\chapter{Dark Energy}
\label{makereference2}

\section{Observational evidence for dark energy}

The first direct evidence of accelerated expansion was provided by SNeIa absolute magnitude versus redshift data \cite{ct:riess:98,ct:perlmutter:99}.
Accelerated expansion is very difficult to explain in GR with only
conventional matter. Equation~\eqref{eq:friedman2} implies that the expansion can
be accelerating only if the Universe is dominated by a component with
negative pressure $P<-\rho/3$. The initial SNeIa measurements were followed by ground-based and space surveys, such as the Hubble Space Telescope (HST) survey \cite{Knop:2003iy}, the supernova legacy survey (SNLS) \cite{Astier:2005qq}, and the ESSENCE survey \cite{Miknaitis:2007jd}. Current SNeIa data extends to $z=1.7$ and provides more than 5$\sigma$ evidence for accelerated expansion.

Positions and amplitudes of acoustic peaks in the CMB anisotropy show that the Universe is very close to spatially flat if DE does not evolve in time. When CMB anisotropy results are
combined with an independent measurement of the Hubble constant they indicate that nonrelativistic matter makes up only about a quarter of the critical density, 
while the energy density of radiation is negligible. This implies that about three quarters of our Universe's energy budget is made of something different than ordinary or dark 
matter and radiation. This extra component must be uniform on cosmological scales and became dominant recently, in order not to adversely effect large-scale structure
formation and big bang nucleosynthesis.

The presence of DE also effects CMB anisotropies directly through the integrated Sachs-Wolf (ISW) effect. In a Universe dominated by DE gravitational wells
decay on large distance scales. This affects the large-angle CMB spectrum, since photons that travel through decaying gravitational potentials gain
energy. If the Universe is dominated by DE there should be cross-correlation between CMB anisotropies and low-redshift matter distribution. Several groups
reported detection of the ISW effect in cross-correlations of the CMB with galaxy and radio-source catalogs \cite{Afshordi:2003xu,Boughn:2004zm,Fosalba:2003ge}.

These results, combined with the BAO peak measurements, LSS tests, and galaxy cluster data, provide fairly overwhelming evidence that
about $70\%$ of the current Universe's budget is in spatially uniform DE with negative pressure.

\section{$\Lambda$CDM model}

All currently available cosmological data is quite consistent with a Universe that is spatially-flat, is dominated by a cosmological constant 
$\Lambda$ with $\Omega_\Lambda\sim0.7$, with the rest of the energy density being in nonrelativistic cold dark matter with $\Omega_{\rm m}\sim0.25$ and
nonrelativistic baryonic matter with $\Omega_{\rm b}\sim0.05$. A cosmological constant has $\omega_\Lambda=-1$ and according to Eq.~\eqref{eq:econsgen} it's
energy density does not change in time. This model is referred to as the ``standard model'' or the spatially-flat $\Lambda$CDM model. 

In the spatially-flat $\Lambda$CDM model the background expansion of the Universe at late times is described by
\begin{equation}
\label{eq:frielcdm}
H=H_0\sqrt{\Omega_{\rm m}(1+z)^3+\Omega_\Lambda}.
\end{equation}
Although the $\Lambda$CDM model is a good fit, the data is not yet tightly
constraining and does not yet require a constant $\Lambda$. To account for
the possibility of a time-varying cosmological
constant the equation of state of DE is sometimes modeled as $P=\omega_{\rm x}\rho$, where $\omega_{\rm x}$ is a negative parameter. This, so called, 
XCDM parametrization lacks physical motivation. XCDM is usually used as an ansatz in data
analysis for the purposes of quantifying time-dependence of DE. 

\section{Problems of $\Lambda$CDM}
\label{sec:lcdmprob}
Although successful in fitting most data, the $\Lambda$CDM model has a number
of theoretical shortcomings. A cosmological constant is difficult to motivate
from fundamental physics. The most plausible candidate for a cosmological constant is vacuum energy. 

Since all Lorentz observers should see the same 
vacuum, it's energy-momentum tensor must be proportional to the metric tensor $T_{\mu\nu}^{\rm vac}=\Lambda g_{\mu\nu}^{\rm vac}$. The vacuum therefore 
has the required equation of state $P^{\rm vac}=\rho^{\rm vac}$. It's very difficult however, to explain the small value of DE energy density based on 
the vacuum energy model.

In the quantum field theory the zero point energy of the vacuum is divergent
\begin{equation}
\label{eq:vacen}
\rho^{\rm vac}\propto\displaystyle\int_0^\infty k^3dk.
\end{equation}
\noindent
Quantum field theory is expected to break down at the Planck scale of around $10^{19}\ \rm  Gev$. If we use this as upper integration limit in Eq.~\eqref{eq:vacen} (instead of $\infty$)
we get a huge number that exceeds the observed dark energy density by 120 orders of magnitude. 

In a supersymmetric model every boson has a fermion of 
equal mass as a supersymmetric partner and the vacuum energies of these partners cancel. Supersymmetry (SUSY), if existent, is believed to be broken at 
an energy of roughly $1\ \rm TeV$ or so. If we cut off the upper integration limit in Eq.~\eqref{eq:vacen} at the energy of SUSY breaking we will still get
a difference of around 60 orders of magnitude. This discrepancy between the small measured value of cosmological constant and the much larger theoretically ``expected''
values of vacuum energy is known as the ``smallness'' problem \cite{Weinberg:1988cp}.

One possible explanation of the ``smallness'' problem is based on anthropic arguments. In string theory, multiple vacuum states with all possible values of
vacuum energy are possible. Different causally disconnected patches of the Universe spontaneously choose vacuum states that are independent of each
other.  If the Universe is infinite there will always be parts of it that have a given value, no matter how unlikely, of the vacuum energy and we just
happen to live in one of those regions with a very small value of vacuum energy density \cite{Susskind:2003kw,Bousso:2004fc}.

Another interesting fact that's difficult to explain in the $\Lambda$CDM model is that today both nonrelativistic matter and DE have comparable energy densities. This is surprising since the matter and DE components scale with redshift differently. For radiation it is $\Omega_{\rm r}\sim(1+z)^4$, for CDM and baryons
$\Omega_{\rm m}\sim(1+z)^3$, and for a cosmological constant $\Omega_\Lambda\sim{\rm const}$. 

\begin{figure}[htb]%t=top, b=bottom, h=here
\begin{center}
    \includegraphics[height=2.5in, viewport=0.5in 2.5in 8in 8in, clip]{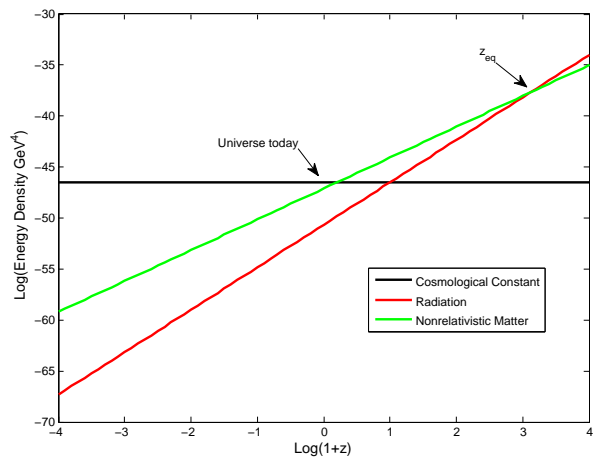}

    \caption[``Coincidence'' problem.]{The time evolution of energy densities of cold dark matter (green line), radiation (red line), and cosmological constant (black line).}

    \label{fig:coincidence}
    \end{center}
\end{figure}

At the beginning of cosmic evolution the Universe was
radiation dominated, today radiation contributes less than 1\% of the total energy density. The contribution of DE was negligible in the past, it has become a
dominant component only recently, and in the future will be the only component driving cosmic expansion as shown in Fig.~\ref{fig:coincidence}.

There is only a short period of time when the energy densities of matter and cosmological constant are comparable. It is unclear why we happen to live
in this narrow window of time. This is called the ``coincidence'' problem and like the ``smallness'' problem can be addressed by the anthropic
arguments.

Besides the two problems mentioned above there are other observational facts that appear to conflict with the predictions of the $\Lambda$CDM model, at possibly more then 2$\sigma$
confidence level. These are

\begin{itemize}
\item{High redshift SNeIa data are consistent with spatially-flat $\Lambda$CDM. It favors however models with $\omega_{\rm DE}<-1$. Initially,
this was thought to be a statistical fluke that would go away as more data accumulated. This discrepancy however still persists even as larger
data sets become available. The discrepancy is caused by high ($z>1$) redshift SNeIa which are systematically brighter than what we would expect in 
the $\Lambda$CDM model \cite{Kowalski:2008ez}.
This could be due to an unknown systematic effects, possibly associated with high-redshift SNeIa evolution, or a statistical effect that will go away with more data. If the
discrepancy persists, it would mean that the Universe in the past was decelerating faster than the $\Lambda$CDM model predicts.}

\item{Large-scale velocity flows have amplitude of $400\ {\rm km/s}$, larger
than what is expected in a $\Lambda$CDM model
\cite{Watkins:2008hf,Kashlinsky:2008us,Kashlinsky:2008ut,Lavaux:2008th,Abate:2009kd}.
Velocity flows extend to $z=0.2$ and could be as large as $1600\ {\rm km/s}$. In $\Lambda$CDM the probability of having velocity
flows with such a large amplitude is less then 1\%. The explanation for this
could be that it is just a big statistical fluctuation or there could be some physical
reason, such as time-dependent Newton's constant, presence of noninflationary perturbations, or a giant void at a distance of few Gpc.}

\item{Cosmological simulations based on the $\Lambda$CDM model predict that large voids should be filled with many dwarf dark matter halos. This
turns out to be true for very large voids (larger than $10\ {\rm Mpc}$). Smaller voids however are observed to be surprisingly empty of dark matter halos
\cite{Peebles:2007qe}. For example, based on $\Lambda$CDM we would expect to observe on average 10 dwarf galaxies in our local void, but there are none. Possible
resolution of this problem could be related to the incompleteness of observational sample, or incorrect bias model that fails to account for specific
environmental properties.}

\end{itemize}

At the moment these discrepancies between observations and theoretical predictions do not have big weight in the overall picture. The $\Lambda$CDM model is
in general a good statistical fit to the combined data. These inconsistencies might go away when new high-quality data become available.

\section{Alternative dark energy models}

At the moment, $\Lambda$CDM is a good fit to available cosmological data,
however, because of the observational issues mentioned in Sec.~\ref{sec:lcdmprob} 
and its inability to naturally account for the small observed energy density of DE, a number of alternative models have been proposed over the years.

One approach is to not introduce a new exotic DE component, but to modify the
equations of GR and consequently Eq.~\eqref{eq:einstein}, \cite{Wei:2008vw,Tsujikawa:2008in,Capozziello:2007zz,Sotiriou:2008rp,Bamba:2008xa}. In this set of models gravity is usually weaker on larger scales and the accelerated expansion is just a gravitational effect. At very early times modified gravity should
act like GR predicts, in order not to conflict with predictions of CMB anisotropy and BBN. In $f(R)$ and scalar-tensor theories, the equivalents
of Eq.~\eqref{eq:einstein} include up to forth order derivatives of the metric tensor. Gravity is not a spin-2 transverse and traceless field anymore. It is
also carried by spin-0 (scalar) and spin-1 (vector) particles. 

Another possibility is that the Universe is not as homogeneous as we thought and the accelerated expansion is an effect of averaging nonlinear spatial
inhomogeneities over cosmological scales \cite{Kolb:2005da,Hirata:2005ei,Rasanen:2005zy,Li:2007ny}. It is not yet clear if the backreaction of structure
formation on the background evolution can be strong enough to mimic accelerated expansion and at the same time be consistent with all other available data.

A third option is to introduce a DE fluid that is close to spatially uniform but unlike a cosmological constant slowly varies in time. This is usually achieved by introducing
a scalar field that is slowly rolling down it's almost flat potential \cite{Peebles:1987ek,Ratra:1987rm}. In the minimal model, the scalar field is coupled
to the rest of the matter only gravitationally, but other couplings have also been considered \cite{Olivares:2007rt,Wei:2007ig,Wu:2007zz,Bonometto:2006jn}. 

In Chapter~\ref{makereference3} I will elaborate more on the scalar field dark energy models.
More detailed discussion of DE theory and the observational situation can be found in recent reviews \cite{Ratra:2007sa,Linder:2008pp,Frieman:2008sn,Uzan:2006mf}.

% +--------------------------------------------------------------------+
% | Sample Chapter 3
% +--------------------------------------------------------------------+

\cleardoublepage

% +--------------------------------------------------------------------+
% | Replace "This is Chapter 3" below with the title of your chapter.
% | LaTeX will automatically number the chapters.
% +--------------------------------------------------------------------+

\chapter{Cosmological Scalar Field}
\label{makereference3}

\section{Scalar field in the expanding universe}

Let's assume that a self-interacting scalar field $\phi$ minimally coupled to gravity is present on cosmological scales. This scalar field is described by the lagrangian density
\begin{equation}
\label{eq:lagrangian}
\mathcal{L}=\frac{1}{2}\partial_\mu\phi\partial^\mu\phi - V(\phi),
\end{equation}
\noindent
where $V(\phi)$ is the potential energy density. The energy momentum tensor of this scalar field is given by
\begin{equation}
\label{eq:energymomentum}
T^\mu_\nu=\frac{\partial\phi}{\partial x_\mu}\frac{\partial\phi}{\partial x^\nu} - g^\mu_\nu\left(\frac{1}{2}\frac{\partial\phi}{\partial x_\alpha}\frac{\partial\phi}{\partial x^\alpha} + V(\phi)\right).
\end{equation}
If the scalar field is close to spatially uniform on cosmological scales we can neglect its spatial derivatives $\phi_{,i}$ compared to time derivatives $\dot{\phi}$ and the individual components of the homogeneous part of the energy momentum tensor can be written as
\begin{align}
\label{eq:enmomuni1}
T^0_i&=0,\\
\label{eq:enmomuni2}
T^i_j&=0 \ (i\neq j),\\
\label{eq:enmomuni3}
T^0_0&\equiv\rho=\frac{1}{2}\dot{\phi}^2 + V(\phi),\\
\label{eq:enmomuni4}
T^i_i&\equiv P=\frac{1}{2}\dot{\phi}^2 - V(\phi).
\end{align}

By looking at Eqs.\ (\ref{eq:enmomuni3}) and (\ref{eq:enmomuni4}) we can see that the equation of state parameter
\begin{equation}
\label{eq:state}
\omega=\frac{\dot{\phi}^2-2V(\phi)}{\dot{\phi}^2+2V(\phi)}
\end{equation}
\noindent
can become negative if the scalar field evolves slowly in time. The equation
of state parameter is always greater than $-1$ and in the limit of the scalar field ``at rest'' ($\dot\phi=0$) we effectively get a cosmological constant with $\omega=-1$.

The time evolution of the scalar field (and consequently the time evolution of the equation of state parameter) is determined by the Klein-Gordon equation. For the metric given by Eq.\ \eqref{eq:metric} the Klein-Gordon equation is
\begin{equation}
\label{eq:clain}
\ddot{\phi}+3\frac{\dot{a}}{a}\dot{\phi}+\frac{\partial V(\phi)}{\partial\phi}=0.
\end{equation}
If the scalar field energy density makes nonnegligible contribution to the total energy budget of the Universe it will affect the evolution of the scale factor and Eq.~(\ref{eq:frie}) should be rewritten as
\begin{equation}
\label{eq:friephi}
H=H_0\sqrt{\frac{\Omega_{\rm m}}{a^3}+\frac{\Omega_{\rm r}}{a^4}+\Omega_\phi(a)+\frac{\Omega_{\rm K}}{a^2}},
\end{equation}
\noindent
where $\Omega_\phi$ is now the time-dependent energy density parameter of the scalar field.

To make specific predictions about how the scalar field affects the Universe's evolution we have to specify the form of the potential $V(\phi)$.
Since the underlying physics is not known at the moment, it's impossible to pick a specific functional form for $V(\phi)$ based on
the considerations of fundamental physics only. 

The only fundamental scalar field in the standard model of particle physics
is the Higgs field. Rest mass of the Higgs boson is larger than about $100\
\rm GeV$ and at low energies it decays into $W$ bosons, $H\rightarrow
W^{+}W^{-}$, therefore Higgs bosons can not be present in large quantities in
the late universe. In the theories that go beyond the standard model, such as
string theory, minimal supersymmetric theory, etc., fundamental scalar fields arise naturally and we have a large number of candidates. The particle physics at very high energies is not well constrained experimentally yet and different theoretical scenarios are possible; because of this uncertainty its impossible to motivate one choice of potential based on fundamental considerations and a number of different models have been proposed over the years. An incomplete list of possible scalar field potentials is shown in Table\ \ref{tab:potential}.

\begin{table}
\begin{center}
\begin{tabular}[c]{|l|l|}
        \hline
        Potential $V(\phi)$& Reference \\
        \hline
         $\phi^{-\alpha},\ \alpha>0$& \cite{Peebles:1987ek}\\
         $\exp(-\lambda\phi)$& \cite{Ratra:1987rm}\\
         $\lambda\phi^4 $& \cite{ct:frieman:95}\\
         $\exp(-\lambda\phi)/\phi^\alpha,\ \alpha>0$& \cite{ct:brax:99}\\
          $\exp(\lambda/\phi)-1$& \cite{ct:zlatev:99}\\
          $(\cosh(\lambda\phi)-1)^n, n>0$& \cite{ct:sahni:00}\\
          $((\phi - C_1)^\lambda + C_2)\exp(-\kappa\phi)$& \cite{ct:albrecht:00}\\
        \hline
    \end{tabular}
    \caption{List of proposed self-interaction potentials for the cosmological scalar field.}
    \label{tab:potential}
\end{center}
\end{table}
 
 In this work I will concentrate on the inverse-power law potential first proposed in Ref.~\cite{Peebles:1987ek}. In this model the potential is assumed to be inversely proportional to a power of the scalar field, $V\propto  1/\phi^\alpha$, where $\alpha$ is a positive constant.
 
 \section{Inverse-power law potential}
 
 If we take scalar fields self-interaction potential to be inversely proportional to a power of $\phi$,
 \begin{equation}
 \label{eq:vphi}
 V(\phi)=\frac{\kappa}{2G}\phi^{-\alpha},
 \end{equation}
 \noindent
  where $\kappa$ and $\alpha$ are nonnegative parameters, Eqs.\ (\ref{eq:friephi}) and (\ref{eq:clain}) can be rewritten as
 \begin{align}
 \label{eq:frie3a}
\left(\frac{\dot{a}}{a}\right)^2=\frac{8\pi G}{3}(\rho_{\rm m} + \rho_{\phi}),\\
 \label{eq:scalar3}
 \ddot{\phi} + 3\frac{\dot{a}}{a}\dot{\phi} - \frac{\kappa}{G}\frac{\alpha}{\phi^{\alpha + 1}}=0,
 \end{align}
 \noindent
where we have set spatial curvature to zero and neglected radiation,
and the energy density of the scalar field is given by
\begin{equation}
\label{eq:scalarenergy}
\rho_\phi=\frac{1}{2}\left(\dot{\phi}^2+\frac{\kappa}{G\phi^\alpha}\right).
\end{equation} 
Parameter $\alpha$ describes the steepness of the scalar field potential. Larger values of $\alpha$ correspond to faster evolution of the scalar field
and vice versa. $\alpha=0$ corresponds to the time-independent cosmological constant.
Parameter $\kappa$ sets the mass scale $M$ of the scalar particle,
\begin{equation}
\label{eq:massscale}
M\sim\left(\frac{\kappa}{2G}\right)^{-(\alpha+4)}.
\end{equation}

In the distant past, when the energy density of the scalar field is much less
than the energy density of other components $\rho_\phi\ll\rho_{\rm cdm}$ , Eq.~\eqref{eq:scalar3} accepts a solution
\begin{equation}
\label{eq:phisol}
\phi\propto a^{3(1+\omega_{\rm cdm})/(\alpha+2)},
\end{equation}
\noindent
and the equation of state parameter of the scalar field is
\begin{equation}
\label{eq:estatesol}
\omega_\phi=\frac{\alpha\omega_{\rm cdm}-2}{\alpha+2}.
\end{equation}
\noindent
For positive $\alpha$, $\rho_\phi$ redshifts less rapidly than the dominant component of energy and at some point eventually starts to dominate. The redshift when cosmic acceleration starts is given by
\begin{equation}
\label{eq:accel}
z_\phi=\frac{\Omega_\phi}{\Omega_{\rm m}}^{(2+\alpha)/6}-1.
\end{equation}
\noindent
As we approach $z_\phi$, the fractional energy density of the scalar field
increases and the approximation $\rho_\phi\ll\rho_{\rm cdm}$ breaks down. Evolution
of the scalar field does not follow Eq.~\eqref{eq:phisol} anymore and to solve for $\phi(a)$ now the set of coupled differential 
equations~\eqref{eq:frie3a} and \eqref{eq:scalar3} must be solved numerically.

\begin{figure}[htb]%t=top, b=bottom, h=here
\begin{center}
    \includegraphics[height=2.5in, viewport=0.5in 2.5in 8in 8in, clip]{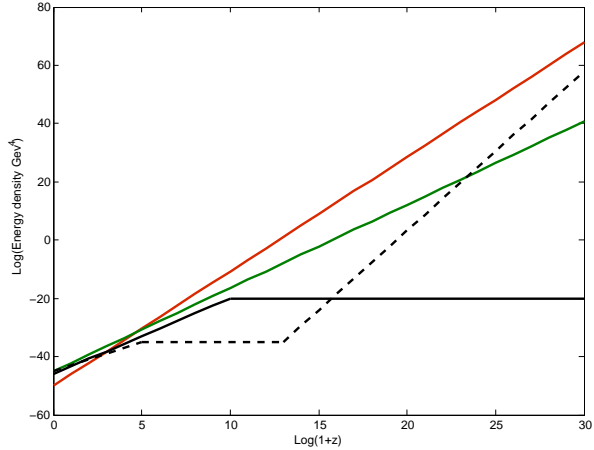}

    \caption[Evolution of scalar field for different initial conditions.]{Evolution of energy densities of nonrelativistic matter (green line), radiation (red line), and scalar field with different initial energies (black lines). Scalar fields that start off from different energies have very similar late-time evolution.}

    \label{fig:attractor}
    \end{center}
\end{figure}

A nice property of Eq.~\eqref{eq:phisol} is that it is an attractor. A wide range of initial conditions on $\phi$ result in solutions that converge to the solution of Eq.~\eqref{eq:phisol}
at some point. Figure~\ref{fig:attractor} shows the evolution of the scalar field for different initial conditions. Although initial conditions on $\rho_\phi$ span
some 70 orders of magnitude, the late-time evolution of the scalar field is similar in all intermediate cases. 

In the scalar field scenario the ``smallness'' and ``coincidence'' problems mentioned in Sec.~\ref{makereference2} are partially solved.
The scalar field can start from a very high energy state, roll down the inverse power law potential, and naturally lead to a $\rho_\phi$ that evolves to the very small measured 
value. Figure~\ref{fig:attractor} shows that the scalar field ``tracks'' the
evolution of the dominant component once it joins the solution of Eq.~\eqref{eq:phisol}. The energy densities of nonrelativistic matter and scalar field are comparable for a longer period of time, and current time becomes less
special. 

The scalar field in this model does not cluster on scales much smaller than the Hubble radius $1/H_0$ as required by CMB and large-scale structure data.

\section{$\phi$CDM model of dark energy}

In our analysis we will assume that the only two constituents affecting the late-time evolution of the Universe are nonrelativistic cold dark
matter and baryons with energy density parameter $\Omega_{\rm m}$ and a
scalar field with inverse power law self-interaction potential $V\propto \phi^{-\alpha}$. We will
also assume that the Universe is spatially-flat which implies $\Omega_{\rm m} + \Omega_\phi =1$.

In this model the background expansion is fully described by two parameters, $\Omega_{\rm m}$ and $\alpha$. The evolution of the scale factor and
scalar field are governed by
\begin{align}
\label{eq:phimain1}
\frac{\dot{a}}{a}=H_0\sqrt{\frac{\Omega_{\rm m}}{a^3} + \Omega_\phi},\\
\label{eq:phimain2}
\ddot{\phi}+3\frac{\dot{a}}{a}\dot{\phi}+\frac{\partial V(\phi)}{\partial\phi}=0.
\end{align}

For $\alpha=0$ the $\phi$CDM model reduces to the spatially-flat $\Lambda$CDM with $\Omega_\Lambda=\Omega_\phi$. For nonzero $\alpha$ predictions of
$\phi$CDM and $\Lambda$CDM differ, for example the scalar field model predicts lower values of luminosity and angular diameter distances for the same value of
Hubble parameter and nonrelativistic matter density. An example of a C++ function that solves Eqs.~\eqref{eq:phimain1} and~\eqref{eq:phimain2} is presented in Appendix~\ref{appendixA}.

Different cosmological tests have been used in the past to constrain the
parameters of the $\phi$CDM model, including counts of galaxies \cite{Podariu:2001yq}, angular sizes of radio sources and
quazars \cite{Podariu:2002jj,Chen:2002dc}, galaxy cluster gas mass fraction data \cite{Chen:2004nqb}, strong gravitational
lensing \cite{Chae:2004jp} and SNeIa absolute magnitude versus redshift data \cite{Podariu:1999ph,Podariu:2000nc,Wilson:2006tr}.
In next chapters I will discuss the constraints on $\phi$CDM model from some current data sets.

For an in depth review of scalar field dark energy see, e.g., Refs.~\cite{Peebles:1987ek,Ratra:1987rm,Peebles:2002gy,Martin:2008qp}.

% +--------------------------------------------------------------------+
% | Sample Chapter 4
% +--------------------------------------------------------------------+

\cleardoublepage

% +--------------------------------------------------------------------+
% | Replace "This is Chapter 4" below with the title of your chapter.
% | LaTeX will automatically number the chapters.
% +--------------------------------------------------------------------+

\chapter{Data Analysis Techniques}
\label{makereference4}

We are interested in observable quantities $X_{\rm i}^{\rm obs}$ measured at redshifts $z_{\rm i}$ (or in redshift bins of width $\Delta z_{\rm i}$).
$X_{\rm i}^{\rm obs}$ could be, e.g., luminosity distance, angular diameter distance, Hubble parameter, or  quantities that implicitly depend on them.
First I will assume that evolution of the Universe is described by the $\phi$CDM model, as presented in Chapter~\ref{makereference3}, with unknown values of
parameters $\alpha$ and $\Omega_{\rm m}$. For given values of model parameters we can compute the theoretical expectations $X_{\rm i}^{\rm th}$ for the
observed quantities. We will find the best-fit parameters $\alpha^*$ and $\Omega_{\rm m}^*$, for which the theoretical predictions $X_{\rm i}^{\rm th}$ are
the closest match to the results of observations $X_{\rm i}^{\rm obs}$ by some measure. We will also construct the confidence level intervals that are
likely to cover the true values of parameters with a specified probability.

We will compare the predictions of $\phi$CDM model to the ones of $\Lambda$CDM and see which ones fit the observations best. In the end we will see if data
favors one model over the other, and quantify the degree of discrepancy between the two models.

\section{$\chi^2$ and likelihood functions}
\label{sec:likelihood}
Let's say we have $N$ measurements $X_{\rm i}^{\rm obs}$ at redshifts $z_{\rm i}$ with measurement errors $\sigma_{\rm i}$, and the $\phi$CDM model
predicts $X_{\rm i}^{\rm th}(\alpha,\Omega_{\rm m})$. The $\chi^2$ function,
\begin{equation}
\label{eq:chi2}
\chi^2(\alpha,\Omega_m)=\displaystyle\sum_{\rm i}\frac{[X^{\rm obs}(z_{\rm i})-X^{\rm th}(z_{\rm i},\alpha,\Omega_{\rm m})]^2}{\sigma_{\rm i}^2},
\end{equation}
\noindent
quantifies the discrepancy between theoretical predictions and observations. Small values of $\chi^2$ indicate a good fit. 
The parameters $\alpha^*$ and $\Omega_{\rm m}^*$ that minimize $\chi^2$ are called the best-fit parameters.
The likelihood function,
\begin{equation}
\label{eq:lik}
\mathcal{L}(\alpha,\Omega_m)=\exp{\left(-\chi^2(\alpha,\Omega_m)/2\right)},
\end{equation}
\noindent
has a local maximum $\mathcal{L}_{\rm ML}$ at ($\alpha^*$, $\Omega_{\rm m}^*$). If the measurements are independent and Gaussian distributed with mean $X_{\rm i}^{\rm obs}$ and
variance $\sigma_{\rm i}$, then the best-fit values of parameters are unbiased estimators of their true values. Values of parameters that result in high
values of the likelihood function are more likely to be the true parameters.

1, 2, and 3$\sigma$ confidence level contours are defined as the pairs of parameters ($\hat{\alpha}$, $\hat{\Omega_{\rm m}}$) for which $-2\ln{\mathcal{L}(\hat{\alpha},\hat{\Omega_{\rm m}})}$ is less by 2.30, 6.18, and 11.83 respectively compared to $\mathcal{L}_{\rm ML}$. If the likelihood function had an exact two dimensional
Gaussian profile,
\begin{equation}
\label{eq:gauss}
\mathcal{L}\propto e^{-(C_1(\alpha-\alpha^*)^2+C_2(\Omega_{\rm m}-\Omega_{\rm
m})^2)/(2\sigma^2)},
\end{equation}
\noindent
N$\sigma$ contours would correspond to the range of parameters $(\alpha^*\pm
N\sigma,\Omega_{\rm m}^*\pm N\sigma)$. In that case the 3$\sigma$ contour
would enclose 99.73\% of integrated likelihood and the parameter values outside that range would be 0.27\% or less likely to be the true parameters.
This is not true for an arbitrary $\mathcal{L}$, however, when it is a sharply peaked function, parameter values outside 3$\sigma$ are usually less
than 1\% likely. 

\section{Nuisance parameters}

Very often $X^{\rm th}(z_i)$ will depend on parameters other than $\alpha$ and $\Omega_{\rm m}$. These ``nuisance'' parameters $\bm\nu$ can be determined
from the data but are of no interest to us. Examples of nuisance parameters in dark energy parameter estimation are the Hubble constant $H_0$ and 
baryonic energy density parameter $\Omega_{\rm b}$.
In this case the likelihood function $\mathcal{L}(\alpha,\omega_{\rm m},\bm\nu)$ will also depend on $\bm\nu$. If we have some idea from other observations
about what the values of $\bm\nu$ are, we can summarize this information in the form of probability distribution function (p.d.f.) $P(\bm\nu)$.
$P(\bm\nu)$ is normalized to one and is usually taken to be a Gaussian peaked at the best-fit value of $\bm\nu^*$ determined from other independent
data sets. If we do not want this implicit use of information from other data sets, we can use the conservative flat prior $P(\bm\nu)=1/(\bm\nu_2 - \bm\nu_1)$
over the range $(\bm\nu_1,\bm\nu_2)$.
We then integrate the likelihood function,
\begin{equation}
\label{eq:intlik}
\mathcal{L}'(\alpha,\Omega_{\rm m})=\displaystyle{\int}\mathcal{L}(\alpha,\Omega_m,\bm{\nu})P(\bm{\nu})d\bm{\nu}.
\end{equation}
\noindent
and determine the best fit values and confidence level contours from $\mathcal{L}'$ as in Sec.~\ref{sec:likelihood}.

Two most frequently used priors on the Hubble constant are the measurement of
Hubble Space Telescope (HST) team \cite{Freedman:2000cf} $H_0=(72\pm8)\ \rm km\ s^{-1}\ Mpc^{-1}$ and
the result from the Wilkinson Microwave Anisotropy Mission (WMAP) 5-year data
\cite{Dunkley:2008ie} $H_0=(73\pm3)\ \rm km\ s^{-1}\ Mpc^{-1}$. 
 The WMAP measurement has less dispersion but is model dependent, unlike the HST measurement which measured $H_0$ directly based on the recessional speeds
 of distant objects. The most conservative option is to integrate over $H_0$ with a flat, noninformative prior. The baryonic matter energy density
 is also well constrained by a number of independent tests. Big bang
nucleosynthesis gives the value $\Omega_{\rm b}h^2=0.0205\pm0.018$
\cite{Fields:2008zz} and the WMAP data results in $\Omega_{\rm b}h^2=0.0223\pm0.0008$.

\section{Constraints on individual cosmological parameters}

One-dimensional confidence level contours and best-fit values can be computed for individual cosmological parameters $\alpha$ and $\Omega_{\rm m}$.
We take the two-dimensional likelihood function from Eq.~\eqref{eq:intlik}
and integrate it with respect to the other parameter with a flat prior.
\begin{align}
\label{eq:1dalpha}
\mathcal{L}_\alpha&=\displaystyle\int \mathcal{L}'(\alpha,\Omega_{\rm m})d\Omega_{\rm m},\\
\label{eq:1domega}
\mathcal{L}_{\rm m}&=\displaystyle\int \mathcal{L}'(\alpha,\Omega_{\rm m})d\alpha.\\
\end{align}
\noindent

For each parameter we find the best-fit value that maximizes the corresponding one-dimensional likelihood function. We define 1, 2, and 3$\sigma$ confidence level intervals $(\alpha_1,\alpha_2)$ and
($\Omega_{\rm m1}, \Omega_{\rm m2}$) so that
\begin{align}
\label{eq:1dsalpha}
s&=\displaystyle\int_{\alpha_1}^{\alpha_2}\mathcal{L}_\alpha d\alpha \bigg{/}\displaystyle\int_{\rm all\ \alpha}\mathcal{L}_\alpha d\alpha,\\
\label{eq:1dsomega}
s&=\displaystyle\int_{\Omega_{\rm m1}}^{\Omega_{\rm m2}}\mathcal{L}_{\Omega_{\rm m}} d\Omega_{\rm m} \bigg{/}\displaystyle\int_{\rm all\ \Omega_{\rm m}}\mathcal{L}_{\Omega_{\rm m}} d\Omega_{\rm m},
\end{align}
\noindent
and $s$ equals to 68.27, 95.45, and 99.73 respectively.
We will choose lower and upper limits on parameter values so that the likelihood is higher everywhere inside the interval than outside.

Parameter $\alpha$ should always be positive. Since the best-fit values are usually close to the $\alpha=0$ line, $\alpha_1$ in Eq.~\eqref{eq:1dsalpha} will
often be equal to zero.

% +--------------------------------------------------------------------+
% | Sample Chapter 5
% +--------------------------------------------------------------------+

\cleardoublepage

% +--------------------------------------------------------------------+
% | Replace "This is Chapter 3" below with the title of your chapter.
% | LaTeX will automatically number the chapters.
% +--------------------------------------------------------------------+

\chapter{Constraints on $\phi$CDM from Hubble Parameter Measurements}
\label{makereference5}

This chapter is based on Refs.~\cite{Samushia:2006fx,Samushia:2007db,Samushia:2009lt}.

\section{Hubble parameter measurements}
The Hubble parameter in spatially flat $\phi$CDM is given by

\begin{equation}
\label{eq:hubble}
H=H_0\sqrt{\Omega_{\rm m} + \Omega_\phi(\alpha)}
\end{equation}
\noindent
and is a function of redshift $z$ and cosmological parameters $\Omega_{\rm m}$ and $\alpha$. In practice, it can be measured indirectly at different redshifts and thus used to constrain cosmological parameters of interest.

First of all let's note that since $a\sim(1+z)^{-1}$, the Hubble parameter can be rewritten as
\begin{equation}
\label{eq:hubble2}
H=\frac{\dot{a}}{a}=-\frac{1}{1+z}\frac{dz}{dt}.
\end{equation}
\noindent
Equation~\ref{eq:hubble2} shows that the Hubble parameter can be determined from the measurement of $dz/dt$. $dz/dt$ itself can be determined from a survey that measures redshifts of galaxies with high precision. The basic idea is that if we have two galaxies very close in redshift space and we can determine their ages with good accuracy, then $dz/dt$ can be approximated by reasonably well with $\Delta z/\Delta T$, where $\Delta z$ is the difference in their redshift and $\Delta T$ is the difference in their age. The most difficult part is to accurately determine ages of galaxies. 

In passively evolving galaxies the star formation rate is low and their ages
can be determined with better accuracy. These galaxies are dominated by the
light from old red main sequence stars. The evolution of those stars is a
well understood process and metallicity can be simulated on computers. The
spectra of galaxies depends mainly on metallicity and age. A computed
spectrum can be compared to the observed one and the galaxy age can be
determined. Stars in galaxies, of course, do not have the same metallicities
and this spread in metallicity introduces a statistical error of about $0.1\
\rm Gyr$. For a detailed description of this and other relevant issues and methods see Refs.~\cite{Jimenez:2001gg}.

\section{Constraints from Hubble parameter data}

Reference~\cite{Simon:2004tf} used Gemini Deep Deep Survey
\cite{Abraham:2004ra} and archival data
\cite{Nolan:2003bt,Treu:2001di,Treu:2001hq,Spinrad:1997md,Dunlop:1996mp} to
estimate ages of 32 passively evolving galaxies. From these age measurements they were able to compute the differential age $dz/dt$ at 9 different redshifts in the redshift range $0.09<z<1.75$. The age measurements are shown in Appendix~\ref{appendixage}, while the Hubble parameter estimates from Ref.~\cite{Simon:2004tf} are shown in Appendix~\ref{appendixhubble}.
These $H(z)$ measurements are independent and the errors are Gaussian distributed. The $\chi^2$ function is
\begin{equation}
\label{eq:chihubble}
\chi^2 = \displaystyle\sum_{i=1}^9\frac{(H_{\rm th}(z_i, \Omega_{\rm m}, \alpha, H_0)-H_{\rm obs})^2}{\sigma^2_i}
\end{equation}
\noindent
where $H_{\rm th}$ is the theoretical value of Hubble parameter computed from
Eq. \eqref{eq:hubble}, and $H_{\rm obs}(z_i)$ and $\sigma_i$ are the observed
values and one sigma uncertainties. In our analysis we assume 12\%
uncertainties on age measurements in
Appendix~\ref{appendixage} \cite{Com:jimenez}.

The Hubble constant (the Hubble parameter now, at zero redshift) is a nuisance parameter and can be integrated over to get a two dimensional likelihood function of cosmological parameters
\begin{equation}
\label{eq:likhubble}
\mathcal{L}(\Omega_{\rm m}, \alpha)= \displaystyle\int\exp(-\chi^2(\Omega_{\rm m}, \alpha, H_0)/2)P(H_0)dH_0
\end{equation}
where $P(H_0)$ is a prior probability distribution function for $H_0$. We
integrate over $H_0$ with a Gaussian WMAP prior with $H_0=(73\pm 3)\ \rm km\ s^{-1}\ Mpc^{-1}$.

\begin{figure}[htb]%t=top, b=bottom, h=here
\begin{center}
    \includegraphics[viewport=0.5in 2.5in 8in 8in, clip]{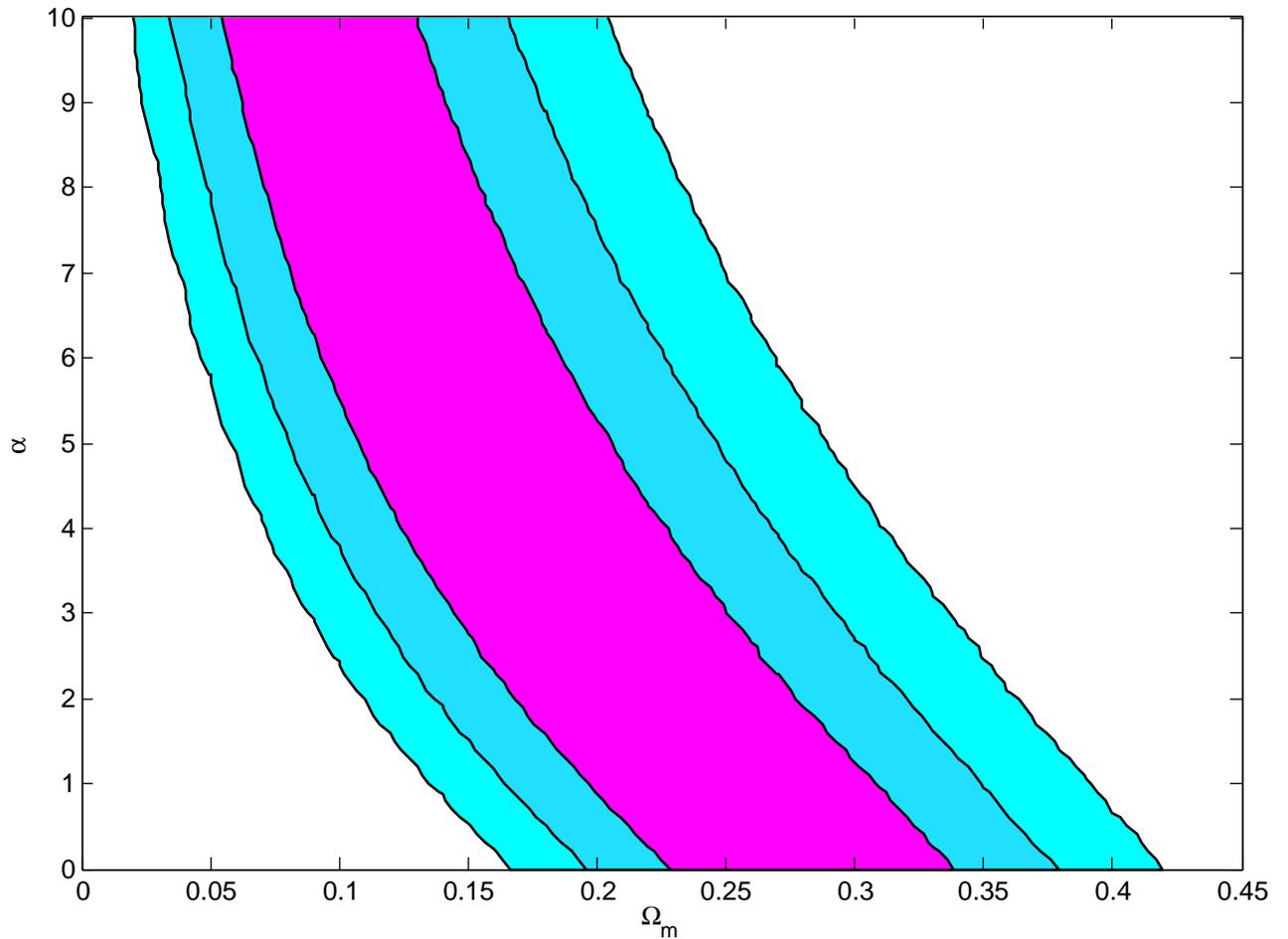}

    \caption[Constraints from Hubble parameter versus redshift data.]{1, 2, and 3$\sigma$ confidence level contours on $\phi$CDM model parameters from Hubble parameter versus redshift data.}

    \label{fig:hubble}
    \end{center}
\end{figure}

\begin{figure}[htb]%t=top, b=bottom, h=here
\begin{center}
    \includegraphics[viewport=0.5in 2.5in 8in 8in, clip]{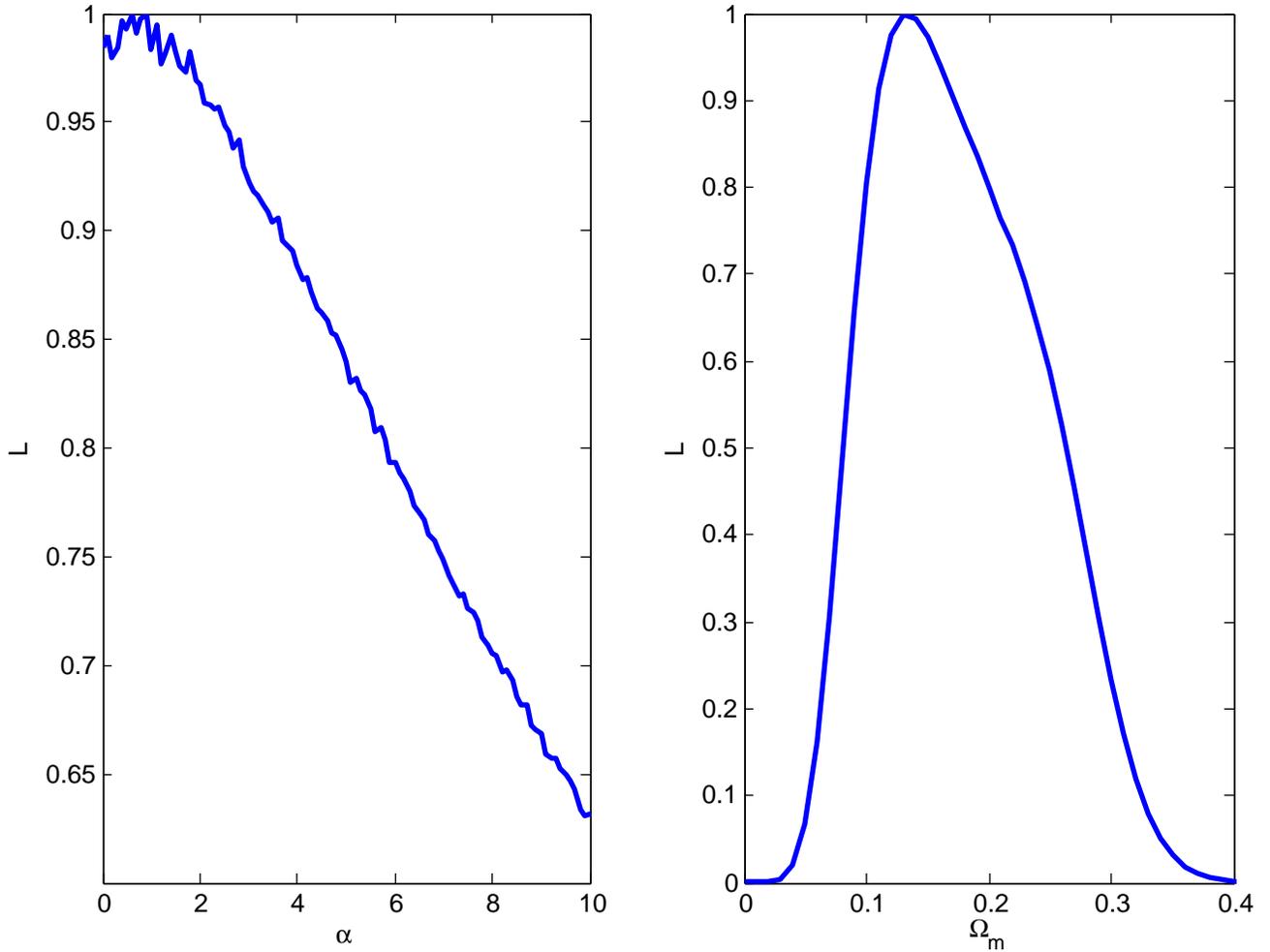}

    \caption[Likelihood functions of individual cosmological parameters from Hubble parameter versus redshift data.]{One dimensional likelihood functions of individual cosmological parameters from Hubble parameter versus redshift data. The maximum likelihood values are normalized to one.}

    \label{fig:hubble1}
    \end{center}
\end{figure}

The 1, 2, and 3$\sigma$ contours on $\phi$CDM are shown in
Fig.~\ref{fig:hubble}. Data constrain the nonrelativistic matter density
parameter to be less than
0.4 at about 3$\sigma$ confidence level. The likelihood in the direction of $\alpha$ is almost flat. The one dimensional likelihood functions on
individual dark energy parameters are shown in Figure~\ref{fig:hubble1}.
$\alpha$ is poorly constrained while the nonrelativistic matter density
parameter is constrained to
be in the range $0.09<\Omega_{\rm m}<0.26$ at 1$\sigma$.

\section{Lookback time}

The data in Appendix~\ref{appendixage} can be used to directly constrain dark energy parameters using the lookback time versus redshift relation \cite{Alcaniz:1999kr,Lima:2000jc,Jain:2005gu,Wei:2007ig,Capozziello:2004jy,Pires:2006rd,Dantas:2006dy,Dantas:2009vs,Yi:2006bw}.
Lookback time at redshift $z$ is defined as the difference between the current age of the Universe and it's age at redshift $z$. In $\phi$CDM the lookback time is given by
\begin{equation}
\label{eq:lb}
t_{\rm lb}(z)=\frac{1}{H_0}\displaystyle\int_0^z \frac{dz}{(1+z)\sqrt{\Omega_{\rm m}(1+z)^3+\Omega_\phi (z)}}.
\end{equation}
For an object at redshift $z$ it's age is the difference between the
Universe's age at $z$ and the one when the object was formed at redshift $z_{\rm f}$.
\begin{equation}
\label{eq:age}
t_{\rm age}(z)=\frac{1}{H_0}\displaystyle\int_z^{z_{\rm f}} \frac{dz}{(1+z)\sqrt{\Omega_{\rm m}(1+z)^3+\Omega_\phi (z)}}=t_{\rm lb}(z_{\rm f})-t_{\rm lb}(z).
\end{equation}
From Eqs.~\eqref{eq:lb} and \eqref{eq:age} the observed lookback time is defined as
\begin{equation}
\label{eq:lbobs}
t_{\rm lb}^{\rm obs}(z)=t_0^{\rm obs} - t_{\rm age}(z) - \tau,
\end{equation}
\noindent
where $t_0^{\rm obs}$ is observed total age of the Universe, $t(z)$ is the objects age at redshift $z$, and $\tau=t_0^{\rm obs}-t_{\rm lb}z_{\rm f}$ 
is a ``delay factor'' that encodes our ignorance of $z_{\rm f}$. 
The ``delay factor'' $\tau$ is different for each
object in the sample.
From CMB experiments the total age of the Universe is estimated to be
$t_0^{\rm obs}=(13.7\pm0.2)\ {\rm Gyr}$ \cite{Dunkley:2008ie}. 

We construct the $\chi^2$ function
\begin{equation}
\label{eq:agechi}
\chi^2(\alpha,\Omega_{\rm
m},H_0,\tau)=\displaystyle\sum_{i=1}^{32}\frac{(t_{\rm
lb}(z_i,\alpha,\Omega_{\rm m},H_0)-t_{\rm lb}^{\rm obs}(z_i))^2}{\sigma_{\rm
t_i}^2}+\frac{(t_{\rm 0}(\alpha,\Omega_{\rm m},H_0) - t_{\rm 0}^{\rm obs})^2}{\sigma_{\rm t_0}^2},
\end{equation}
\noindent
and compute the likelihood function
\begin{equation}
\label{eq:agelik}
\mathcal{L}(\alpha,\Omega_{\rm m},\tau)\propto\exp(-\chi^2(\alpha,\Omega_{\rm m},\tau)/2).
\end{equation}
We account for our ignorance of $\tau$ by marginalizing over it by integrating the likelihood function in Eq.~\eqref{eq:agelik} analytically over $\tau$ with a flat prior.
\begin{equation}
\label{eq:tauint}
\mathcal{L}'(\alpha,\Omega_{\rm m},H_0)=\displaystyle\int \mathcal{L} d\tau=\sqrt{\frac{\pi}{2C}}{\rm erfc}\left(\frac{B}{2C}\right)\exp\left(\frac{B^2-AC-DC}{2C}\right).
\end{equation}
\noindent
where
\begin{align}
A &= \displaystyle\sum_{i=1}^{32}\frac{(t_{\rm lb}(z_i)-t_{\rm lb}^{\rm
obs}(z_i))^2}{\sigma_{\rm t_i}^2},\\
B &= \displaystyle\sum_{i=1}^{32}\frac{(t_{\rm lb}(z_i)-t_{\rm lb}^{\rm
obs}(z_i))}{\sigma_{\rm t_i}^2},\\
C &= \displaystyle\sum_{i=1}^{32}\frac{1}{\sigma_{\rm t_i}^2},\\
D &= \frac{(t_{\rm 0}- t_{\rm 0}^{\rm obs})^2}{\sigma_{\rm t_0}^2},
\end{align}
\noindent
and $\rm erfc(x)$ is the complementary error function. Finally we integrate
over the Hubble constant with a Gaussian WMAP prior with $H_0=(73\pm 3) \ \rm km\ s^{-1}\ Mpc^{-1}$ to get the constraints on $\phi$CDM model
parameters $\alpha$ and $\Omega_{\rm m}$.

\section{Constraints from lookback time versus redshift data}

\begin{figure}[htb]%t=top, b=bottom, h=here
\begin{center}
    \includegraphics[viewport=0.5in 2.5in 8in 8in, clip]{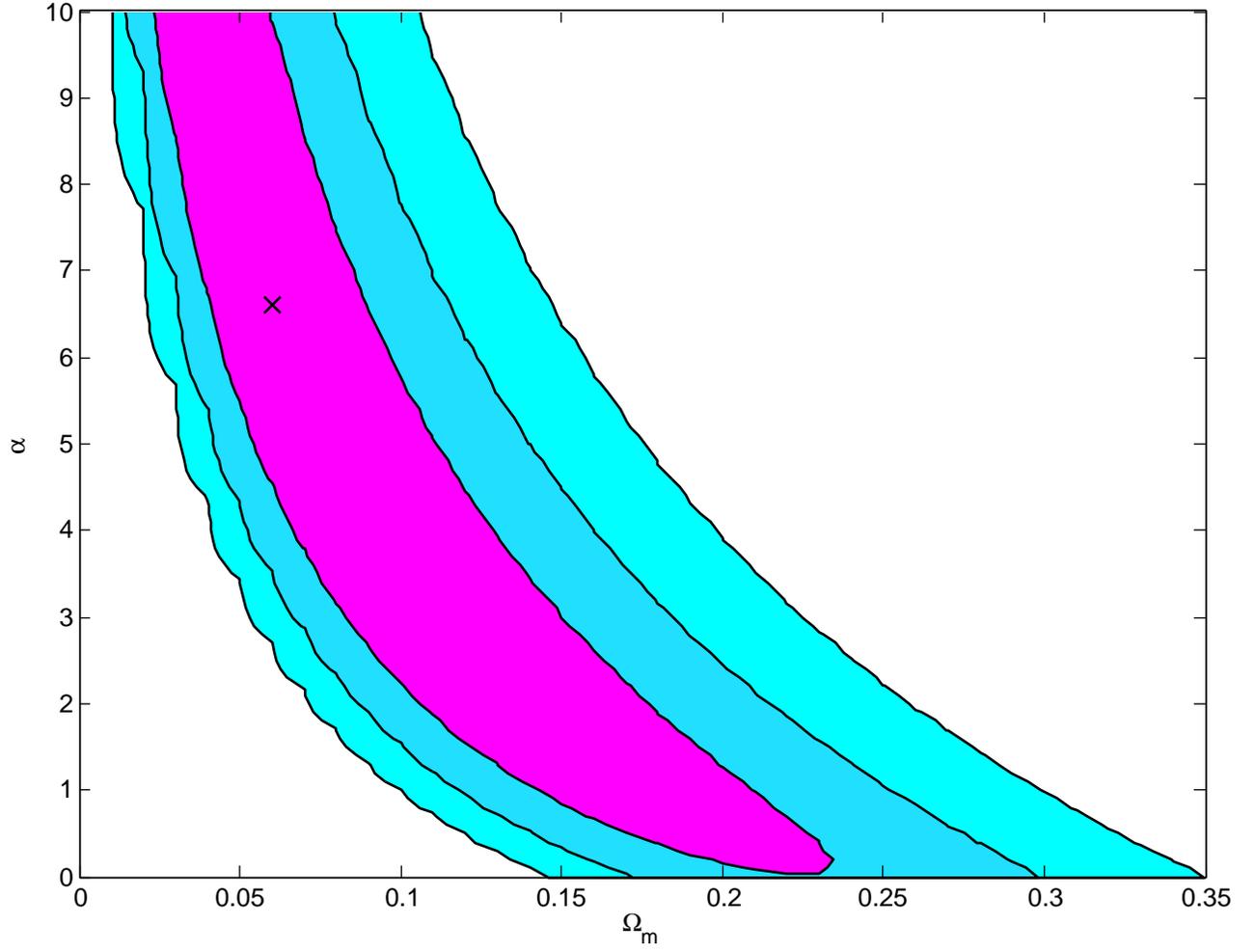}

    \caption[Constraints from lookback time versus redshift data.]{1, 2, and
3$\sigma$ confidence level contours on $\phi$CDM model parameters lookback
time versus redshift data. Cross denotes the best-fit parameters at
$\Omega_{\rm m}^*=0.09$ and $\alpha^*=0.60$ with $\chi^2=20.54$ for $30$ degrees of freedom.}

    \label{fig:lbtime}
    \end{center}
\end{figure}

The constraints on $\phi$CDM from the lookback time method are shown in Fig.~\ref{fig:lbtime}.

Lookback time versus redshift data constrains the nonrelativistic matter density to be less than $\Omega_{\rm m}=0.35$ at about 3$\sigma$ confidence level.
The $\alpha$ parameter is not well constrained and values as large as $\alpha=10$ are allowed at 1$\sigma$ confidence level. The best fit
parameters from the two dimensional likelihood function are $\Omega_{\rm m}^*=0.09$ and $\alpha^*=6.60$.

\begin{figure}[htb]%t=top, b=bottom, h=here
\begin{center}
    \includegraphics[viewport=0.5in 2.5in 8in 8in, clip]{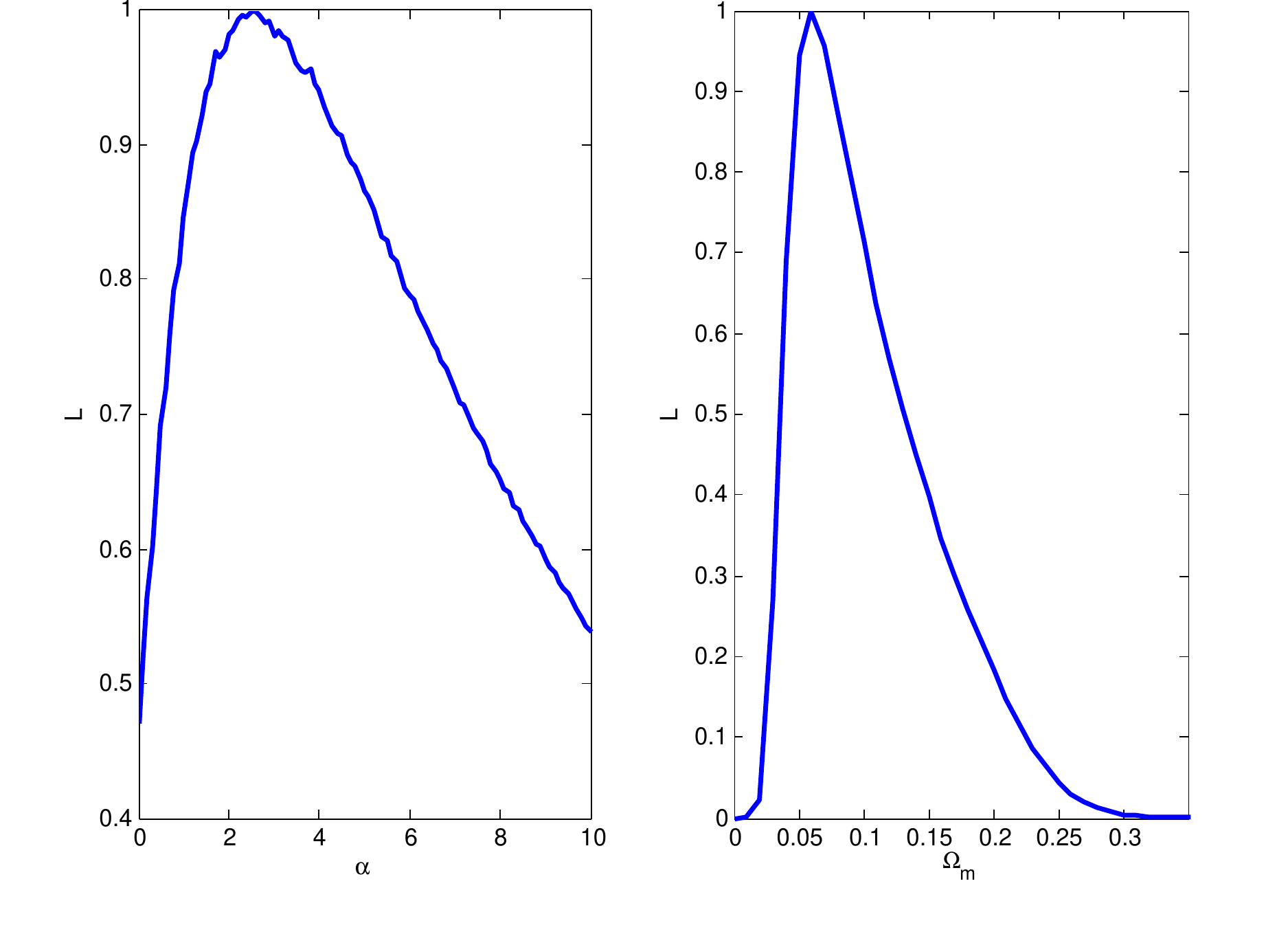}

    \caption[Likelihood functions of individual cosmological parameters from lookback time versus redshift data.]{One dimensional likelihood functions of individual cosmological parameters from lookback time versus redshift data. The maximum likelihood values are normalized to one.}

    \label{fig:lbtime1}
    \end{center}
\end{figure}

The one-dimensional likelihoods of individual cosmological parameters are shown in Fig.~\ref{fig:lbtime1}. The best fit values are $\alpha^*=2.3$ and $\Omega_{\rm m}^*=0.06$. The 1$\sigma$ intervals are $0.3<\alpha<8.9$ and $0.04<\Omega_{\rm m}<0.14$.

At the moment lookback time versus redshift and Hubble parameter versus
redshift measurements can not provide strong constraints on cosmological
parameters. Both data sets are expected to improve significantly in the near
future \cite{Com:jimenez}. The updated data sets in combination with other measurements might prove to be very useful in constraining cosmological parameters.

% +--------------------------------------------------------------------+
% | Sample Chapter 6
% +--------------------------------------------------------------------+

\cleardoublepage

% +--------------------------------------------------------------------+
% | Replace "This is Chapter 3" below with the title of your chapter.
% | LaTeX will automatically number the chapters.
% +--------------------------------------------------------------------+

\chapter{Constraints on $\phi$CDM from SNeIa and GRB Data}
\label{makereference6}

This chapter is based on Ref.~\cite{Samushia:2008wq,Samushia:2009ib}.

\section{Supernova Type Ia as standard candles}

Supernovae (SNe) are highly luminous objects. At the peak of their activity SNe can be as bright as an average galaxy.
They reach maximum brightness shortly after the stellar explosion and then slowly fade away in several weeks.
The shock wave from explosion pushes a spherical shell of gas and dust into the interstellar medium and leaves behind a visible SNe remnant.
SNe are classified based on the absorption lines of chemical elements in their spectra. SNe that do not have hydrogen atom Balmer lines in their spectra
are classified as Type I, while those that have Balmer lines belong to type II. SN can be farther divided into subtypes.
SNe type Ia (SNeIa) for example lack hydrogen lines but have a singly ionized silicon (Si II) line at $615\ \rm nm$.
SNeIa are the most useful for cosmological parameter estimation. First of all, they are astonishingly bright and can be seen at large distances.
Secondly, they are very homogeneous in spectra, light-curve shapes, and peak absolute magnitude and can be easily calibrated and used as standard candles.

SNeIa are believed to be carbon-oxygen white dwarfs in a binary system. The white dwarf accretes matter from its companion until it reaches
the Chandrasekhar mass limit of 1.4 solar masses. As the density and
temperature inside the star increases, the mass in the core undergoes
nuclear fusion, which results in a SNeIa explosion. The masses of white dwarfs
at the time of explosion are very similar  and this could explain why SNeIa
have similar absolute magnitudes. The inhomogeneities in absolute magnitude
of SNeIa are correlated with other observables. For example, brighter SNeIa tend to fade slower. This relationship can be used to further reduce the dispersion in absolute magnitude of SNeIa and make them a highly accurate standardizable candle.

The measurement of absolute magnitudes of high-redshift SNeIa gave the first direct evidence of the accelerated expansion of the Universe 
(see Refs.\ \cite{ct:riess:98, ct:perlmutter:99}). They were followed by more
high-quality observations of SNeIa (see, e.g., Refs.\
\cite{Tonry:2003zg,Knop:2003iy,Barris:2003dq,Astier:2005qq,Riess:2006fw,Miknaitis:2007jd}).
We now have more than 400 well calibrated, 
high-redshift SNeIa, going up to redshift of 1.7, that can be used to determine cosmological parameters.

Ref.\ \cite{Kowalski:2008ez} used results of several new and old observations to compile a new SNeIa data set. They used a single, consistent
analysis on all subsamples and implemented a new procedure to reject
outliers. This compiled ``Union'' data set includes 307 SNeIa in the redshift
range of $z=0.015$ to $1.551$. This data is consistent with the
spatially-flat $\Lambda$CDM model and gives best-fit value
$\Omega_\Lambda=0.731^{+0.027}_{-0.029}(\text{stat})^{+0.036}_{-0.039}(\text{sys})$
in this model. The union data by themselves give more than 5$\sigma$ evidence for the accelerated expansion of the Universe. It has been also used to constrain parameters of other dark energy and modified gravity models (see, e.g., \cite{Rubin:2008wq}).

\section{Constraints on $\phi$CDM from Supernova Type Ia data}

In astronomy, for historical reasons, flux (power per unit area) is usually
given as apparent magnitude
\begin{equation}
\label{eq:flux}
m(z)=-2.5\log\left(\frac{\phi}{\phi_0}\right),
\end{equation}
\noindent
where $m(z)$ is the apparent magnitude of an object at redshift $z$, $\phi$ is its flux, and $\phi_0$ is some arbitrarily chosen reference flux.
Absolute magnitude $M$ measures the objects intrinsic brightness and can be computed from apparent magnitude if the distance to the object is known.
In cosmology fluxes are normalized so that the absolute magnitude of a SNeIa
is equal to the apparent magnitude it would have if it were at a distance
of $100\ \text{Kpc}$. Apparent magnitude then can be expressed through luminosity distance as

\begin{equation}
\label{eq:magnitude}
m(z)=M+25+5\log(d_L(z)).
\end{equation}

Redshifts, distance moduli $\mu = m - M$, and measurement errors for SNeIa in
the Union data set are given in Appendix~\ref{appendixb}. We can compute
luminosity distance at different redshifts in the $\phi$CDM model and compare theoretical predictions with the measurements. The luminosity distance 
depends on the parameter $\alpha$, the nonrelativistic matter energy density
$\Omega_{\rm m}$, and the Hubble constant $H_0$. $H_0$ can be determined from
the data by minimizing the $\chi^2$ function, or it can be marginalized over with a prior probability distribution function $P(H_0)$.

\begin{figure}[htb]%t=top, b=bottom, h=here
\begin{center}
    \includegraphics[viewport=0.5in 2.5in 8in 8in, clip]{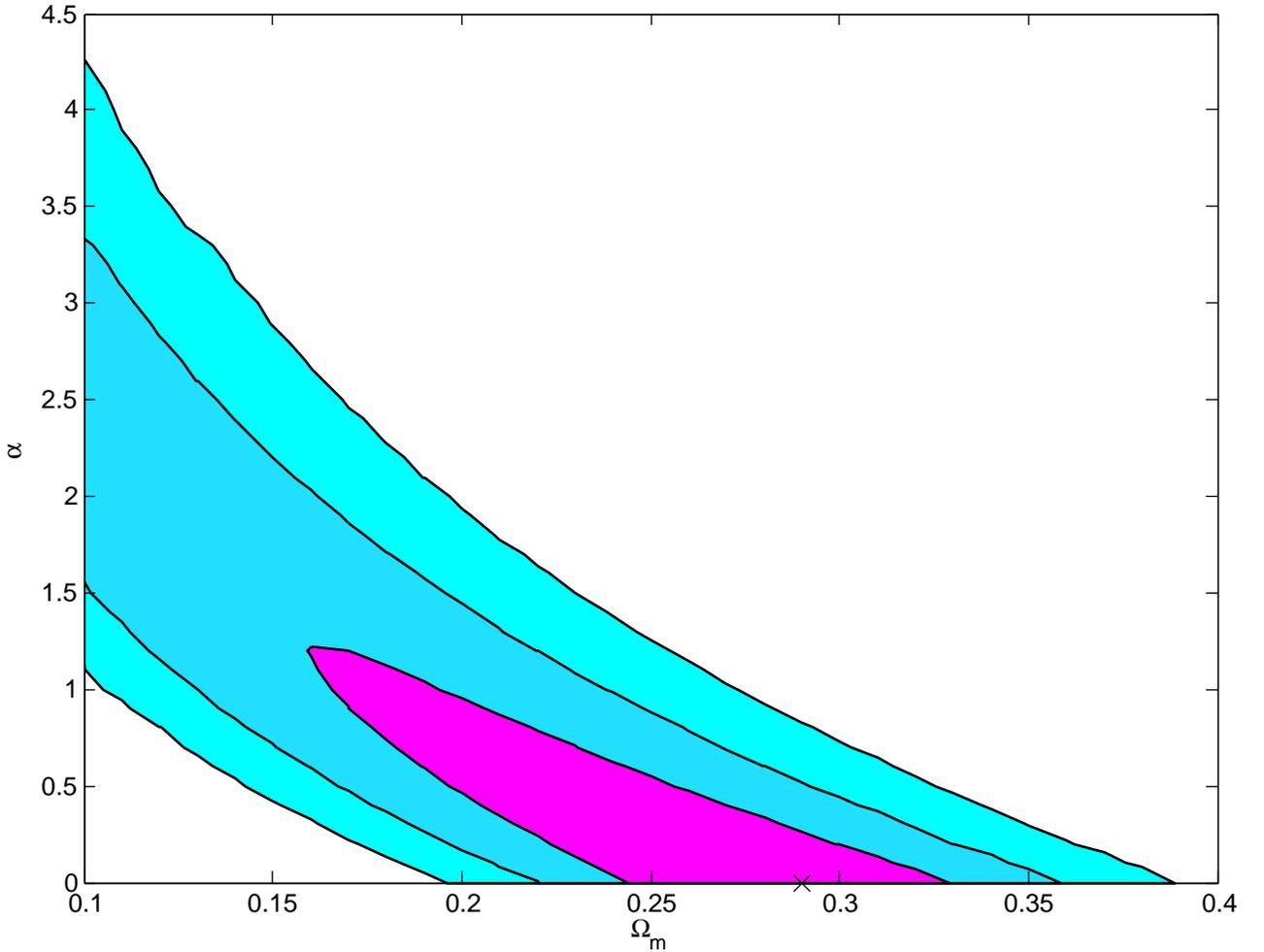}

    \caption[Constraints from SNeIa data.]{1, 2, and 3$\sigma$ confidence
level contours on $\phi$CDM model parameters from SNeIa absolute magnitude
versus redshift data. Cross denotes the best-fit value at $\Omega_{\rm
m}^*=0.29$ and $\alpha^*=0$ with $\chi^2=320$ for $305$ degrees of freedom.}

    \label{fig:sn}
    \end{center}
\end{figure}

Constraints from SNeIa data on the $\phi$CDM model are shown in Fig.~\ref{fig:sn}. SNeIa data constrains the nonrelativistic matter density parameter to 
be less than 0.4 at about 3$\sigma$ confidence level. The $\alpha$ parameter is constrained to be less than 4.5 at about 3$\sigma$ confidence level.

\begin{figure}[htb]%t=top, b=bottom, h=here
\begin{center}
    \includegraphics[viewport=0.5in 2.5in 8in 8in, clip]{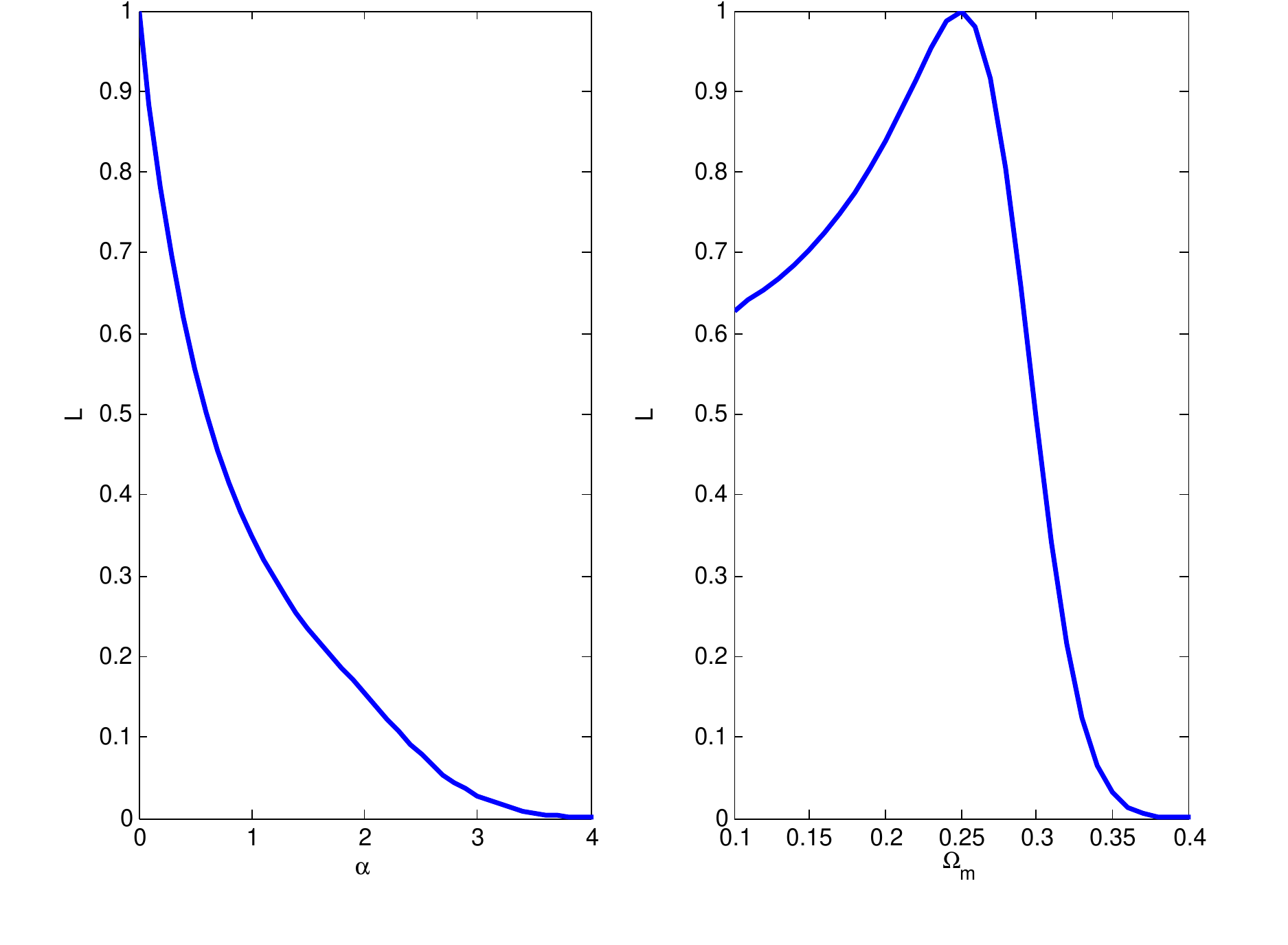}

    \caption[Likelihood functions of individual cosmological parameters from SNeIa data.]{One dimensional likelihood functions of individual cosmological parameters from SNeIa absolute magnitude versus redshift data. The maximum likelihood values are normalized to one.}

    \label{fig:sn1}
    \end{center}
\end{figure}

The one dimensional likelihood functions for individual cosmological parameters are shown in Fig.~\ref{fig:sn1}. The best fit values are $\alpha^*=0$ and $\Omega_{\rm m}^*=0.25$. The 1$\sigma$ intervals are $0<\alpha<1.1$ and $0<\Omega_{\rm m}<0.29$.

\section{GRB as standard candles}

One way of improving our understanding of how dark energy behaves is to study
the evolution of the Universe at redshifts higher than those probed by SNeIa.
This requires standard candles that are visible at greater distances. Gamma-ray bursts (GRBs) could in principle serve as such high redshift standardizable candles. They are the most luminous events in the Universe today and can be seen to beyond $z=8$. If it is definitely established that GRBs are standardizable candles, their visibility at high redshift should prove to be very useful in discriminating between $\Lambda$CDM and time-varying dark energy models.

With the intention of getting cosmological constraints from GRB observations a number of GRB calibrations  have been used so far \cite{Schaefer:2007pq}. One that gives least scatter and therefore most information is 
\begin{equation}
\label{EE}
\log\left(\frac{E_{\gamma}}{1\ \rm erg}\right) = A_1 +  B_1 \log\left(\frac{E_{\rm peak}(1+z)}{300\ \rm keV}\right),
\end{equation}
\noindent
a relation that connects the total burst energy of the GRB ($E_{\gamma}$) to the peak energy of 
the GRB spectrum ($E_{\rm peak}$) \cite{Ghirlanda:2004me}. Regrettably, we do not yet have a model-independent way of computing the coefficients $A_1$ and $B_1$. A better understanding of physical processes that result in the burst, or observations of nearby GRBs (to which distances can be measured independently), could in principle help us to calibrate the $E_{\gamma}$-$E_{\rm peak}$ relation without any prior assumptions. Until recently the only way to extract cosmological information from GRBs was to recalibrate them for every dark energy model considered (at each set of parameter values). This is time consuming and also results in large statistical uncertainties and hence GRB cosmological constraints that are poor. 

Recently, two methods of calibrating GRBs in cosmology-independent manners have been proposed and used to constrain some dark energy models. One method externally calibrates GRBs, by using SNeIa measurements \cite{Kodama:2008dq,Liang:2008kx,Wei:2008kq,Liang:2008hp}, while the other uses GRB measurements alone \cite{Wang:2008vja}. While externally calibrating GRBs results in tighter cosmological constraints, internally calibrated GRB data may be straightforwardly combined with other data when deriving cosmological constraints. The resulting cosmological constraints are still loose, but in the future when more high precision GRB observations become available this could provide a strong test of dark energy.

Reference~\cite{Wang:2008vja} recently used data of 69 GRBs \cite{Schaefer:2007pq} to construct a model-independent distance measure that can be used to constrain cosmological models. When this method is used to constrain $\Lambda$CDM the GRB data favor lower values of both cosmological constant energy density ($\Omega_\Lambda$) and nonrelativistic matter energy density ($\Omega_{\rm m}$) than do the SNeIa data. The GRB data by themself are unable to strongly constrain cosmological parameters, for example in spatially-flat $\Lambda$CDM the GRB data require $\Omega_{\rm m}=0.25^{+0.12}_{-0.11}$ at 1$\sigma$ confidence \cite{Wang:2008vja}.

For early discussions of the use of GRBs as a cosmology probe see, e.g., Refs.~\cite{Ghirlanda:2004me,Friedman:2005cz,Firmani:2005gs,Xu:2005uv,Mortsell:2005yb}. More recent studies may be traced back through Refs.~\cite{MosqueraCuesta:2007xv,Basilakos:2008tp,Capozziello:2008tc,Tsutsui:2008cu}. For a review of GRB physics see, e.g., Ref.~\cite{Meszaros:2006rc}.

\section{Constraints from GRB data}

Besides the $E_{\rm peak}$-$E_\gamma$ relation, Eq.\ \eqref{EE}, four other
calibrations for GRBs are used in Ref.~\cite{Wang:2008vja}. Reference~\cite{Wang:2008vja} placed each of the 69 GRBs in the redshift range $z=0.17$ to $z=6.6$ at a luminosity distance that minimized a combined $\chi^2$ that took weighted account of all five calibration relations. From this \cite{Wang:2008vja} computed the cosmology-independent distance measure
\begin{equation}
\label{dm}
\bar r_{\rm p}=\frac{r_{\rm p}(z)}{r_{\rm p}(0.17)},
\end{equation}
\noindent
where 
\begin{equation}
\label{rp}
r_{\rm p}(z)=\frac{H_0}{hc}\frac{1}{z(1+z)^{1/2}}d_L(z),
\end{equation}
\noindent
and $d_L(z)$ is the luminosity distance at redshift $z$, and $c$ is the speed of light. The ratio in Eq.\ (\ref{dm}) does not depend on the Hubble constant and does not require information about the absolute calibration of GRBs (which are unknown). 

Reference~\cite{Wang:2008vja} computed the cosmology-independent distance measure $\bar r_{\rm p}$ in six redshift bins $\bar r_{\rm p}(z_{\rm i})$, $i=1,2,\dots6$. The values of $\bar r_{\rm p}(z_{\rm i})$ are shown in Appendix~\ref{appendixgrb} and the normalized covariance matrix is

\begin{equation}
\label{eq:covmat}
S=\left(\begin{array}{cccccc}
           1.0000 & 0.7056 & 0.7965 & 0.6928 & 0.5941 & 0.5169\\
           0.7056 & 1.0000 & 0.5653 & 0.6449 & 0.4601 & 0.4376\\
           0.7965 & 0.5653 & 1.0000 & 0.5521 & 0.5526 & 0.4153 \\
           0.6928 & 0.6449 & 0.5521 & 1.0000 & 0.4271 & 0.4242 \\
           0.5941 & 0.4601 & 0.5526 & 0.4271 & 1.0000 & 0.2999 \\
           0.5169 & 0.4376 & 0.4153 & 0.4242 & 0.2999 & 1.0000 \\
           \end{array}
           \right)
\end{equation}
 For currently viable cosmological models, these $\bar r_{\rm p}(z_{\rm i})$
are almost completely independent of the cosmological model and so provide a
useful summary of current GRB data \cite{Wang:2008vja}. This information can be used to constrain any dark energy model and the resulting GRB data constraints can be straightforwardly combined with other constraints.

We compute the difference between the theoretical prediction and the measured
value at each of the 6 redshifts,
\begin{equation}
\Delta(z_{\rm i}) = \bar r_{\rm p}^{\rm data}(z_{\rm i}) - \bar r_{\rm p}^{\rm theory}(z_{\rm i}),
\end{equation}
\noindent
and the $\chi^2$ function
\begin{equation}
\chi^2(\Omega_{\rm m}, p) = \Delta(z_{\rm i}) \sigma_{\rm i} (S^{-1})_{\rm i j} \sigma_{\rm j} \Delta(z_{\rm j}),
\end{equation}
\noindent
where $S_{\rm i j}$ is the normalized covariance matrix given by
Eq.~\eqref{eq:covmat} and summation over repeated indexes is assumed. Here
$\sigma_{\rm i}$ is $\sigma_{\rm i}^+$ if $\Delta(z_{\rm i})>0$ and
$\sigma_{\rm i}^-$ if $\Delta(z_{\rm i})<0$, see Appendix~\ref{appendixgrb}.

\begin{figure}[htb]%t=top, b=bottom, h=here
\begin{center}
    \includegraphics[viewport=0.5in 2.5in 8in 8in, clip]{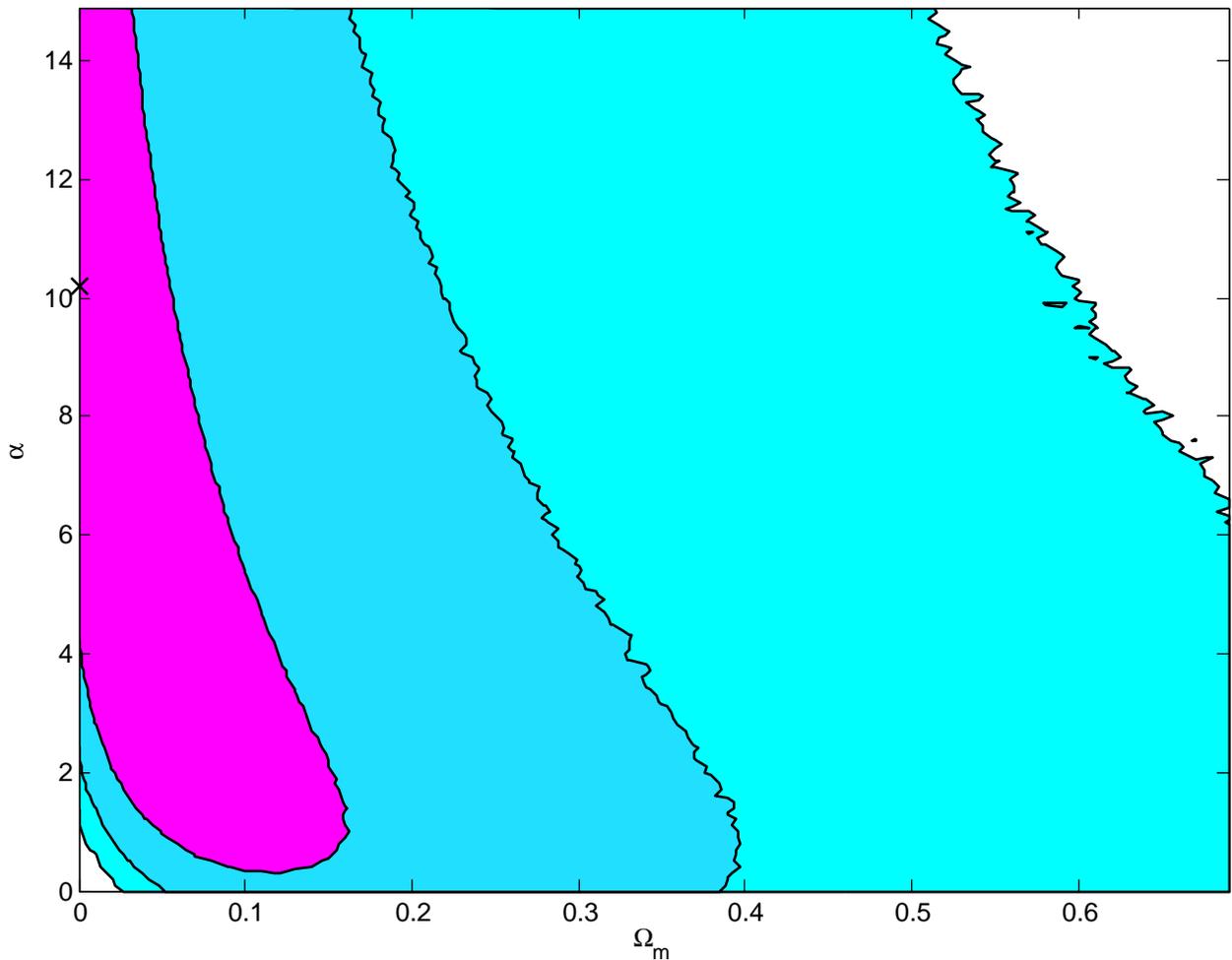}

    \caption[Constraints from GRB data.]{1, 2, and 3$\sigma$ confidence level contours on $\phi$CDM model parameters from GRB data. Cross denotes the best-fit value at $\Omega_{\rm m}^*=0.29$ and $\alpha^*=0$.}

    \label{fig:grb}
    \end{center}
\end{figure}

The constraints on $\phi$CDM from GRB data are shown in Fig.~\ref{fig:grb}. Current GRB data can not constrain cosmological parameters well and a wide
range of values are at 3$\sigma$ confidence level. The best-fit parameter values are $\alpha^*=10.2$ and $\Omega_{\rm m}^*=0.0$ with $\chi^2=1.39$ for 4 degrees of freedom.

\begin{figure}[htb]%t=top, b=bottom, h=here
\begin{center}
    \includegraphics[viewport=0.5in 2.5in 8in 8in, clip]{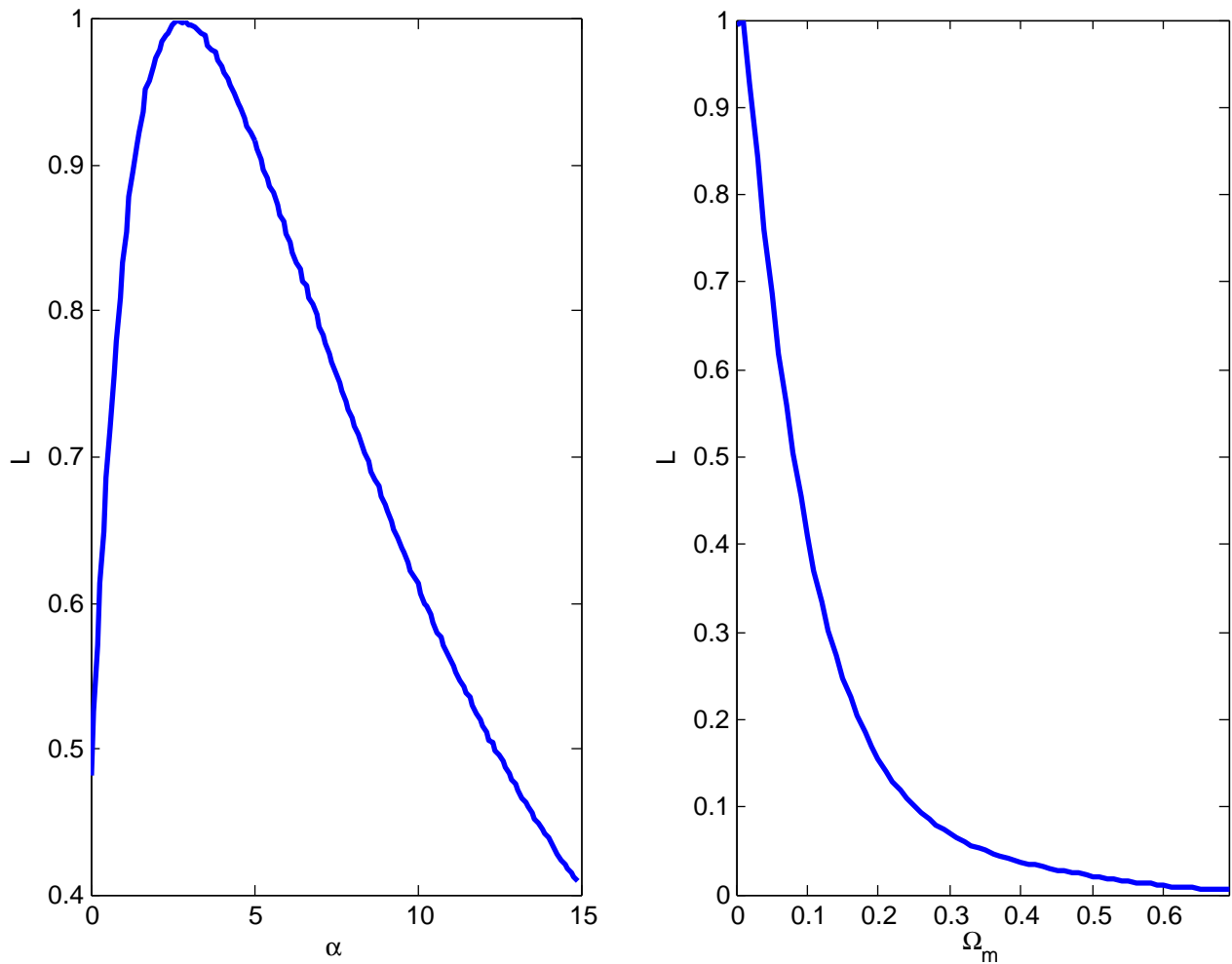}

    \caption[Likelihood functions of individual cosmological parameters from GRB data.]{One dimensional likelihood functions of individual cosmological parameters from GRB data. The maximum likelihood values are normalized to one.}

    \label{fig:grb1}
    \end{center}
\end{figure}

The one dimensional likelihood functions for individual parameters are shown in Fig.~\ref{fig:grb1}. The best fit parameter values are
$\alpha^*=2.3$ and $\Omega_{\rm m}^*=0$. One standard deviation intervals are $0<\alpha<12.3$ and $0<\Omega_{\rm m}<0.17$.

GRB data by themselves do not provide tight constraints on cosmological parameters. Moreover, while not greatly significant, current GRB data favor cosmological parameter values that are at odds with what other data favor. When used in combination with some of the highest quality current data (e.g., SNeIa and BAO peak measurements) current GRB data only slightly change the results. This is mainly because in the absence of an absolute calibration of GRBs they, as standard candles, have big measurement uncertainties. This is however quite likely to change as more and better-quality GRB measurements become available and with improvements in how to calibrate GRBs. GRBs could potentially provide a very strong test of the time variation of dark energy as they can be observed up to redshifts beyond 8, at distances where other standard candles cannot be detected. 

% +--------------------------------------------------------------------+
% | Sample Chapter 7
% +--------------------------------------------------------------------+

\cleardoublepage

% +--------------------------------------------------------------------+
% | Replace "This is Chapter 3" below with the title of your chapter.
% | LaTeX will automatically number the chapters.
% +--------------------------------------------------------------------+

\chapter{Constraints from Cluster Gas Mass Fraction Data}
\label{makereference7}

This chapter is based on Ref.~\cite{Samushia:2008fk}.

\section{Using galaxy clusters as cosmological probes}

Galaxy clusters are the biggest gravitationally bound objects in the
Universe. Old, relaxed, rich ones should provide a fair sample of the matter
content of the Universe.
The ratio of baryonic mass to total nonrelativistic mass in the clusters should be close to the ratio of cosmological parameters
$\Omega_{\rm b}/\Omega_{\rm m}$, the baryonic fraction $f$. More than $80\%$ of clusters' baryonic mass is in hot X-ray emitting intergalactic gas. Estimates of gas mass fraction
in clusters can be combined with the priors on $\Omega_{\rm b}$ from WMAP or BBN to get tight constrains on nonrelativistic matter density.
During cosmic evolution the ratio $\Omega_{\rm b}/\Omega_{\rm m}$ should
remain the same, but clusters are observed at different redshifts and since
the
reconstructed $f_{\rm gas}$ depends on the assumed distance to the cluster, this data can be used to constrain cosmic evolution and hence the dark energy.

To get a good estimate of $f_{\rm gas}$, the gas mass fraction, large, relaxed clusters in thermal equilibrium should be used. Clusters with high core
temperature ($kT>5\ \rm keV$) are better for the analysis, because in computer simulations and observations they have less systematic scatter in gas mass
fraction compared to ``cold'' clusters.

The gas mass fraction is given by \cite{Allen:2007ue}
\begin{equation}
\label{eq:fraction}
f_{\rm gas}(z)=\frac{KA\gamma b(z)}{1+s(z)}\frac{\Omega_{\rm b}}{\Omega_{\rm m}}\left(\frac{d_{\rm A}^{\rm ref}(z)}{d_{\rm A}(z)}\right)^{1.5}.
\end{equation}
\noindent
Here $K$ is a calibration constant that accounts for effects, such as
accuracy of instrument calibration and X-ray modelling. A conservative $10\%$
Gaussian uncertainty $K=1.0\pm0.1$ is used in Ref.~\cite{Allen:2007ue}. 
The factor $A$ accounts for the change in the angle subtended by the cluster
between the reference cosmology and
the cosmology of interest and is always very close to $1$. $\gamma$ models non-thermal pressure due to subsonic motion in the gas, cosmic rays,
magnetic fields, etc. In the analysis of Ref.~\cite{Allen:2007ue} uniform
prior with $1.1<\gamma<1.2$ is assumed. $b$ is the ``bias'' factor, the ratio by which the baryon fraction
in the cluster differs from the same mean ratio in the Universe. The ``bias''
factor is modeled as $b=b_0(1+\alpha_{\rm b}z)$ \cite{Allen:2007ue} and cosmological
simulations suggest the uniform priors $0.65<b_0<1.0$ and $-0.1<\alpha_{\rm
b}<0.1$, \cite{Allen:2007ue}. The parameter $s=s_0(1+\alpha_{\rm s}z)$ models the baryon gas mass fraction
in stars. We use the uniform prior with $-0.2<\alpha_{\rm s}<0.2$ and the Gaussian
prior with $s_0=0.13\pm0.01$, \cite{Allen:2007ue}. $d_{\rm A}^{\rm ref}$ is the angular diameter distance
computed in a reference, spatially-flat $\Lambda$CDM model with
$\Omega_\Lambda=0.7$, and $d_{\rm A}$ is the angular diameter distance
computed
in the $\phi$CDM model.

\section{Many dimensional integration}

We use the modified technique discussed below to perform some of the
integrals to account for these nuisance parameters, which have usually veen
done numerically, as discussed next.

Let's say we want to integrate a function $f(x)$ on the interval $(a,b)$.
If this cannot be done analytically, different numerical
integration algorithms may be used to compute the integral to a given precision.

If the problem is one dimensional and the integrand is reasonably well-behaved, $(a,b)$ can be subdivided into $N$ segments of equal length; in each segment
$f(x)$ can be interpolated by a polynomial; and the integral can be
approximated by a sum \cite{Press}. The simplest method of this sort uses the trapezoidal
Newton-Cotes formula
\begin{equation}
\label{eq:nc}
\displaystyle\int_a^b f(x)dx \simeq \frac{b-a}{N}\left(\frac{f(a) + f(b)}{2} + \displaystyle\sum_{k=1}^{N-1} f\left(a + k\frac{b-1}{N}\right)\right)
\end{equation}
The error on the integral computed using Eq.~\eqref{eq:nc} can be approximated by
\begin{equation}
\label{eq:ncerror}
\sigma_{\rm I} \sim \frac{1}{2N}\sup|f'(x)| 
\end{equation}
\noindent
where prime denotes the derivative. The error could be smaller than this if the function is sufficiently smooth.

This method can be applied to multi-dimensional integrals also, by
transforming them into repeated one-dimensional integrals. The number of function evaluations
required to get the same precision, however, will grow exponentially in this case and so will the required computational time. To sample a unit interval
in one dimensional space with a distance between adjacent points of 0.01, only hundred evenly-spaced points are required. To do the same thing
in 10 dimensional space will require $10^{20}$ points.

To cut down on computational time Monte-Carlo integration techniques can be
employed. Monte-Carlo algorithms randomly pick $N$ points in the
multi-dimensional space. By the law of large numbers, the error on an integral computed using Monte-Carlo method scales as $1/\sqrt{N}$ independent of dimensionality.

The gas mass fraction in Eq.~\eqref{eq:fraction} depends on eight ``nuisance'' parameters. These are $K$, $\gamma$, $b_0$, $\alpha_{\rm b}$, $s_0$, 
$\alpha_{\rm s}$, $\Omega_{\rm b}$, and $H_0$. To perform this eight-dimensional integration we use a method that in this particular case works
faster than Monte-Carlo sampling.

Five out of eight nuisance parameters enter Eq.~\eqref{eq:fraction} as a
redshift independent combination $\varGamma=K\gamma b_0 (\Omega_{\rm
b}h^2)/h^{1/2}$. We can think of each nuisance parameter as a random number
with a given probability distribution function (PDF). Then $\varGamma$ will
also be a 
random variable with some PDF.

We use the following statistics results \cite{Riley}. If two random variables $a$ and $b$ are independent with PDFs $P_{\rm a}(x)$ and $P_{\rm b}(x)$ then variables
$c=ab$, $d=a/b$ and $f=F(a)$ are also random with PDFs
\begin{equation}
\label{eq:stat1}
P_{\rm c}=\displaystyle\int\int P_{\rm a}(x')P_{\rm b}(x'')\delta(x'x''-x)dx'dx''=\displaystyle\int\frac{1}{|x'|}P_{\rm a}(x')P_{\rm b}(x/x')dx',
\end{equation}
\begin{equation}
\label{eq:stat2}
P_{\rm d}=\displaystyle\int\int P_{\rm a}(x')P_{\rm b}(x'')\delta(x'/x''-x)dx'dx''=\displaystyle\int|x'|P_{\rm a}(xx')P_{\rm b}(x')dx',
\end{equation}
\begin{equation}
\label{eq:stat3}
P_{\rm f}=\left|\frac{F^{-1}(x)}{dx}\right|P_{\rm a}(F^{-1}(x)).
\end{equation}
We first numerically compute the PDF for $\varGamma$ using
Eqs.~(\ref{eq:stat1}-\ref{eq:stat3}) and replace the the five-fold
integration of the likelihood
function by a one-dimensional integral. This reduces computational time significantly.

\section{Constraints from $f_{\rm gas}$ data}

We use the measurements of gas mass fraction of 42 hot, X-ray luminous,
relaxed galaxy clusters in the redshift range of $z=0.01$ to $1.1$
\cite{Allen:2007ue}. The cluster gas mass fraction data is presented in
Appendix~\ref{appendixc}. We integrate the likelihood over the Hubble
constant with the Gaussian WMAP prior $h=0.73\pm0.03$, and baryonic mass
energy density with a Gaussian prior $\Omega_{\rm b}=(0.0223\pm0.0008)h^{-2}$.

\begin{figure}[htb]%t=top, b=bottom, h=here
\begin{center}
    \includegraphics[viewport=0.5in 2.5in 8in 8in, clip]{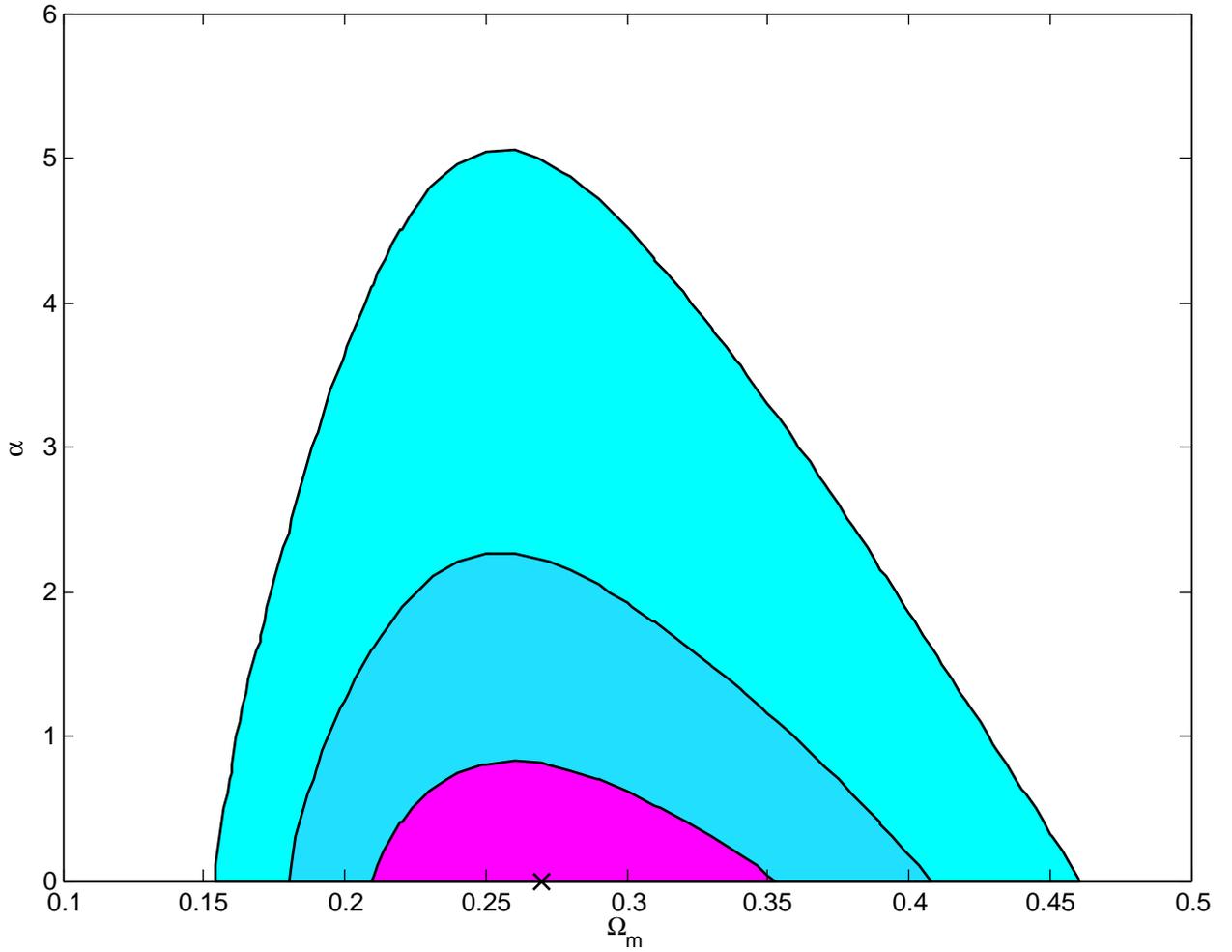}

    \caption[Constraints from galaxy cluster gas mass fraction data.]{1, 2, and 3$\sigma$ confidence level contours on $\phi$CDM model parameters from galaxy cluster gas mass fraction data. Cross denotes the best-fit value at $\Omega_{\rm m}=0.27$ and $\alpha=0.0$.}

    \label{fig:clusters}
    \end{center}
\end{figure}

The constraints on $\phi$CDM parameters are shown in Fig.~\ref{fig:clusters}. Brown, light blue and deep blue areas correspond to 1, 2,
and 3$\sigma$ confidence levels respectively. The black cross denotes best-fit parameters
$\Omega_{\rm m}^*=0.27$ and $\alpha^*=0.0$, with $\chi^2=43.5$ for
40 degrees of freedom. The nonrelativistic matter density is constrained to be in the range of $0.15<\Omega_{\rm m}<0.45$ at about 3$\sigma$ confidence
level and $\alpha$ is constrained to be less than 5 at about 3$\sigma$ confidence level.

\begin{figure}[htb]%t=top, b=bottom, h=here
\begin{center}
    \includegraphics[viewport=0.5in 2.5in 8in 8in, clip]{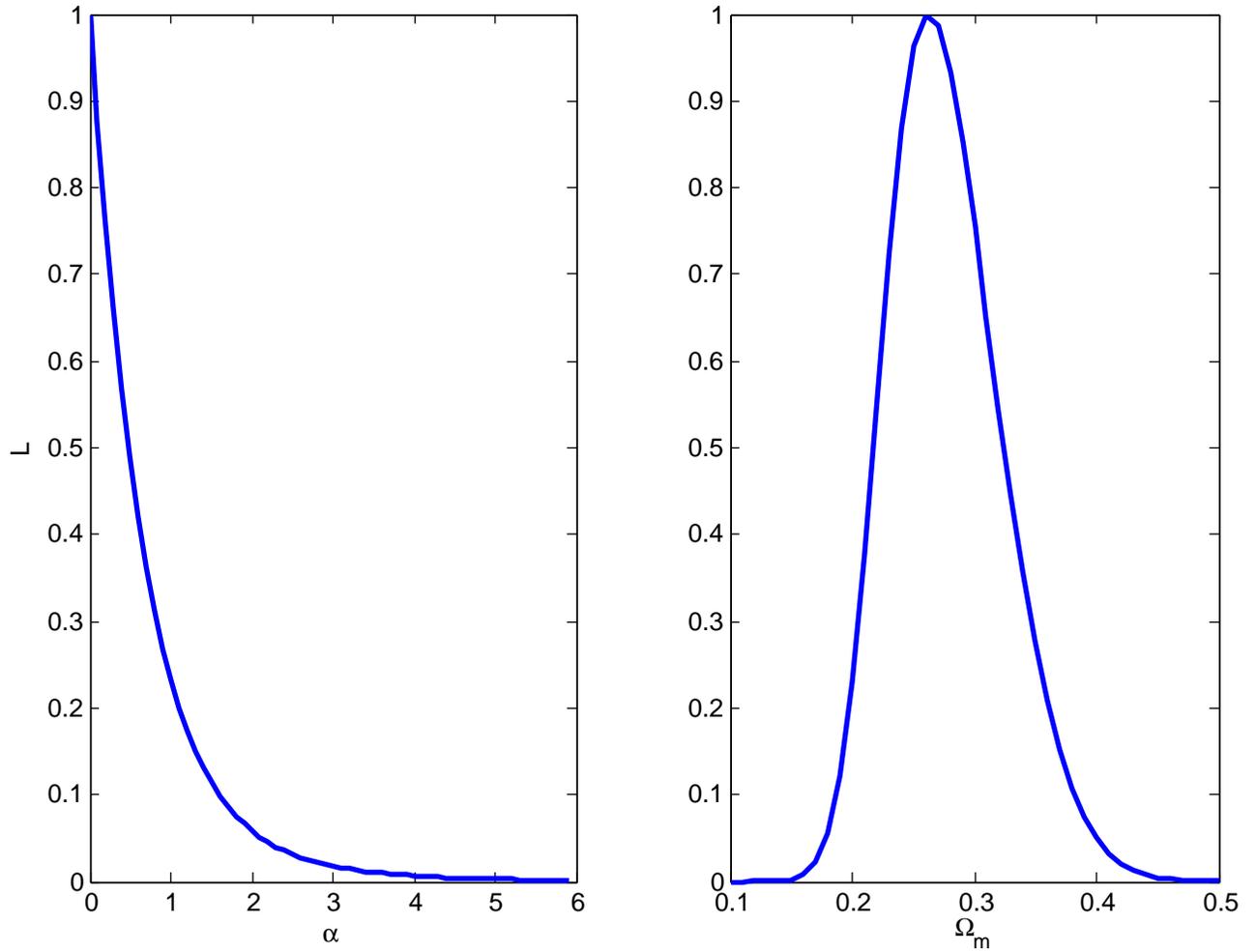}

    \caption[Likelihood functions of individual cosmological parameters from galaxy cluster gas mass fraction data.]{One dimensional likelihood functions of individual cosmological parameters from galaxy cluster gas mass fraction data. The maximum likelihood values are normalized to one.}

    \label{fig:clusters1}
    \end{center}
\end{figure}

One dimensional likelihood functions for individual parameters are shown in
Fig.~\ref{fig:clusters1}. The best-fit values from the one dimensional
likelihood functions are $\alpha^*=0.0$ and $\Omega_{\rm m}^*=0.27$.
1$\sigma$ confidence level intervals from the one dimensional likelihood
functions are
$0<\alpha<1.1$ and $0.22<\Omega_{\rm m}<0.33$.

% +--------------------------------------------------------------------+
% | Sample Chapter 8
% +--------------------------------------------------------------------+

\cleardoublepage

% +--------------------------------------------------------------------+
% | Replace "This is Chapter 3" below with the title of your chapter.
% | LaTeX will automatically number the chapters.
% +--------------------------------------------------------------------+

\chapter{Constraints on $\phi$CDM from Baryon Acoustic Peak Measurements}
\label{makereference8}

This chapter is based on Refs.~\cite{Samushia:2008wq,Samushia:2008pi}.

\section{Transverse baryon acoustic peak measurements}

Before recombination baryons and photons are tightly coupled and gravity and pressure gradients induce sub-acoustic-Hubble-radius oscillations in the baryon-photon fluid \cite{Sunyaev:1970eu,Peebles:1970ag}. These transmute into the acoustic peaks observed now in the CMB anisotropy angular power spectrum, which provide very useful information on various cosmological parameters. The baryonic matter gravitationally interacts with the dark matter and so the matter power spectrum should also exhibit these ``baryon acoustic" wiggles. Because the baryonic matter is a small fraction of the total matter the amplitudes of the BAO wiggles are small. The BAO peak length scale is set by the sound horizon at decoupling, $\sim 10^2 \ {\rm Mpc}$, and so detecting the BAO peak in a real space correlation function requires observationally sampling a large volume. The BAO peak in the galaxy correlation function has recently been detected by using SDSS data \cite{Eisenstein:2005su,Hutsi:2005qv} and by using 2dFGRS data \cite{Cole:2005sx}. For more recent discussions of the observational situation see Refs.~\cite{Padmanabhan:2006ku,Percival:2006gs,Percival:2007yw}.

The sound horizon at decoupling can be computed from relatively well-measured quantities by using relatively well-established physics. Consequently it is a standard ruler and can be used to trace the universe's expansion dynamics \cite{Blake:2003rh,Linder:2003ec,Seo:2003pu, Hu:2003ti}. A measurement of the BAO peak length scale at redshift $z$ fixes a combination of the angular diameter distance and Hubble parameter at that redshift. More precisely, what is determined \cite{Eisenstein:2005su} is the distance
\begin{equation}
\label{eq:distancem}
D_V(z)=\left[(1+z)^2d_A^2(z)z/H(z)\right]^{1/3},
\end{equation}
\noindent
where $H(z)$ is the Hubble parameter and $d_A$ the angular diameter distance.
$D_V(z)$ depends on the cosmological parameters of the model, including those which describe dark energy, so we can constrain these parameters by comparing the predicted $D_V(z)$ to the measurements.

We examine the constraints on $\phi$CDM from two measurements of the BAO peak. The first is from the BAO peak measured at $z=0.35$ in the correlation function of luminous red galaxies in the SDSS \cite{Eisenstein:2005su}. This measurement results in $A(0.35)=0.469\pm0.017$ (one standard deviation error), where the dimensionless and $H_0$-independent function
\begin{equation}
A(z)=D_V(z)\frac{\sqrt{\Omega_{\rm m}H_0^2}}{z}
\end{equation}
\noindent
and $D_V(z)$ is the distance measure defined in Eq.~(\ref{eq:distancem}). The measured value of $A(0.35)$ does not depend on the dark energy model and only weakly depends on the baryonic energy density. The measurement also has a weak dependence on parameters like the spectral index of primordial scalar energy density perturbations (the assumed value is $n=0.98$) and the sum of the neutrino masses, but this is not strong enough to have significant effect on the final result. To constrain cosmological model parameters in this case we perform a standard $\chi^2$ analysis.

The second BAO peak measurement we use is from the correlation function of galaxy samples drawn from the SDSS and 2dFGRS at two different redshifts, $z=0.2$ and $z=0.35$, as determined by Ref.~\cite{Percival:2007yw}. This analysis includes the SDSS luminous red galaxies, so the the two BAO peak measurements are not statistically independent.

This measurement gives the correlated values $r_s/D_V(0.2)=0.1980\pm0.0058$ and $r_s/D_V(0.35)=0.1094\pm0.0033$ (one standard deviation errors), where $r_s$ is the comoving sound horizon at recombination. These two measurements are correlated, with the inverse of the correlation matrix given b
\begin{equation}
V^{-1}=\left( \begin{array}{c c}
35059& -24031\\
-24031& 108300\\
\end{array}
\right).
\nonumber
\end{equation}
\noindent
To compute $r_s$ we first compute the angular diameter distance to the surface of last scattering, $d_A(1089)$. We then use the WMAP measurement of the apparent acoustic horizon angle in the CMB anisotropy data \cite{Spergel:2006hy} to determine the sound horizon $r_s=[(1+z)d_A(z)]|_{z=1089}\times 0.0104$ (where we ignore the WMAP measurement uncertainty and assume that $r_s$ is known perfectly). The use of the WMAP prior on the apparent acoustic horizon angle results in very tight constraints on the spatial curvature. When this measurement is not used, these measurements alone can not tightly constrain the dark energy parameters.

To constrain cosmological parameters in this case we follow Ref.~\cite{Percival:2007yw} and first compute
\begin{equation}
X(\Omega_{\rm m}, \alpha)=\left( \begin{array}{c}
 r_s/D_V(0.2,\Omega_{\rm m},\alpha)-0.1980\\
r_s/D_V(0.35,\Omega_{\rm m},\alpha)-0.1094\\ 
\end{array} \right),
\end{equation}
\noindent
where for definiteness we consider the $\phi$CDM model. We then compute the $\chi^2$ function
\begin{equation}
\chi^2(\Omega_{\rm m}, \alpha)=X^{-1}V^{-1}X.
\end{equation}
\noindent
and the likelihood function
\begin{equation}
L(\Omega_{\rm m}, \alpha)\propto{\rm exp}(-\chi^2(\Omega_{\rm m}, \alpha)/2).
\end{equation}

\section{Constraints from transverse BAO peak measuremnts}

\begin{figure}[htb]%t=top, b=bottom, h=here
\begin{center}
    \includegraphics[viewport=0.5in 2.5in 8in 8in, clip]{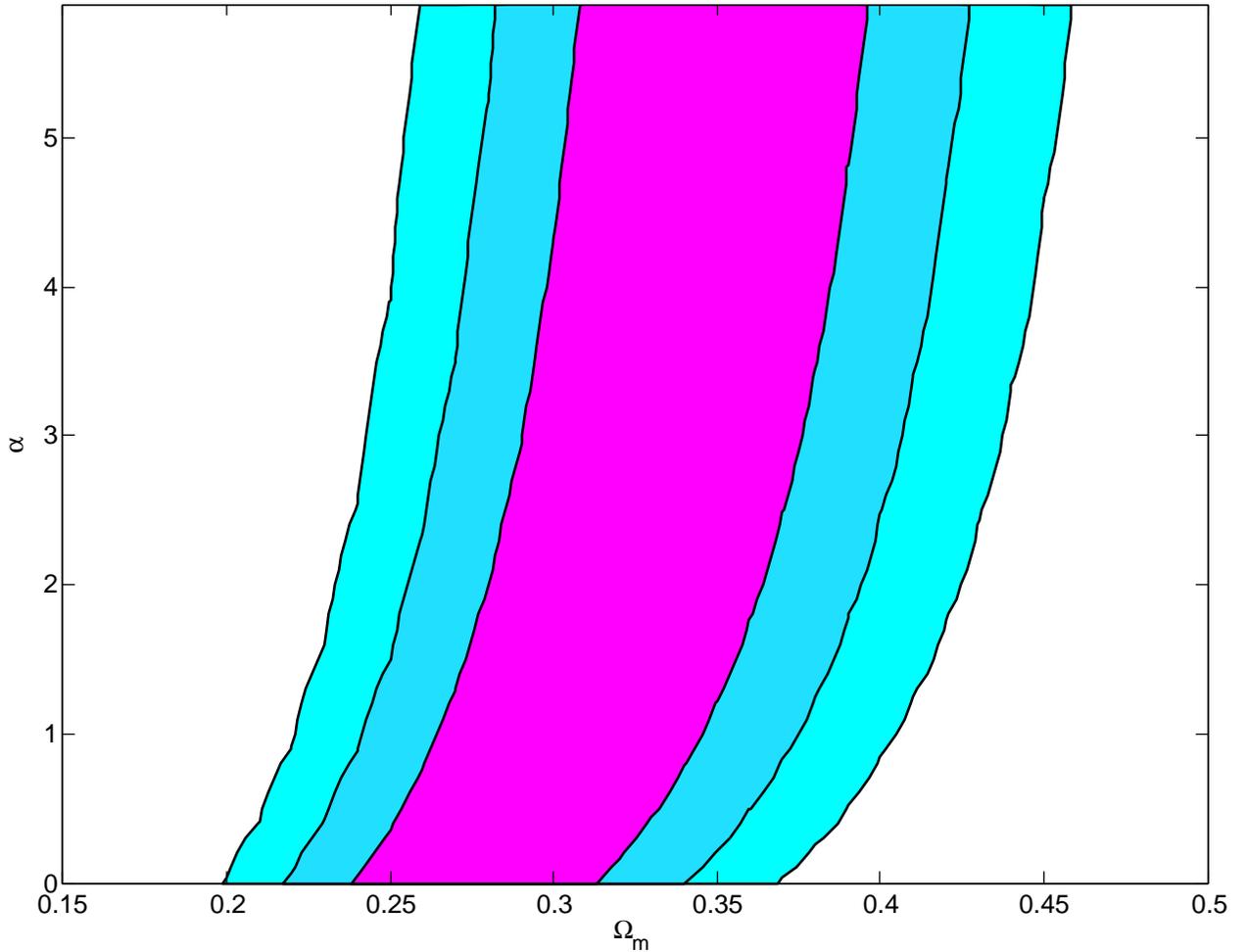}

    \caption[Constraints from BAO peak measurement.]{1, 2, and 3$\sigma$ confidence level contours on $\phi$CDM model parameters from BAO peak measurement of Ref.~\cite{Eisenstein:2005su}.}

    \label{fig:eisenst}
    \end{center}
\end{figure}

\begin{figure}[htb]%t=top, b=bottom, h=here
\begin{center}
    \includegraphics[viewport=0.5in 2.5in 8in 8in, clip]{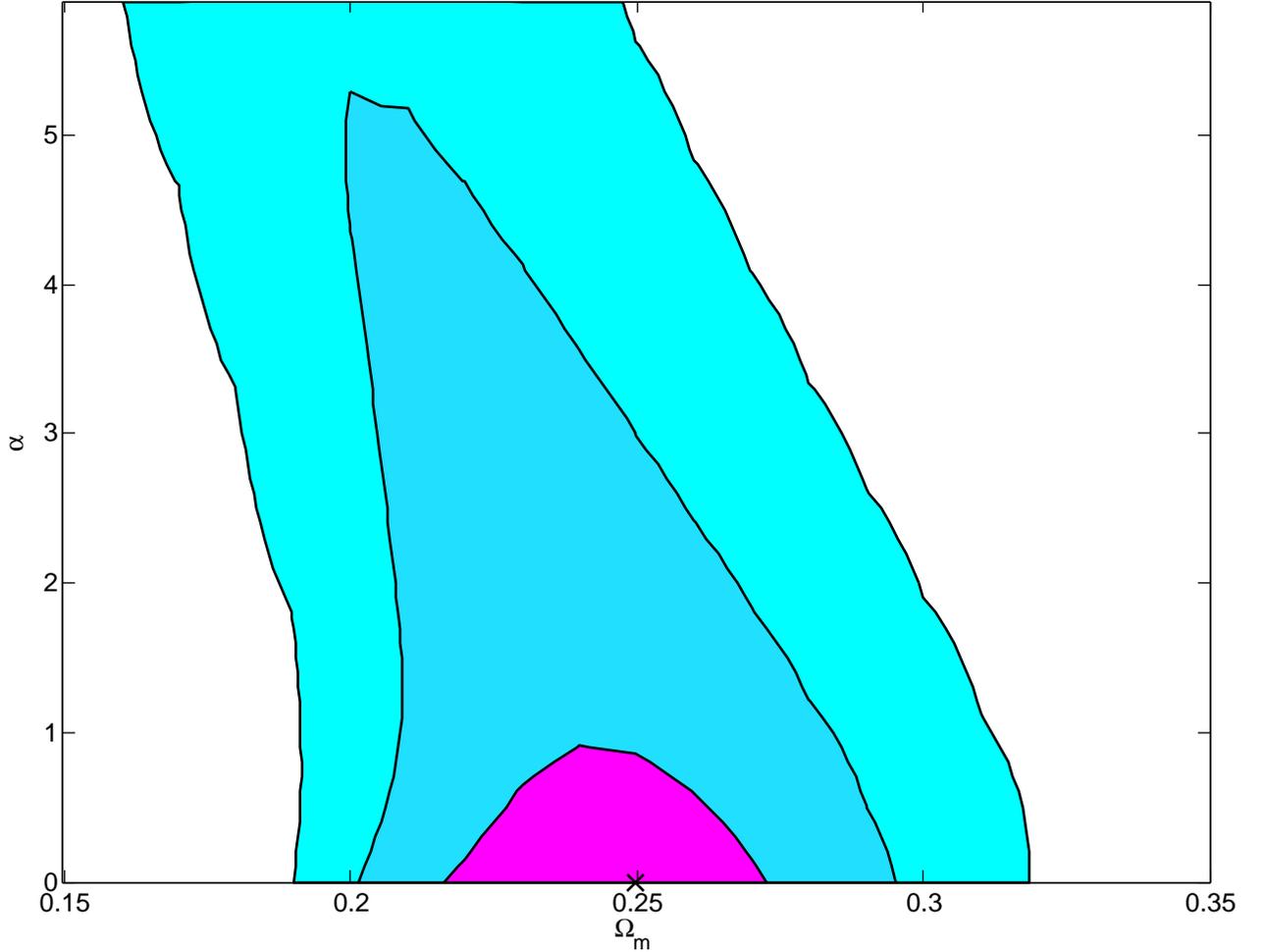}

    \caption[Constraints from BAO peak measurement.]{1, 2, and 3$\sigma$
confidence level contours on $\phi$CDM model parameters from BAO peak
measurement of Ref.~\cite{Percival:2007yw}. Cross denotes the best-fit value
at $\Omega_{\rm m}^*=0.25$ and $\alpha^*=0$.}

    \label{fig:perc}
    \end{center}
\end{figure}

\begin{figure}[htb]%t=top, b=bottom, h=here
\begin{center}
    \includegraphics[viewport=0.5in 2.5in 8in 8in, clip]{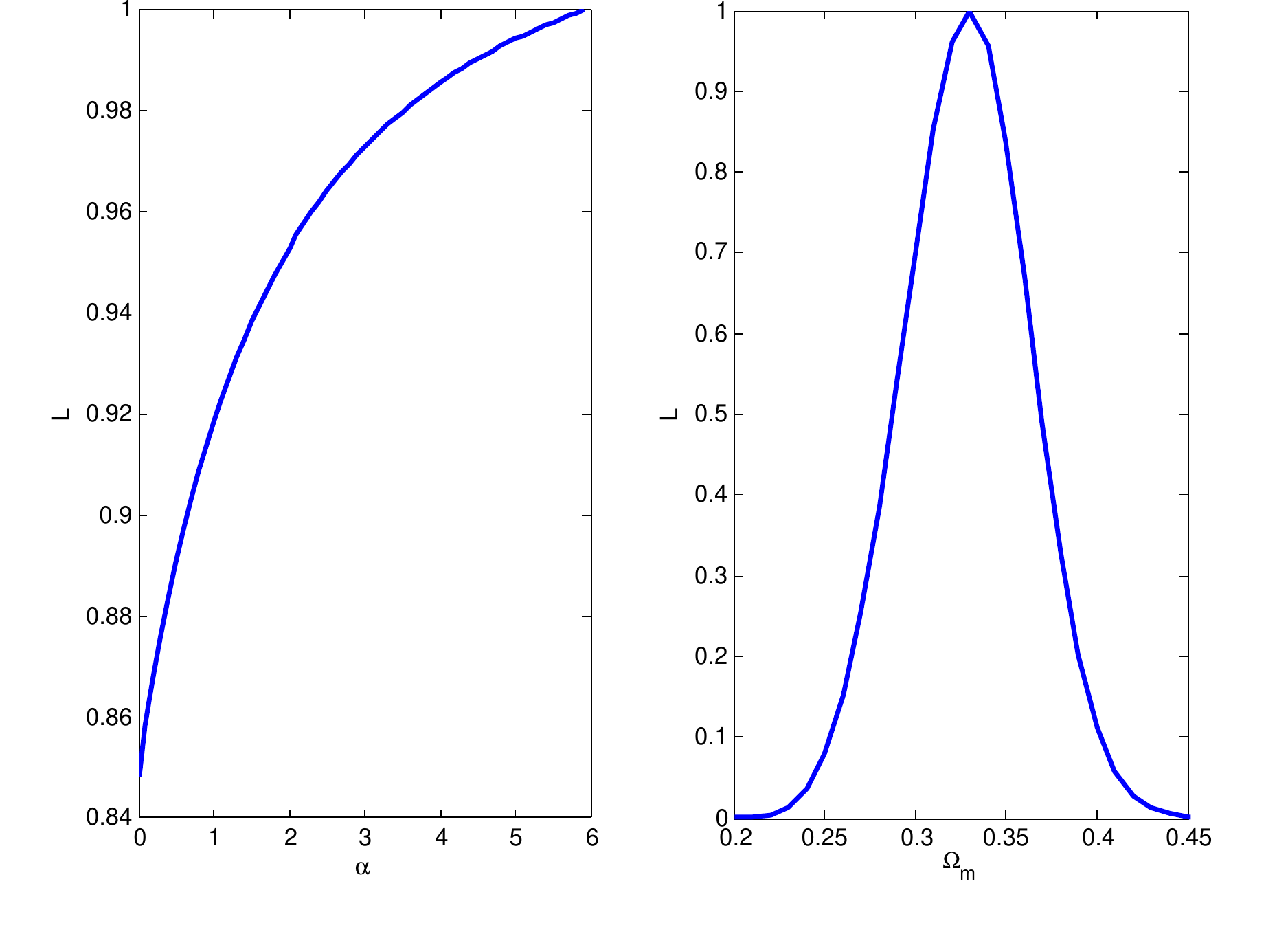}

    \caption[Likelihood functions of individual cosmological parameters from BAO peak measurement.]{One dimensional likelihood functions of individual cosmological parameters from BAO peak measurement of Ref.~\cite{Eisenstein:2005su}. The maximum likelihood values are normalized to one.}

    \label{fig:eisent1}
    \end{center}
\end{figure}

\begin{figure}[htb]%t=top, b=bottom, h=here
\begin{center}
    \includegraphics[viewport=0.5in 2.5in 8in 8in, clip]{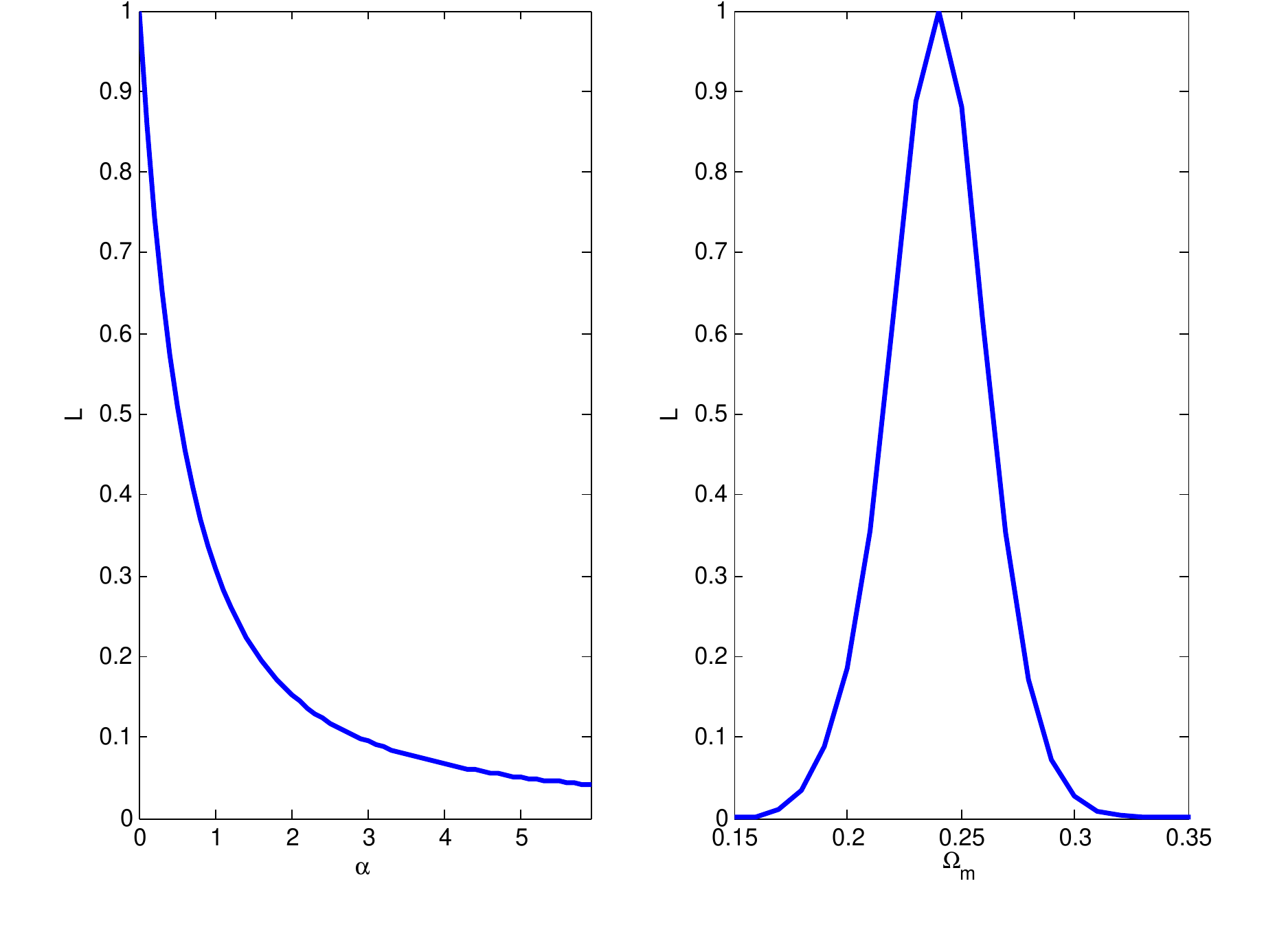}

    \caption[Likelihood functions of individual cosmological parameters from BAO peak measurement.]{One dimensional likelihood functions of individual cosmological parameters from BAO peak measurement of Ref.~\cite{Percival:2007yw}. The maximum likelihood values are normalized to one.}

    \label{fig:perc1}
    \end{center}
\end{figure}

The constraints on $\phi$CDM model from BAO data presented in Ref.~\cite{Eisenstein:2005su} are shown in Fig.~\ref{fig:eisenst}. Constrains from
BAO data presented in Ref.~\cite{Percival:2007yw} are shown in Fig.~\ref{fig:perc}. 
The one dimensional individual likelihoods are shown in Figs.~\ref{fig:eisent1} and \ref{fig:perc1} respectively.
The best fit values of individual parameters from Fig.~\ref{fig:perc1} are
$\alpha^*=0$ and $\Omega_{\rm m}^*=0.24$, with 1$\sigma$ intervals 
$0<\alpha<4.5$ and $0.24<\Omega_{\rm m}<0.26$. 

These measurements were made at only two redshifts and by themselves do not
provide a robust and reliable test of dark energy models, but in combination
with other data they do  provide useful constraints on cosmological
parameters. In addition, a number of surveys are planned in next few years that will measure the BAO scale accurately and at a variety of redshifts up to $z=1.2$. This upcoming BAO data, especially when combined with other data, will prove very useful in tightly constraining dark energy parameters \cite{Wang:2008hgb}.

\section{Radial BAO peak measurements}

Recently it was argued that the above measurements of the BAO scale were
essentially measurements orthogonal to the line of sight and so statistically
independent from a line of sight measurement of the BAO scale, even if the
same galaxy catalog is used for both measurements.
Reference~\cite{Gaztanaga:2008xz} used the SDSS data to compute the line of
sight or radial BAO scale in redshift space for two ranges of redshift and
showed the resulting constraints on the spatially-flat XCDM parameterization
\cite{Gaztanaga:2008de}. These constraints from the radial BAO scale data are
quite similar to the constraints derived from earlier ``transverse'' BAO
scale measurements. The data are consistent with spatially-flat $\Lambda$CDM.
However, these current radial BAO measurements (like current non-radial BAO
measurements) can not tightly constrain time-varying dark energy by
themselves (although the situation is anticipated to improve in the next few
years), as discussed next.

In a spherically symmetric Universe the two-point correlation function is a
function of two variables, $\xi=\xi(\sigma,\pi)$, where $\sigma$ is the
separation along the line of sight and $\pi$ is the separation on the sky. It
can also be expressed as a function of absolute separation
$r=\sqrt{\sigma^2+\pi^2}$ and the cosine of the angle between the line of
sight and the direction of separation, $\mu=\pi/r$. The correlation function
can then be decomposed into multipole moments,
\begin{equation}
\xi_l(r)=\int_{-1}^{+1}{\xi(r,\mu)P_l(\mu)d\mu},
\end{equation}
\noindent
where $P_l$ is the $l^{\rm {th}}$ order Legendre polynomial. Multipole moments of different orders can be related to each other if one has a complete theory of linear and nonlinear evolution. Although high multipoles that describe the ``shape'' of baryon acoustic oscillation imprints on the matter distribution are very difficult to measure in practice, theoretically they are independent of the monopole and could provide additional structure formation tests. 

Initial work considered only the averaged over direction monopole part of the correlation function,
\begin{equation}
\xi_0(r)=\frac{1}{2}\int_{-1}^{+1}{\xi(r,\mu)d\mu},
\end{equation}
\noindent
and found a BAO peak signal at a comoving distance of $r\approx110 h^{-1}\rm{ Mpc}$. This measurement was however mostly transverse to the line of sight direction $\pi$; the weight of separation along the line of sight  contributes less then $1\%$. Consequently, it is fair to assume that the radial baryon acoustic peak scale measurement in the line of sight direction from $\xi(\sigma)$ is statistically independent from that measured from $\xi(r)\approx\xi(\pi)$, even if the same galaxy sample is used for both measurements.

Reference.~\cite{Gaztanaga:2008xz} used SDSS data to measure the radial baryon acoustic scale in two redshift ranges $z\sim0.15-0.30$ with radial BAO peak scale $\Delta z=0.0407 \pm 0.0014$ and $z\sim0.40-0.47$ with $\Delta z=0.0442 \pm 0.0016$ (both one standard deviation errors). Theoretically the radial BAO peak scale is given by
\begin{equation}
\Delta z=H(z)r_s(z_{\rm d})/c
\end{equation}
\noindent
where $H(z)$ is the Hubble parameter at redshift $z$, $r_s(z_{\rm d})$ is the sound horizon size at the drag redshift $z_{\rm d}$, at which baryons were released from photons, and $c$ is the speed of light. $H(z)$ can be easily computed in a given cosmological model and depends on model parameters such as the non-relativistic matter density and the time dependence of dark energy. 

$r_s$ can be computed with two different methods. One is to use the ratio $l_s$ between the distance to the last-scattering surface and $r_s$ measured by CMB anisotropy experiments and compute the sound horizon at photon decoupling from
 \begin{equation}
 \label{horizon1}
 r_s(z_*)=\frac{\pi(1+z_*)d_A(z_*)}{l_s}.
 \end{equation}
\noindent
Here $z_*$ is the redshift at photon decoupling and $d_A$ is the angular diameter distance. Alternatively, one can use priors on the fractional energy density parameters of baryonic matter, $\Omega_{\rm b}$, nonrelativistic matter, $\Omega_{\rm m}$, and relativistic matter, $\Omega_{\rm r}$, from, e.g., CMB anisotropy measurements, and compute the sound horizon at the drag redshift from
\begin{equation}
\label{horizon2}
r_s(z_{\rm d})=\frac{c}{H_0\sqrt{3\Omega_{\rm m}}}\int_{0}^{a(z_{\rm d})}{\frac{da}{\sqrt{(a+1.69\Omega_{\rm r}/\Omega_{\rm m})(1+a0.75\Omega_{\rm b}/\Omega_{\rm r})}}}.
\end{equation}
\noindent
Both options have similar drawbacks. One has to assume priors on ``nuisance'' parameters like $l_s$ or various energy densities. CMB anisotropy measurements themselves have measurement errors that must be accounted for, otherwise the errors on the estimates of dark energy model parameters of interest will be underestimated. Also, the best fit values for nuisance parameters given by CMB anisotropy data are in general different for every cosmological model and also depend on model parameter values. To be fully consistent when using priors one would have to reanalyze CMB experiments for each cosmological model (and model parameter value) instead of using a single set of values for $l_s$, $\Omega_{\rm b}$, $\Omega_{\rm m}$, and $\Omega_{\rm r}$.  

At present, however, the BAO scale is measured only in two redshift ranges and does not provide very tight parameter constraints compared to other observational tests. Hence, as long as we are interested in preliminary constraints on dark energy from BAO scale measurements we may use the simplified approach of Ref.~\cite{Gaztanaga:2008xz}, keeping in mind that when more and better quality BAO scale measurements become available a more complete, careful, and time-consuming analysis will be warranted.

\section{Constraints from radial BAO peak measurements}

\begin{figure}[htb]%t=top, b=bottom, h=here
\begin{center}
    \includegraphics[viewport=0.5in 2.5in 8in 8in, clip]{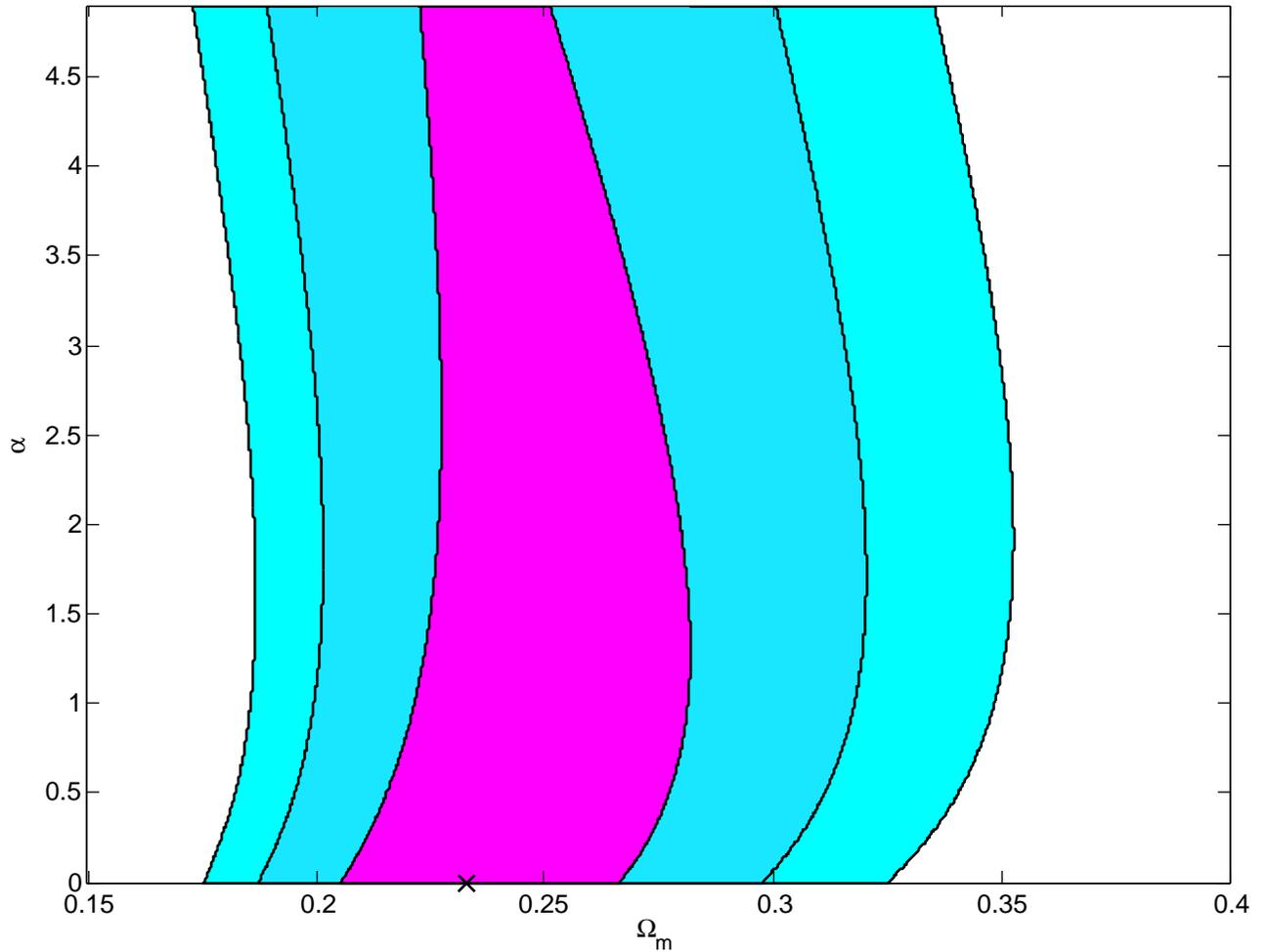}

    \caption[Constraints from radial BAO peak measurement.]{1, 2, and
3$\sigma$ confidence level contours on $\phi$CDM model parameters from radial
BAO peak measurements using the technique of Eq.~\eqref{horizon1}. The best
fit values are $\alpha^*=0$ and $\Omega_{\rm m}^*=0.23$.}

    \label{fig:rbao}
    \end{center}
\end{figure}

\begin{figure}[htb]%t=top, b=bottom, h=here
\begin{center}
    \includegraphics[viewport=0.5in 2.5in 8in 8in, clip]{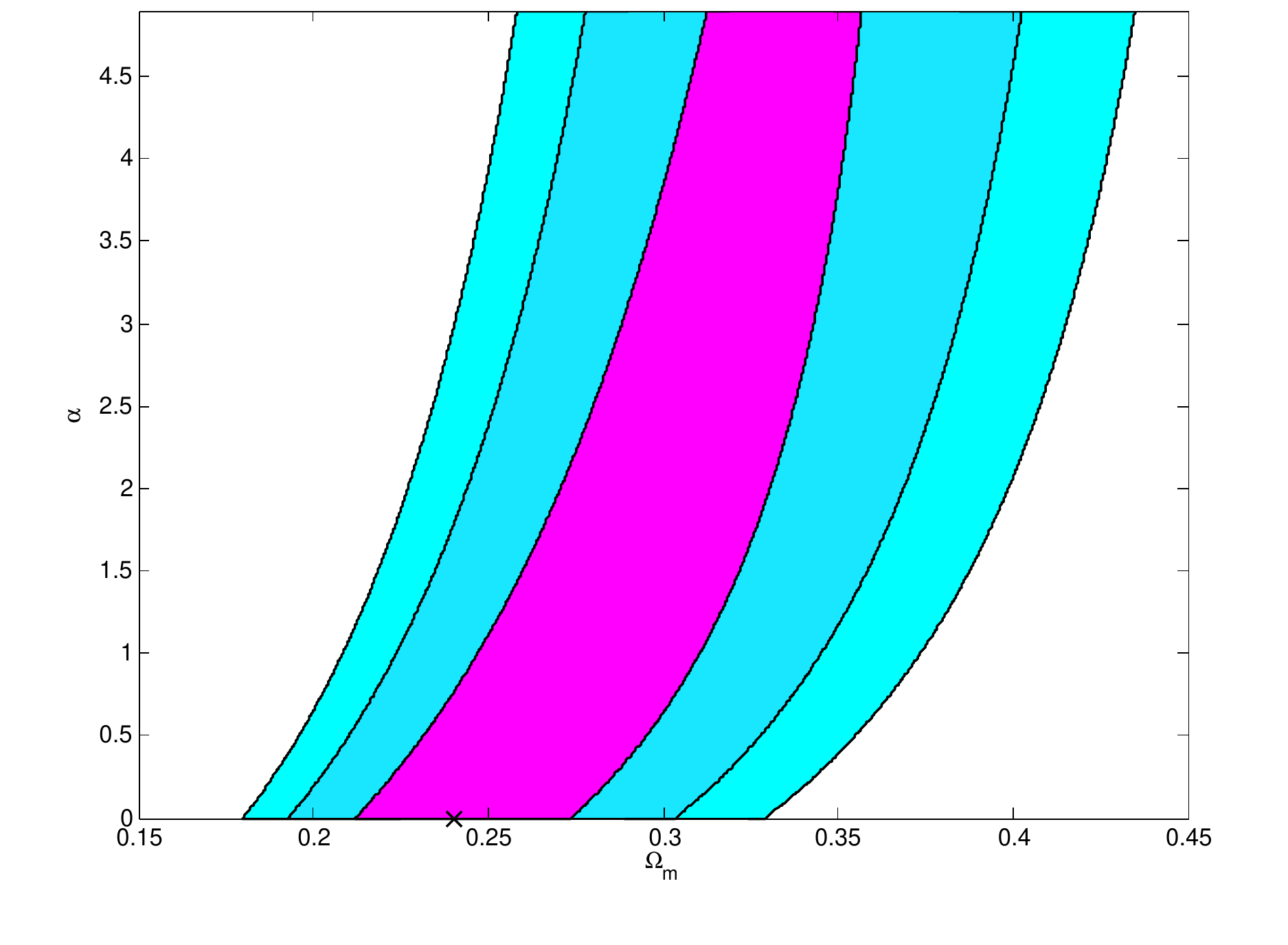}

    \caption[Constraints from radial BAO peak measurement.]{1, 2, and
3$\sigma$ confidence level contours on $\phi$CDM model parameters from radial
BAO peak measurements using the technique of Eq.~\eqref{horizon2}. The best
fit values are $\alpha^*=0$ and $\Omega_{\rm m}^*=0.25$.}
    \label{fig:rrbao}
    \end{center}
\end{figure}

\begin{figure}[htb]%t=top, b=bottom, h=here
\begin{center}
    \includegraphics[viewport=0.5in 2.5in 8in 8in, clip]{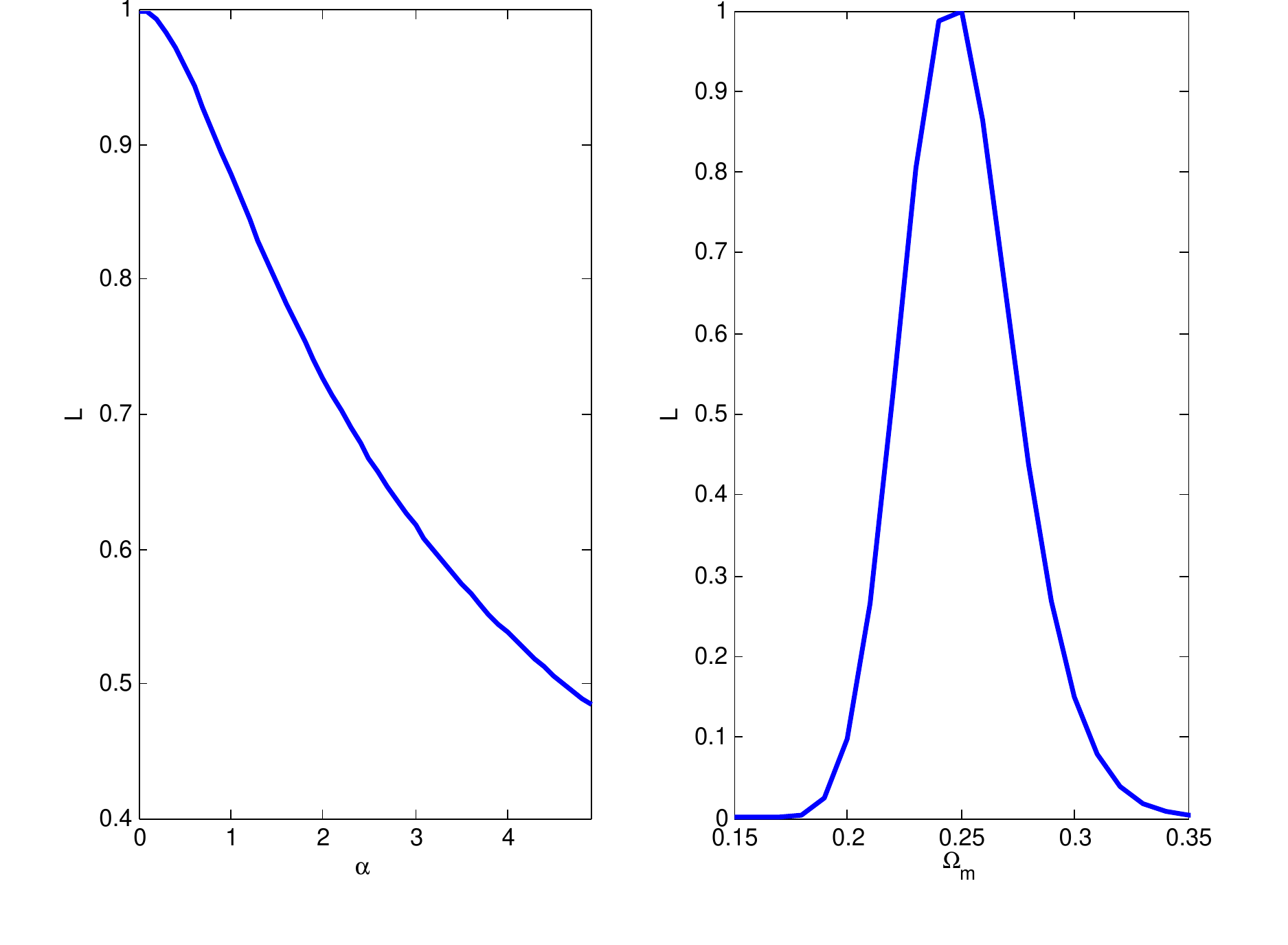}

    \caption[Likelihood functions of individual cosmological parameters from
radial BAO peak measurement.]{One dimensional likelihood functions of
individual cosmological parameters from radial BAO peak measurements and
using the technique of Eq.~\eqref{horizon1}. The maximum likelihood values are normalized to one.}

    \label{fig:rbao1}
    \end{center}
\end{figure}

\begin{figure}[htb]%t=top, b=bottom, h=here
\begin{center}
    \includegraphics[viewport=0.5in 2.5in 8in 8in, clip]{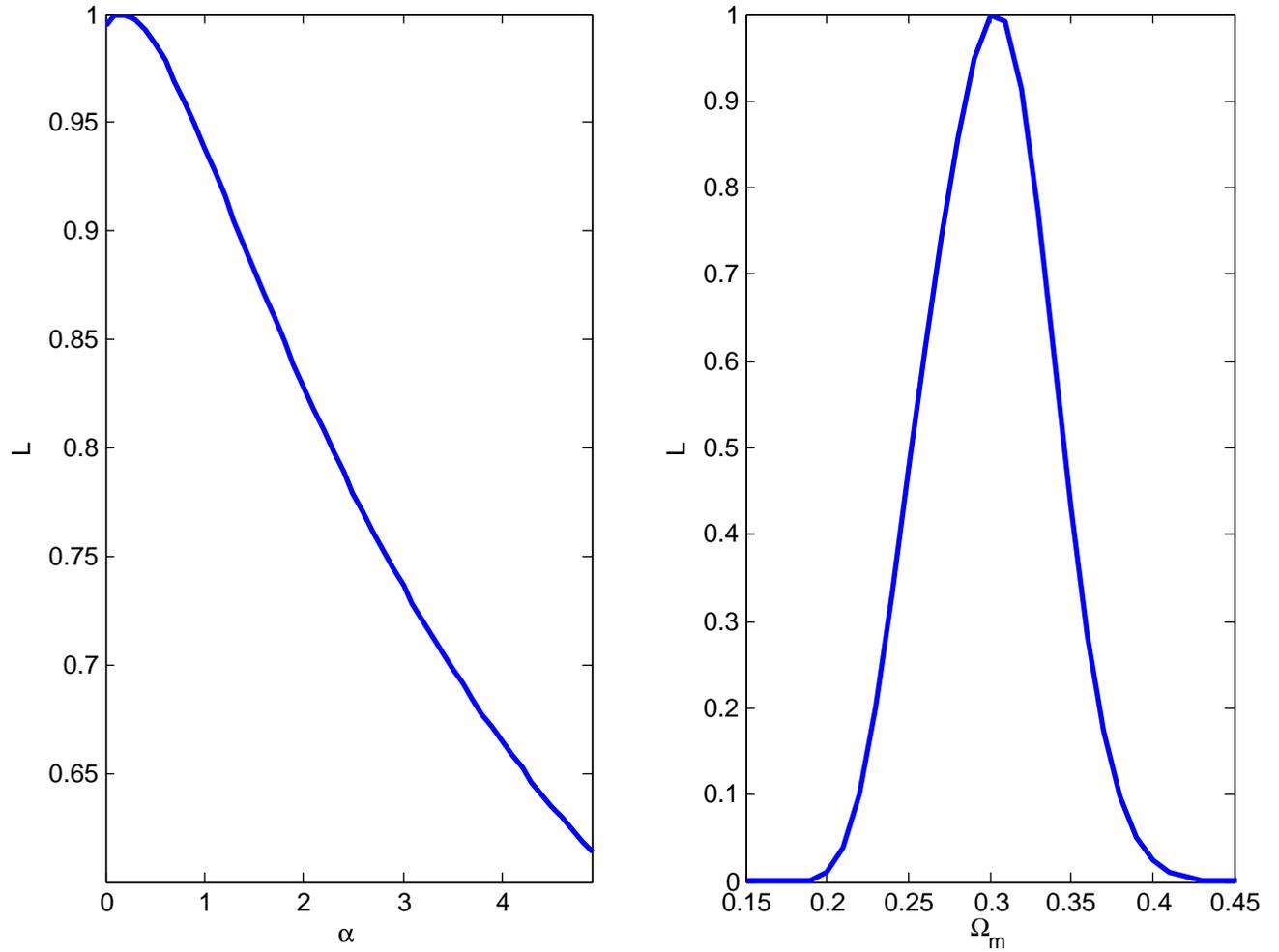}

    \caption[Likelihood functions of individual cosmological parameters from
radial BAO peak measurement.]{One dimensional likelihood functions of
individual cosmological parameters from radial BAO peak measurements and
using the technique of Eq.~\eqref{horizon2}. The maximum likelihood values are normalized to one.}

    \label{fig:rrbao1}
    \end{center}
\end{figure}

The confidence level contours for the spatially-flat $\phi$CDM model  are
shown in Fig.~\ref{fig:rbao} and \ref{fig:rrbao}. Corresponding one
dimensional likelihood functions for individual cosmological parameters are
shown in Fig.~\ref{fig:rbao1} and \ref{fig:rrbao1}. Here, the radial BAO
measurements constrain $\Omega_{\rm m}$ to be between $0.15$ and $0.4$ at
about 3$\sigma$, but the $\alpha$ parameter is not constrained well and large
values of $\alpha$ (relatively rapidly evolving dark energy) are not ruled
out, although the likelihood peaks at $\alpha=0$. These results are similar
to the ones derived earlier using the non-radial BAO peak scale measurements.

% +--------------------------------------------------------------------+
% | Sample Chapter 9
% +--------------------------------------------------------------------+

\cleardoublepage

% +--------------------------------------------------------------------+
% | Replace "This is Chapter 3" below with the title of your chapter.
% | LaTeX will automatically number the chapters.
% +--------------------------------------------------------------------+

\chapter{Conclusion}
\label{makereference9}

\section{Joint constraints}

Chapters~\ref{makereference5} to \ref{makereference8} present constraints on $\phi$CDM model parameters using different available cosmological data sets.
The lookback time versus redshift and the GRB data cannot at the moment constrain dark energy strongly. Likelihoods resulting from these two data sets
are flat in the direction of $\alpha$ and give preference to cosmological
models with low nonrelativistic mass energy density $\Omega_{\rm m}<0.2$,
which is at odds with other observations, but this is not very significant. 

In both cases this happens because the measurements are available only at a small number of redshifts (32 for lookback time data and 6 for GRB) and the measurement
uncertainties are large. This should change when more and better quality data become available. In the future these two methods could prove to be
very useful in constraining dark energy.

Constraints from SNeIa absolute magnitude versus redshift data, galaxy cluster gas mass fraction versus redshift data, and BAO peak measurements are
considerably tighter. BAO peak measurements are not very effective in
constraining the $\alpha$ parameter at the moment, but they give very good constraints on
nonrelativistic matter energy density. The power of the constraints seems surprising given that the BAO peak measurements have been made at only two
redshifts so far. SNeIa and cluster gas mass fraction data are not that
effective in constraining $\Omega_{\rm m}$ but provide upper limits on $\alpha$.
Current data favors the time-independent cosmological constant, but there is
a large part of time-evolving dark energy ($\phi$CDM) parameter space that is still not ruled out.

\begin{figure}[htb]%t=top, b=bottom, h=here
\begin{center}
    \includegraphics[viewport=0.5in 2.5in 8in 8in, clip]{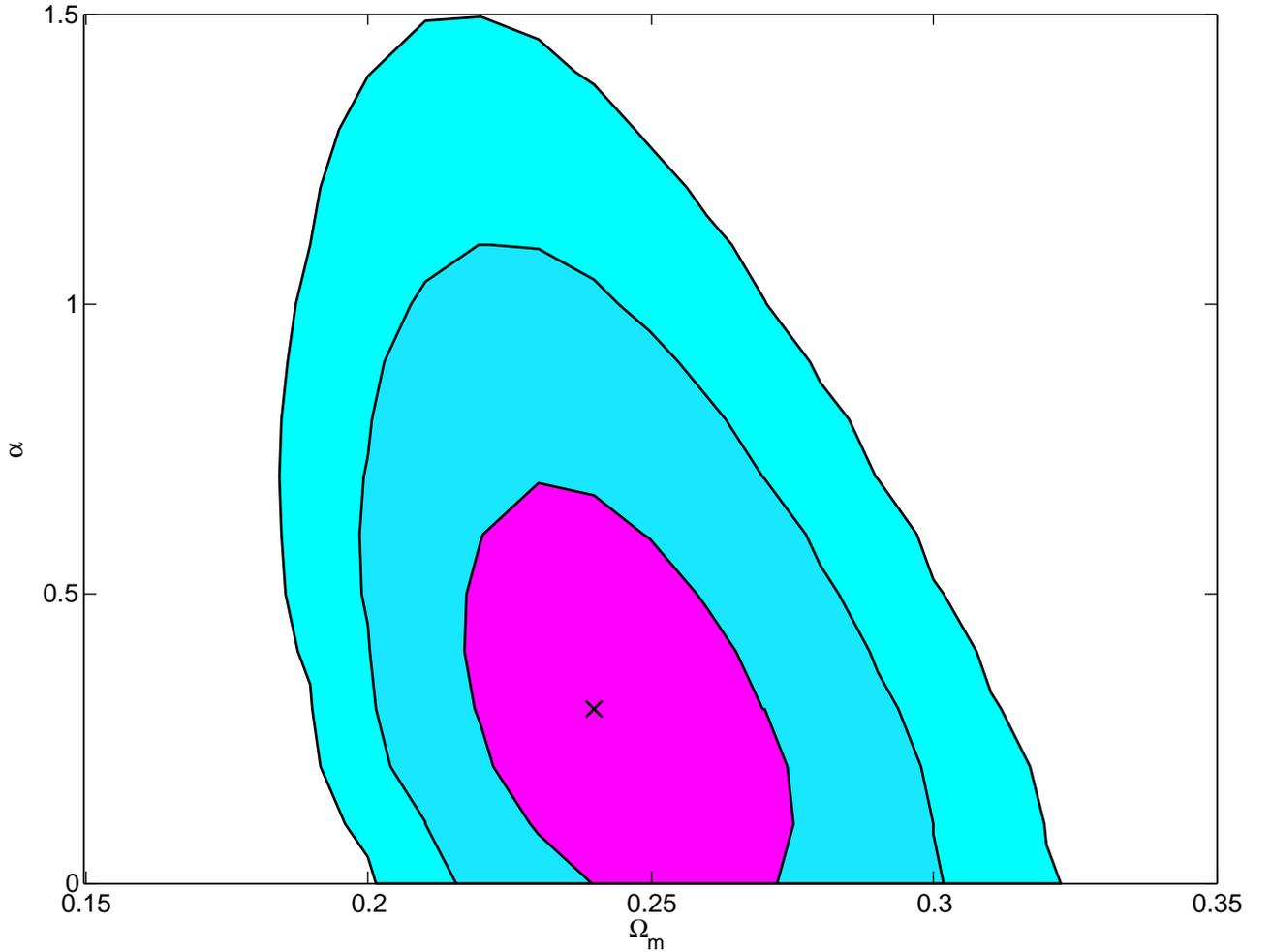}

    \caption[Joint constraints from SNeIa data and BAO peak measurements.]{1,
2, and 3$\sigma$ confidence level contours on $\phi$CDM model parameters from
joint analysis of SNeIa Union data and transverse BAO peak measurements. The
best-fit parameters are $\alpha^*=0.3$ and $\Omega_{\rm m}^*=0.24$ with
$\chi^2=321$ for $307$ degrees of freedom.}

    \label{fig:joint}
    \end{center}
\end{figure}

Since none of the available data sets alone can simultaneously tightly
constrain both cosmological parameters $\alpha$ and $\Omega_{\rm m}$, we
derive tighter constraints by performing joint analyses of observational data. We get the best
results when using SNeIa and BAO peak measurements; adding other data to the
mix does not significantly improve the constraints. Since these two are
independent we define the joint likelihood as the product of the
individual likelihoods
\begin{equation}
\label{eq:joint}\
\mathcal{L}_{\rm tot}=\mathcal{L}_{\rm SN}\mathcal{L}_{\rm BAO}.
\end{equation}
The constraints derived from this joint likelihood function are shown in
Fig.~\ref{fig:joint}. The joint analyses results in much tighter constraints.
The nonrelativistic matter density is in the range of $0.18<\Omega_{\rm
m}<0.32$ and $\alpha$ is constrained to be less than 1.5 at about 3$\sigma$
confidence. The best-fit values $\alpha^*=0.27$ and $\Omega_{\rm m}^*=0.24$ correspond to slowly-varying dark energy but the time-independent cosmological
constant also fits the data very well.

\begin{figure}[htb]%t=top, b=bottom, h=here
\begin{center}
    \includegraphics[viewport=0.5in 2.5in 8in 8in, clip]{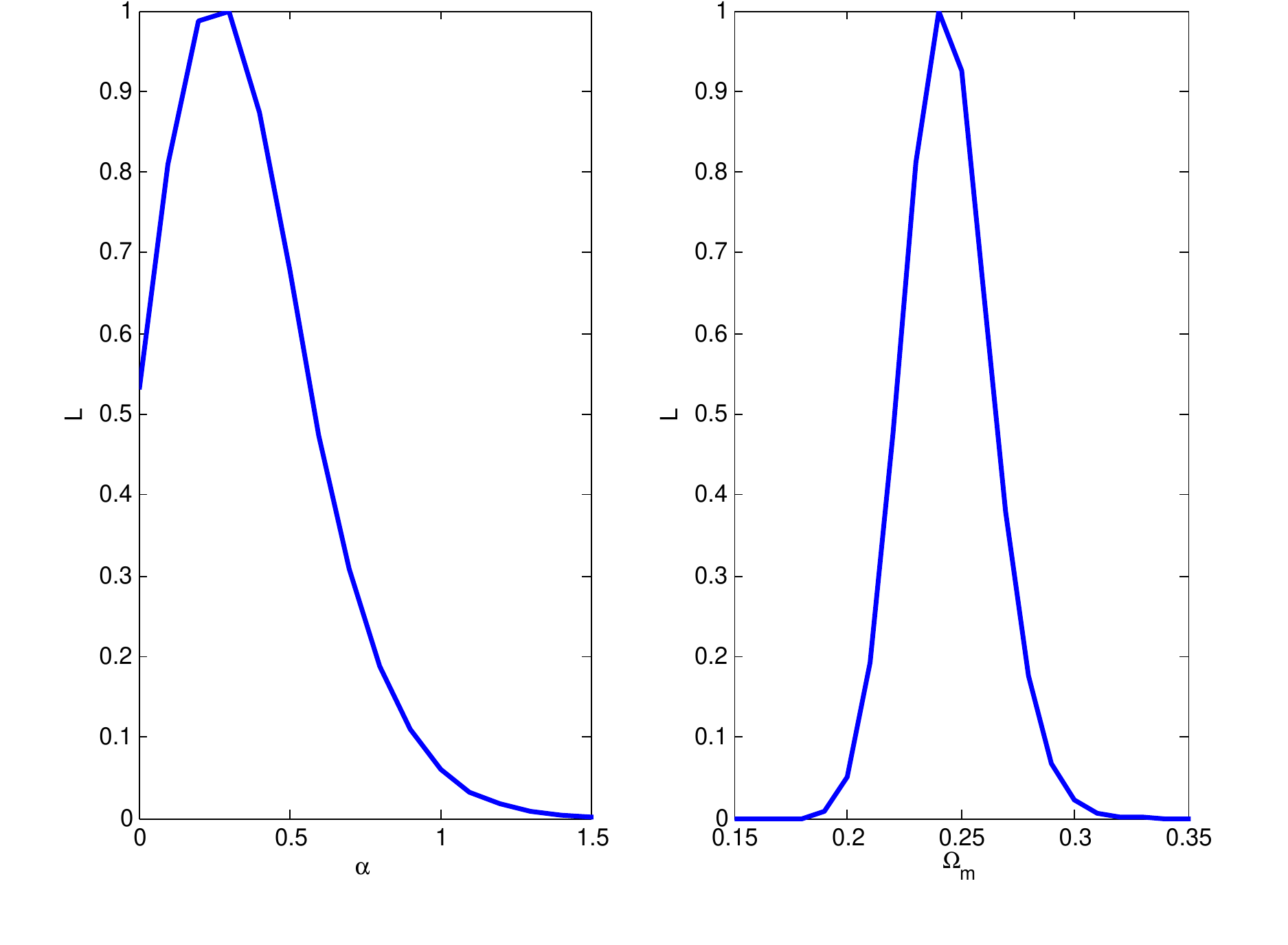}

    \caption[Likelihood functions of individual cosmological parameters from
joint analysis of SNeIa data and BAO peak measurements.]{One dimensional
likelihood functions of individual cosmological parameters from the joint
analysis of SNeIa Union data and transverse BAO peak measurements. The maximum likelihood values are normalized to one.}
    \label{fig:joint1}
    \end{center}
\end{figure}

The likelihood functions of individual cosmological parameters are shown in Fig.~\ref{fig:joint1}. The best-fit values from one dimensional likelihood
functions are $\alpha^*=0.27$ and $\Omega_{\rm m}^*=0.24$. The 1$\sigma$ intervals are $0.02<\alpha<0.54$ and $0.22<\Omega_{\rm m}<0.26$.

\section{Future prospects of detecting time-evolving dark energy}

Joint analysis of currently available cosmological data already results in
strong constraints on the $\phi$CDM model. $\Omega_{\rm m}$ is estimated to be
in between 0.2 and 0.3 and $\alpha$ is constrained to be less than 1.5.

The quality and quantity of available cosmological data sets is expected to
increase significantly in the next decade. This will result in much tighter
constraints on
$\phi$CDM model parameters. Some of the uncertainty in $\alpha$ and $\Omega_{\rm m}$ comes from the fact that we do not know precise values of nuisance
parameters, such as $H_0$ and $\Omega_{\rm b}$. Ongoing surveys (such as HST)
and new missions (such as PLANCK satellite) will measure these with greater accuracy, which will
translate into better constraints on $\alpha$ and $\Omega_{\rm m}$.

From the joint analysis of near future SNeIa, GRB, galaxy cluster gas mass
fraction and BAO measurements we should be able to either detect the
time-dependence of
dark energies energy density at a high confidence level or to constrain it to
 small, physically uninteresting values.

% +-------------------------------------------------------------------------+
% | References                                                              |
% +-------------------------------------------------------------------------+

% +-------------------------------------------------------------------------+
% | In order for WinEDT to index references correctly, it has to know where |
% | the file resides.  The following command is prefaced by %, and will be  |
% | ignored completely by LaTeX.  However, WinEDT will use this line to     |
% | access the external .bib bibliography file.  Also note that WinEDT can  |
% | read file path names with either "\" or "/" - LaTeX, however, doesn't   |
% | like "\", so it's easier to store a path name in the "Unix" style.      |
% +-------------------------------------------------------------------------+

%Included for Gather Purpose only.  Do NOT uncomment:
%input "references.bib"

% +--------------------------------------------------------------------+
% | This template uses the BibTeX program to format references.  The
% | 3 lines below create a separate Bibliography section and add
% | an entry for "Bibliography" to the Table of Contents.  The actual
% | data for your references (author, title, journal, date, etc.) are
% | entered in the references.bib file.  See that file for information
% | on how to enter references.
% +--------------------------------------------------------------------+

\bibdata{references}
\bibliography{references}
\addcontentsline{toc}{chapter}{Bibliography}

% +--------------------------------------------------------------------+
% | Finally, we generate the appendix.  To add or delete appendices,
% | add or remove the line
% |
% |     \input{appendixX.tex}
% |
% | where "X" is the letter designation of the Appendix (A, B, C, etc.)
% | You should have one \input{appendixX.tex} line and a corresponding
% | file appendixX.tex for each appendix.                                 |
% +--------------------------------------------------------------------+

\appendix
% +--------------------------------------------------------------------+
% | Appendix A Page (Optional)                                         |
% +--------------------------------------------------------------------+

\cleardoublepage

\chapter{C++ function for a numerical solution of the scalar field equations.}
\label{appendixA}

% +--------------------------------------------------------------------+
% | Enter text for your Appendix page in the space below this box.     |
% |                                                                    |
% +--------------------------------------------------------------------+
\begin{lstlisting}

#include<iostream>
#include<fstream>
#include<cmath>

using namespace std;

double fy(double a, double y, double dy, double kmp, double alpha);
double fa(double a, double y, double dy, double kmp, double alpha);

//----------------------------------------------------------------------
// This function takes as an input the value of alpha, nonrelativistic 
// matter density Omega_m, and redshift z.
// It solves ODEs using 4th order Runge-Kutta method and returns
// comoving distance.
//----------------------------------------------------------------------

double distance(double alpha, double OmegaM, double z)  
{
  const double t0 = 0.01;
  const double dt = 0.001;
  
  double t = t0;
  double A = 2.0/3.0*alpha*(alpha + 2.0);
  double kmp = 8.0/3.0*(alpha + 4.0)/(alpha + 2.0)*pow(A , alpha/2.0);

  double afin, dafin;
  double sum = 0.0;

  // Initial conditions on scale factor - a, and scalar field - y.
  
  double a = pow(t0 , 2.0/3.0);
  double y = sqrt(A)*pow(t0 , 2.0/(alpha + 2.0));
  double dy = 
          sqrt(8.0/3.0*alpha/(alpha + 2.0))/pow(t0 , alpha/(alpha + 2.0));
  double OM = 
          4.0/9.0/a/a/a/(4.0/9.0/a/a/a + (dy*dy + kmp/pow(y , alpha))/12.0);

  double k11, k12, k13, k21, k22, k23, k31, k32, k33, k41, k42, k43;

  while(OM > omegam)
  {
    k11 = fa(a, y, dy, kmp, alpha)*dt;
    k12 = dy*dt;
    k13 = fy(a, y, dy, kmp, alpha)*dt;
    k21 = fa(a + 0.5*k11, y + 0.5*k12, dy + 0.5*k13, kmp, alpha)*dt;
    k22 = (dy + 0.5*k12)*dt;
    k23 = fy(a + 0.5*k11, y + 0.5*k12, dy + 0.5*k13, kmp, alpha)*dt;
    k31 = fa(a + 0.5*k21, y + 0.5*k22, dy + 0.5*k23, kmp, alpha)*dt;
    k32 = (dy + 0.5*k22)*dt;
    k33 = fy(a + 0.5*k21, y + 0.5*k22, dy + 0.5*k23, kmp, alpha)*dt;
    k41 = fa(a + k31, y + k32, dy + k33, kmp, alpha)*dt;
    k42 = (dy + k32)*dt;
    k43 = fy(a + k31, y + k32, dy + k33, kmp, alpha)*dt;
    
    a += k11/6.0 + k21/3.0 + k31/3.0 + k41/6.0;
    y += k12/6.0 + k22/3.0 + k32/3.0 + k42/6.0;
    dy += k13/6.0 + k23/3.0 + k33/3.0 + k43/6.0;
    
    t = t + dt;
    
    OM = 4.0/9.0/a/a/a/(4.0/9.0/a/a/a + (dy*dy + kmp/pow(y , alpha))/12.0);
  }
  
  afin = a;
  dafin = fa(a, y, dy, kmp, alpha);

  t = t0;
  a = pow(t0 , 2.0/3.0);
  y = sqrt(A)*pow(t0 , 2.0/(alpha + 2.0));
  dy = sqrt(8.0/3.0*alpha/(alpha + 2.0))/pow(t0 , alpha/(alpha + 2.0));
  OM = 4.0/9.0/a/a/a/(4.0/9.0/a/a/a + (dy*dy + kmp/pow(y , alpha))/12.0);

  while(OM > omegam)
  {
    k11 = fa(a, y, dy, kmp, alpha)*dt;
    k12 = dy*dt;
    k13 = fy(a, y, dy, kmp, alpha)*dt;
    k21 = fa(a + 0.5*k11, y + 0.5*k12, dy + 0.5*k13, kmp, alpha)*dt;
    k22 = (dy + 0.5*k12)*dt;
    k23 = fy(a + 0.5*k11, y + 0.5*k12, dy + 0.5*k13, kmp, alpha)*dt;
    k31 = fa(a + 0.5*k21, y + 0.5*k22, dy + 0.5*k23, kmp, alpha)*dt;
    k32 = (dy + 0.5*k22)*dt;
    k33 = fy(a + 0.5*k21, y + 0.5*k22, dy + 0.5*k23, kmp, alpha)*dt;
    k41 = fa(a + k31, y + k32, dy + k33, kmp, alpha)*dt;
    k42 = (dy + k32)*dt;
    k43 = fy(a + k31, y + k32, dy + k33, kmp, alpha)*dt;
    
    a += k11/6.0 + k21/3.0 + k31/3.0 + k41/6.0;
    y += k12/6.0 + k22/3.0 + k32/3.0 + k42/6.0;
    dy += k13/6.0 + k23/3.0 + k33/3.0 + k43/6.0;
    
    t = t + dt;
    OM = 4.0/9.0/a/a/a/(4.0/9.0/a/a/a + (dy*dy + kmp/pow(y , alpha))/12.0);
    
    if(a/afin > 1.0/(1.0 + z))
    {
      sum += 1.0/a;
    }
  }
  
  sum *= dt;
  sum *= dafin;
  
  return sum; 
}

double fy(double a, double y, double dy, double kmp, double alpha)
{
  return (-3.0*sqrt(4.0/9.0/a/a/a + 
           1.0/12.0*(dy*dy + kmp/pow((y) , alpha)))*
           dy + kmp*alpha/2.0/pow((y) , alpha + 1.0));
}

double fa(double a, double y, double dy, double kmp, double alpha)
{
  return (sqrt(4.0/9.0/(a) + (a)*(a)/12.0*(dy*dy + kmp/pow(y , alpha))));
}

\end{lstlisting}

% +--------------------------------------------------------------------+
% | Appendix B Page (Optional)                                         |
% +--------------------------------------------------------------------+

\cleardoublepage

\chapter{Ages of passively evolving galaxies.}
\label{appendixage}

% +--------------------------------------------------------------------+
% | Enter text for your Appendix page in the space below this box.     |
% |                                                                    |
% +--------------------------------------------------------------------+
  \begin{center}
    \begin{longtable}[c]{|c|c|}
    \caption[Ages of passively evolving galaxies.]{Ages of passively evolving galaxies versus redshift data.}\\

\hline \multicolumn{1}{|c|}{$z$} & \multicolumn{1}{c|}{Age (Gyr)} \\ \hline 
\endfirsthead

\multicolumn{2}{c}%
{{\bfseries \tablename\ \thetable{} -- continued from previous page}} \\
\hline \multicolumn{1}{|c|}{$z$} &
\multicolumn{1}{c|}{Age (Gyr)} \\ \hline 
\endhead

\hline \multicolumn{2}{|r|}{{Continued on next page}} \\ \hline
\endfoot

\hline \hline
\endlastfoot

0.117100 & 10.2000\\
 0.117400 & 10.0000\\ 
 0.222000 & 9.00000\\
 0.231100 & 9.00000\\
 0.355900 & 7.60000\\ 
 0.452000 & 6.80000\\ 
  0.575000 & 7.00000\\ 
 0.644000 & 6.00000\\ 
 0.676000 & 6.00000\\ 
 0.833000 & 6.00000\\ 
 0.836000 & 5.80000\\ 
 0.922000 & 5.50000\\ 
 1.17900 & 4.60000\\ 
  1.22200 & 3.50000\\ 
 1.22400 & 4.30000\\ 
 1.22500 & 3.50000\\ 
 1.22600 & 3.50000\\ 
 1.34000 & 3.40000\\ 
 1.38000 & 3.50000\\ 
 1.38300 & 3.50000\\ 
 1.39600 & 3.60000\\ 
  1.43000 & 3.20000\\ 
 1.45000 & 3.20000\\ 
 1.48800 & 3.00000\\ 
 1.49000 & 3.60000\\
 1.49300 & 3.20000\\ 
 1.51000 & 2.80000\\ 
 1.55000 & 3.00000\\ 
 1.57600 & 2.50000\\ 
  1.64200 & 3.00000\\ 
 1.72500 & 2.60000\\ 
 1.84500 & 2.50000\\

     \hline
    \end{longtable}
    \label{table1}
   \end{center}

% +--------------------------------------------------------------------+
% | Appendix B Page (Optional)                                         |
% +--------------------------------------------------------------------+

\cleardoublepage

\chapter{Hubble parameter versus redshift data}
\label{appendixhubble}

% +--------------------------------------------------------------------+
% | Enter text for your Appendix page in the space below this box.     |
% |                                                                    |
% +--------------------------------------------------------------------+
  \begin{center}
    \begin{longtable}[c]{|c|c|c|}
    \caption[Hubble parameter measurements.]{Hubble parameter versus redshift data with 1$\sigma$ measurement uncertainties.}\\

\hline \multicolumn{1}{|c|}{$z$} & \multicolumn{1}{c|}{$H(z)$} & \multicolumn{1}{c|}{$\sigma_H$}\\ \hline 
\endfirsthead

\multicolumn{3}{c}%
{{\bfseries \tablename\ \thetable{} -- continued from previous page}} \\
\hline \multicolumn{1}{|c|}{$z$} &
\multicolumn{1}{c|}{$H(z)$} &
\multicolumn{1}{c|}{$\sigma_H$} \\ \hline 
\endhead

\hline \multicolumn{3}{|r|}{{Continued on next page}} \\ \hline
\endfoot

\hline \hline
\endlastfoot
0.09 & 69 & 12\\
0.17 & 83 & 8.3\\
0.27 & 70 & 14\\
0.4 & 87 & 17.4\\
0.88 & 117 & 23.4\\
1.3 & 168 & 13.4\\
1.43 & 177 & 14.2\\
1.53 & 140 & 14\\
1.75 & 202 & 40.4\\
     \hline
    \end{longtable}
    \label{table1}
   \end{center}

% +--------------------------------------------------------------------+
% | Appendix B Page (Optional)                                         |
% +--------------------------------------------------------------------+

\cleardoublepage

\chapter{SNeIa ``union'' data set}
\label{appendixb}

% +--------------------------------------------------------------------+
% | Enter text for your Appendix page in the space below this box.     |
% |                                                                    |
% +--------------------------------------------------------------------+
  \begin{center}
    \begin{longtable}[c]{|l|l|l|}
    \caption[SNeIa ``union'' data set.]{SNeIa ``union'' data set. The redshift $z$, distance modulii $\mu$, 
             and 1$\sigma$ statistical measurement errors on the measurement of $\mu$.}\\

\hline \multicolumn{1}{|c|}{$z$} & \multicolumn{1}{c|}{$\mu$} & \multicolumn{1}{c|}{$\sigma_\mu$} \\ \hline 
\endfirsthead

\multicolumn{3}{c}%
{{\bfseries \tablename\ \thetable{} -- continued from previous page}} \\
\hline \multicolumn{1}{|c|}{$z$} &
\multicolumn{1}{c|}{$\mu$} &
\multicolumn{1}{c|}{$\sigma_\mu$} \\ \hline 
\endhead

\hline \multicolumn{3}{|r|}{{Continued on next page}} \\ \hline
\endfoot

\hline \hline
\endlastfoot

0.050043 & 36.7730018543 & 0.153148680782 \\
0.052926 & 36.8241423261 & 0.148454214927 \\
0.02513 & 35.1678175291 & 0.165420933733 \\
0.070086 & 37.567740156 & 0.15403068043 \\
0.062668 & 37.5515804571 & 0.145317136417 \\
0.087589 & 38.1915656352 & 0.159495013964 \\
0.078577 & 37.5212710132 & 0.147429338762 \\
0.017227 & 34.6546665523 & 0.1880259565 \\
0.042233 & 36.3586532798 & 0.152312134352 \\
0.045295 & 36.6566139461 & 0.148758275509 \\
0.019599 & 34.5182384466 & 0.179081303451 \\
0.100915 & 38.5088119689 & 0.145279087383 \\
0.027342 & 35.2300574166 & 0.171939536277 \\
0.074605 & 37.6663645268 & 0.147931348064 \\
0.026489 & 35.5199700817 & 0.182864910719 \\
0.049922 & 36.6999653546 & 0.146530609404 \\
0.030604 & 35.5404531616 & 0.161624590233 \\
0.016345641 & 34.0924538801 & 0.144775220876 \\
0.0154363 & 34.0329851021 & 0.150597548663 \\
0.030529 & 35.5818149937 & 0.0920067053122 \\
0.024525 & 34.9577174302 & 0.109085696057 \\
0.023953 & 34.935520479 & 0.115625212493 \\
0.026038 & 35.3566237332 & 0.106784897071 \\
0.048948 & 36.6391120177 & 0.1766982111 \\
0.024314 & 35.0794913195 & 0.18643839963 \\
0.015166 & 34.2041114924 & 0.225518080121 \\
0.03572 & 36.1050539158 & 0.188094161827 \\
0.048818 & 36.3507343053 & 0.166961986591 \\
0.0219800059146 & 34.8811235819 & 0.196258768364 \\
0.0275 & 35.6735856723 & 0.181507304221 \\
0.1244 & 38.9521645786 & 0.168017789501 \\
0.036 & 35.8260342723 & 0.173778570484 \\
0.01673 & 34.1592687583 & 0.209352704395 \\
0.016321 & 34.1144805357 & 0.209884301455 \\
0.021793 & 34.9048499767 & 0.282608881266 \\
0.01645 & 34.1818298266 & 0.292212075551 \\
0.023208 & 35.117280195 & 0.276622195571 \\
0.036457 & 36.1142750974 & 0.267250650207 \\
0.019264 & 35.1559953619 & 0.285695464272 \\
0.017605 & 34.3943547917 & 0.308882749558 \\
0.031528 & 35.7346061594 & 0.270713039098 \\
0.023536 & 35.1870287376 & 0.278885118559 \\
0.016743 & 33.8878121745 & 0.292567467485 \\
0.05371 & 36.4559044522 & 0.268351164818 \\
0.016991 & 34.2677976869 & 0.290910806654 \\
0.027865 & 35.0340198507 & 0.273008042132 \\
0.017173 & 34.1860410198 & 0.293258906218 \\
0.029955 & 35.9589514911 & 0.273866480307 \\
0.016559 & 34.3764044862 & 0.292368203484 \\
0.015 & 34.0959952821 & 0.145904476604 \\
0.0544 & 36.9686138698 & 0.0771934613908 \\
0.1561 & 39.294455039 & 0.0599663857105 \\
0.0393 & 36.3055280023 & 0.0717497761757 \\
0.1241 & 38.8100134766 & 0.0715971658275 \\
0.1441 & 39.0528219919 & 0.0634435595707 \\
0.1299 & 38.9956124177 & 0.0631822941145 \\
0.0784 & 37.7810061412 & 0.0681153647931 \\
0.62 & 43.2066240148 & 0.374321656282 \\
0.57 & 42.6299495836 & 0.329397155099 \\
0.3 & 40.9068557079 & 0.322661461524 \\
0.38 & 41.981448045 & 0.294675002057 \\
0.43 & 42.2865323661 & 0.353944288913 \\
0.24 & 40.92039446 & 0.604691512396 \\
0.3 & 40.152454556 & 0.463180764839 \\
0.24 & 40.8941528415 & 0.281508606621 \\
0.44 & 42.0189541221 & 0.290912412354 \\
0.5 & 42.4251183643 & 0.285032220696 \\
0.97 & 43.0841865629 & 0.742582169563 \\
0.479 & 42.2593258058 & 0.305353355969 \\
0.83 & 43.4934896902 & 0.372098016098 \\
0.416 & 42.042208973 & 0.43961209764 \\
0.581 & 41.9804788609 & 0.422581723678 \\
0.45 & 41.7965293235 & 0.367900106831 \\
0.579 & 43.1102730617 & 0.562403228541 \\
0.32 & 41.1885505108 & 0.339690792497 \\
0.657 & 43.0812935967 & 0.549444460563 \\
0.472 & 42.0990717933 & 0.469994746225 \\
0.374 & 43.1153409984 & 0.694417590782 \\
0.526 & 42.013309316 & 0.402473462081 \\
0.763 & 44.3657126034 & 0.885466302481 \\
0.58 & 42.9398798711 & 0.395142768132 \\
0.43 & 41.8899803061 & 0.393293335469 \\
0.45 & 42.2825129374 & 0.422650328365 \\
0.828 & 44.203018183 & 0.611918160502 \\
0.656 & 43.181554143 & 0.528708119909 \\
0.495 & 42.1205013485 & 0.39827409355 \\
0.49 & 41.8217131258 & 0.355251221615 \\
0.57 & 42.742574543 & 0.395438890913 \\
0.388 & 42.0969738461 & 0.370734294336 \\
0.45 & 42.2128189449 & 0.377256221564 \\
0.48 & 42.1266258961 & 0.403490417945 \\
0.615 & 42.6108744805 & 0.520864721236 \\
0.4 & 42.1077687105 & 0.382043588034 \\
0.655 & 42.3305104738 & 0.496246176513 \\
0.498 & 43.0296153325 & 0.451179870261 \\
0.465 & 41.8375675989 & 0.677740680801 \\
0.453 & 42.6855854693 & 0.388547543287 \\
0.425 & 41.3010835176 & 0.366453610612 \\
0.278 & 40.6729770427 & 0.205733686158 \\
0.477 & 42.0071669413 & 0.149503540949 \\
0.95 & 43.6972613338 & 0.274437861244 \\
1.057 & 44.0787405468 & 0.211584422417 \\
0.816 & 43.7951226482 & 0.423980625959 \\
0.455 & 42.3522141356 & 0.239111591348 \\
0.514 & 42.4275455763 & 0.427580299971 \\
0.423 & 41.6141933692 & 0.239548120652 \\
0.946 & 43.1597477534 & 0.806741116211 \\
0.859 & 44.022811801 & 0.304111603744 \\
1.031 & 42.1761580597 & 1.28660173268 \\
0.936 & 44.148887081 & 0.518330924179 \\
0.528 & 42.3528772215 & 0.240748040733 \\
0.645 & 44.0353749092 & 0.964152598465 \\
0.978 & 43.6526492518 & 0.312966122482 \\
0.885 & 44.1789952676 & 0.466461712419 \\
0.815 & 44.1347089556 & 0.919484299622 \\
0.568 & 42.5610152006 & 0.293309253351 \\
0.711 & 43.2491247935 & 0.248908077637 \\
0.3396 & 41.1022712766 & 0.259553392222 \\
0.3965 & 41.4084095732 & 0.240552944224 \\
0.812 & 43.8133209255 & 0.282936944717 \\
0.799 & 43.1683351205 & 0.27444888856 \\
0.882 & 42.9333230617 & 0.324847417172 \\
0.833 & 43.407367536 & 0.347349605053 \\
0.874 & 43.3865810553 & 0.462308366531 \\
0.772 & 43.4354874701 & 0.263416996827 \\
0.543 & 42.3919207564 & 0.135200439255 \\
0.75 & 43.167266596 & 0.135523948437 \\
0.64 & 42.7302552042 & 0.225082191179 \\
0.43 & 42.202845059 & 0.143459564521 \\
0.64 & 43.1881838551 & 0.186543501751 \\
0.497 & 42.4090348089 & 0.158488764413 \\
0.44 & 41.9964410659 & 0.127761277223 \\
0.355 & 41.516716698 & 0.130966220782 \\
0.78 & 43.4907628672 & 0.151115530925 \\
0.54 & 42.4339111537 & 0.121366799166 \\
0.86 & 43.7451019144 & 0.156150591229 \\
1.02 & 44.1993977338 & 0.223075604603 \\
1.14 & 44.2839104237 & 0.280516457089 \\
0.854 & 43.9153676425 & 0.219112489318 \\
1.37 & 44.7910462695 & 0.324513218653 \\
0.975 & 44.2284945859 & 0.200606912017 \\
0.97 & 44.5842379887 & 0.240421078134 \\
0.74 & 43.417393486 & 0.196331774539 \\
1.39 & 44.8274132597 & 0.591264643225 \\
0.46 & 42.1615041182 & 0.190529350474 \\
1.02 & 43.9810540565 & 0.241216980878 \\
1.12 & 44.2495126581 & 0.241680223529 \\
1.23 & 44.852264567 & 0.236250346674 \\
1.19 & 44.1207308635 & 0.245756009721 \\
0.839 & 43.5971981016 & 0.211771806558 \\
1.01 & 44.4466207888 & 0.195202527359 \\
0.521 & 42.4304300932 & 0.194708817687 \\
0.475 & 42.1851461163 & 0.216179583718 \\
0.95 & 43.7698442025 & 0.212959739832 \\
1.3 & 44.9491044223 & 0.351060245309 \\
1.305 & 44.4157049301 & 0.292031859582 \\
0.526 & 43.0211775986 & 0.234516092908 \\
0.735 & 43.1143549007 & 0.18304597653 \\
1.14 & 44.0289616697 & 0.770661375912 \\
1.551 & 45.3317386715 & 0.395907584039 \\
1.265 & 44.6312640583 & 0.231966862578 \\
1.34 & 44.953151304 & 0.252440440679 \\
0.84 & 43.5620294075 & 0.195661579434 \\
0.468 & 42.4934212383 & 0.156341752732 \\
0.84 & 43.3513315259 & 0.326964462638 \\
0.96 & 43.5868914421 & 0.427260522942 \\
0.8218 & 43.5365504755 & 0.348162814934 \\
0.93 & 44.6348635492 & 0.689119011672 \\
0.451 & 41.8469440412 & 0.136544105548 \\
0.61 & 42.8826976033 & 0.146106297403 \\
0.83 & 44.6922378489 & 0.457964640542 \\
0.707 & 43.320290922 & 0.250954794869 \\
0.415 & 41.7696832434 & 0.12867036751 \\
0.557 & 42.6475499276 & 0.146252717965 \\
0.791 & 43.334857994 & 0.19213955201 \\
0.695 & 43.1559541077 & 0.177154136049 \\
0.633 & 43.200350196 & 0.204733134997 \\
0.2486 & 40.6291787524 & 0.144170347426 \\
0.532 & 42.6510205265 & 0.198544372744 \\
0.331 & 40.9259072634 & 0.124868930577 \\
0.346 & 41.2816358992 & 0.133990721793 \\
0.961 & 44.0860127956 & 0.541232742301 \\
0.613 & 42.9577096308 & 0.147055302427 \\
0.3402 & 41.2627973103 & 0.124216129606 \\
0.983 & 43.9686072633 & 0.873805060546 \\
0.71 & 43.03031003 & 0.184976659207 \\
0.73 & 43.3112856025 & 0.172061162401 \\
0.47 & 42.2950363729 & 0.149449923436 \\
0.62 & 43.0958537208 & 0.150521756378 \\
0.521 & 42.2813323021 & 0.171703576261 \\
0.369 & 41.5099445593 & 0.127128442179 \\
0.571 & 42.4563923477 & 0.1847896727 \\
0.604 & 42.4422921595 & 0.142594201708 \\
0.9271 & 44.7438233798 & 0.616675765778 \\
0.285 & 40.7117195315 & 0.128805577715 \\
0.2912 & 40.7936031593 & 0.128325808339 \\
0.548 & 42.684557193 & 0.163119668307 \\
0.868 & 43.9551353049 & 0.697041426681 \\
0.496 & 42.2194365099 & 0.160441226732 \\
0.811 & 44.0557629679 & 0.356535950064 \\
0.756 & 43.6720421887 & 0.2041097647 \\
0.817 & 43.5678420781 & 0.305034089394 \\
0.752 & 43.1434994817 & 0.179046394457 \\
0.5516 & 42.4072918995 & 0.146846411569 \\
0.3578 & 41.3894284101 & 0.125749074506 \\
1.01 & 44.6896175735 & 0.910412577811 \\
0.741 & 43.4820220235 & 0.161255047619 \\
0.43 & 41.8372805299 & 0.138494545284 \\
0.526 & 42.6736883867 & 0.151797295017 \\
0.592 & 42.524194556 & 0.197075800472 \\
0.905 & 43.686095579 & 0.424695030791 \\
0.949 & 43.4679668471 & 0.467379485606 \\
0.4607 & 41.9741739893 & 0.2613454679 \\
0.3709 & 41.5023209499 & 0.132033007561 \\
0.8 & 44.4199835555 & 0.537991937963 \\
0.679 & 43.4170080051 & 0.175601683429 \\
0.5817 & 42.8374976585 & 0.184185753799 \\
0.55 & 42.3999711394 & 0.158230175273 \\
0.81 & 43.7278491047 & 0.491922061811 \\
0.95 & 43.780780395 & 0.421021094207 \\
0.3373 & 41.1812895877 & 0.124867747916 \\
0.91 & 44.4079361607 & 0.394461837584 \\
0.263 & 40.6399477676 & 0.123324515918 \\
0.643 & 43.0120301671 & 0.161285559543 \\
0.691 & 43.1407175735 & 0.170861512594 \\
0.357 & 41.3588695068 & 0.129068537648 \\
0.721 & 43.1285593884 & 0.185537803758 \\
0.581 & 42.7918076033 & 0.147505707094 \\
0.6268 & 42.7073384153 & 0.145364365689 \\
0.818 & 43.5731368588 & 0.562299153124 \\
0.449 & 41.9045845939 & 0.150334954079 \\
0.688 & 43.0446792667 & 0.179014967059 \\
0.87 & 44.0828430243 & 0.496674095444 \\
0.5043 & 42.3045856196 & 0.141778819311 \\
0.591 & 43.3138714564 & 0.267379474385 \\
0.426 & 41.806666569 & 0.204450457968 \\
0.329 & 41.3757739861 & 0.25327455415 \\
0.531 & 43.0629635087 & 0.301071564198 \\
0.583 & 42.5993396847 & 0.461245287391 \\
0.519 & 43.0898571036 & 0.385243880567 \\
0.401 & 41.6214142647 & 0.259723821426 \\
0.34 & 41.1002984666 & 0.22330741712 \\
0.436 & 41.8355461879 & 0.219396208675 \\
0.363 & 41.3991790454 & 0.206870861224 \\
0.436 & 41.9186081822 & 0.214365467471 \\
0.309 & 41.2854634774 & 0.20604264325 \\
0.342 & 41.3335602868 & 0.20174070167 \\
0.332 & 41.2428570494 & 0.223960203004 \\
0.469 & 42.0869431125 & 0.219485296548 \\
0.239 & 40.4663605358 & 0.193168326915 \\
0.352 & 41.4402007529 & 0.222632458264 \\
0.612 & 42.2601378946 & 0.819156755697 \\
0.631 & 42.5665239435 & 0.270098723797 \\
0.645 & 42.8315316986 & 0.22803549973 \\
0.429 & 41.8321619796 & 0.217936090167 \\
0.497 & 42.0900771628 & 0.23056454709 \\
0.539 & 42.3621361424 & 0.235072095671 \\
0.561 & 42.5988124145 & 0.254763512704 \\
0.41 & 41.3480927299 & 0.217232818458 \\
0.412 & 41.7758022341 & 0.2949457048 \\
0.599 & 42.8544760765 & 0.290620357209 \\
0.619 & 42.9246968389 & 0.246627072238 \\
0.422 & 41.6678602675 & 0.238231296443 \\
0.54 & 42.5656285356 & 0.254777678043 \\
0.401 & 42.072813335 & 0.312075794951 \\
0.218 & 40.3179435116 & 0.198433228858 \\
0.633 & 42.2256486429 & 0.296301030677 \\
0.383 & 41.592556266 & 0.228094106236 \\
0.302 & 41.5660169963 & 0.259288152895 \\
0.34 & 41.1438764437 & 0.199964450355 \\
0.51 & 41.9188202391 & 0.258309489516 \\
0.421 & 41.965762911 & 0.228780937104 \\
0.399 & 41.7087506407 & 0.28376407894 \\
0.493 & 42.122916001 & 0.229109806841 \\
0.687 & 42.832217021 & 0.279368496827 \\
0.502 & 41.8992528194 & 0.313490292477 \\
0.687 & 42.8216812976 & 0.250279752764 \\
0.495 & 42.2021639825 & 0.245006206755 \\
0.603 & 42.6354160597 & 0.273892147864 \\
0.421 & 42.0443812883 & 0.280749747008 \\
0.348 & 41.5494690313 & 0.204875174321 \\
0.213 & 40.3399930846 & 0.204646940237 \\
0.344 & 41.0679683623 & 0.200210760141 \\
0.271 & 40.5536369049 & 0.206465230307 \\
0.564 & 42.3759756991 & 0.400947441397 \\
0.274 & 40.7740469723 & 0.216212348503 \\
0.582 & 43.2137150031 & 0.278992592748 \\
0.68 & 42.8318958632 & 0.247902597912 \\
0.401 & 41.9301329836 & 0.235368495661 \\
0.416 & 41.7521888571 & 0.295167259169 \\
0.286 & 41.387427719 & 0.334584228292 \\
0.314 & 41.0209048393 & 0.287974294878 \\
0.581 & 43.5994710152 & 0.377088903672 \\
0.463 & 41.9546009246 & 0.265451577145 \\
0.341 & 41.092049721 & 0.209704216095 \\
0.671 & 42.1936025518 & 0.331542575911 \\
0.631 & 42.9261055377 & 0.232187948068 \\
0.522 & 42.6690298106 & 0.304891267733 \\
0.368 & 41.4066563292 & 0.204741881488 \\
0.309 & 40.8507297646 & 0.207725830155 \\
0.528 & 42.4787440903 & 0.305372522825 \\
0.268 & 40.7695452633 & 0.197555926698 \\
0.695 & 42.9207400196 & 0.41358331096 \\
0.284 & 40.7849823164 & 0.199874464698 \\
0.508 & 42.1375516415 & 0.212951542267 \\
0.781 & 43.3825529133 & 0.341139699918 \\
0.613 & 42.4302887204 & 0.333098849656 \\
     \hline
    \end{longtable}
    \label{table1}
   \end{center}

% +--------------------------------------------------------------------+
% | Appendix B Page (Optional)                                         |
% +--------------------------------------------------------------------+

\cleardoublepage

\chapter{GRB data}
\label{appendixgrb}

% +--------------------------------------------------------------------+
% | Enter text for your Appendix page in the space below this box.     |
% |                                                                    |
% +--------------------------------------------------------------------+
  \begin{center}
    \begin{longtable}[c]{|c|c|c|c|}
    \caption[GRB data.]{GRB distance measures with 1$\sigma$ upper and lower uncertainties.}\\

\hline \multicolumn{1}{|c|}{$z$} & \multicolumn{1}{c|}{$\bar{r}_{\rm p}(z)$} & \multicolumn{1}{c|}{$\sigma^+$} & \multicolumn{1}{c|}{$\sigma^-$}\\ \hline 
\endfirsthead

\multicolumn{4}{c}%
{{\bfseries \tablename\ \thetable{} -- continued from previous page}} \\
\hline \multicolumn{1}{|c|}{$z$} &
\multicolumn{1}{c|}{$H(z)$} &
\multicolumn{1}{c|}{$\sigma_H$} &
\multicolumn{1}{c|}{$\sigma_H$} \\ \hline 
\endhead

\hline \multicolumn{4}{|r|}{{Continued on next page}} \\ \hline
\endfoot

\hline \hline
\endlastfoot
0.17 & 1.0000 &  &  \\
1.036 & 0.9416 & 0.1688 & 0.1710\\
1.902 & 1.0011 & 0.1395 & 0.1409\\
2.768 & 0.9604 & 0.1801 & 0.1785\\
3.634 & 1.0598 & 0.1907 & 0.1882\\
4.500 & 1.0163 & 0.2555 & 0.2559\\
6.600 & 1.0862 & 0.3339 & 0.3434\\
     \hline
    \end{longtable}
    \label{table1}
   \end{center}

% +--------------------------------------------------------------------+
% | Appendix B Page (Optional)                                         |
% +--------------------------------------------------------------------+

\cleardoublepage

\chapter{Galaxy cluster gas mass fraction data set}
\label{appendixc}

% +--------------------------------------------------------------------+
% | Enter text for your Appendix page in the space below this box.     |
% |                                                                    |
% +--------------------------------------------------------------------+
  \begin{center}
    \begin{longtable}[c]{|l|l|l|}
    \caption[Galaxy cluster gas mass fraction data set.]{Galaly cluster gas mass fraction data set. The cluster ID, redshift $z$, 
             and the measurement of $f_{\rm gas}$ with 1$\sigma$ error.}\\

\hline \multicolumn{1}{|c|}{Cluster} & \multicolumn{1}{c|}{$z$} & \multicolumn{1}{c|}{$f_{\rm gas}\pm \sigma_{\rm f}$} \\ \hline 
\endfirsthead

\multicolumn{3}{c}%
{{\bfseries \tablename\ \thetable{} -- continued from previous page}} \\
\hline \multicolumn{1}{|c|}{Cluster} &
\multicolumn{1}{c|}{$z$} &
\multicolumn{1}{c|}{$f_{\rm gas}\pm 1\sigma_{\rm f}$} \\ \hline 
\endhead

\hline \multicolumn{3}{|r|}{{Continued on next page}} \\ \hline
\endfoot

\hline \hline
\endlastfoot

Abell 1795             &  0.063 & $0.1074\pm0.0075$  \\
Abell 2029             &  0.078 &     $0.1117\pm0.0042$  \\
Abell 478              &  0.088 &    $0.1211\pm0.0053$  \\
PKS0745-191            &  0.103 &    $0.1079\pm0.0124$  \\
Abell 1413             &  0.143 &    $0.1082\pm0.0058$  \\
Abell 2204             &  0.152 &    $0.1213\pm0.0116$  \\
Abell 383              &  0.188 &   $0.0903\pm0.0080$  \\
Abell 963              &  0.206 &    $0.1144\pm0.0102$  \\
RXJ0439.0+0521         &  0.208 &    $0.0917\pm0.0127$  \\
RXJ1504.1-0248         &  0.215 &    $0.1079\pm0.0111$  \\
Abell 2390             &  0.230 &    $0.1257\pm0.0110$  \\
RXJ2129.6+0005         &  0.235 &    $0.1299\pm0.0299$  \\
Abell 1835             &  0.252 &   $0.1197\pm0.0082$  \\
Abell 611              &  0.288 &    $0.1020\pm0.0133$  \\
Zwicky 3146            &  0.291 &     $0.0943\pm0.0163$  \\
Abell 2537             &  0.295 &     $0.0949\pm0.0147$  \\
MS2137.3-2353          &  0.313 &   $0.1106\pm0.0061$  \\
MACSJ0242.6-2132       &  0.314 &   $0.1268\pm0.0131$  \\
MACSJ1427.6-2521       &  0.318 &   $0.1052\pm0.0220$  \\
MACSJ2229.8-2756       &  0.324 &    $0.1452\pm0.0265$  \\
MACSJ0947.2+7623       &  0.345 &    $0.1048\pm0.0196$  \\
MACSJ1931.8-2635       &  0.352 &     $0.1193\pm0.0266$  \\
MACSJ1115.8+0129       &  0.355 &      $0.0925\pm0.0283$  \\
MACSJ1532.9+3021       &  0.363 &      $0.1280\pm0.0162$  \\
MACSJ0011.7-1523       &  0.378 &      $0.1067\pm0.0125$  \\
MACSJ1720.3+3536       &  0.391 & $0.1153\pm0.0151$  \\
MACSJ0429.6-0253       &  0.399 & $0.1375\pm0.0154$  \\
MACSJ0159.8-0849       &  0.404 &   $0.1097\pm0.0160$  \\
MACSJ2046.0-3430       &  0.423 &     $0.1253\pm0.0398$  \\
MACSJ1359.2-1929       &  0.447 & $0.0845\pm0.0290$  \\
MACSJ0329.7-0212       &  0.450 &   $0.1262\pm0.0129$  \\
RXJ1347.5-1144         &  0.451 &    $0.0923\pm0.0078$  \\
3C295                  &  0.461 &   $0.1067\pm0.0096$  \\
MACSJ1621.6+3810       &  0.461 &     $0.0954\pm0.0172$  \\
MACS1427.3+4408        &  0.487 &    $0.1201\pm0.0294$  \\
MACSJ1311.0-0311       &  0.494 &      $0.1066\pm0.0168$  \\
MACSJ1423.8+2404       &  0.539 &      $0.1141\pm0.0086$  \\
MACSJ0744.9+3927       &  0.686 &     $0.1151\pm0.0140$  \\
MS1137.5+6625          &  0.782 &    $0.0716\pm0.0235$  \\
ClJ1226.9+3332         &  0.892 &     $0.0769\pm0.0198$  \\
CL1415.2+3612          &  1.028 &      $0.1086\pm0.0262$  \\
3C186                  &  1.063 &  $0.1340\pm0.0777$  \\

     \hline
    \end{longtable}
    \label{table1}
   \end{center}

\end{document}